\documentclass[twocolumn,aps,amsmath,amssymb,prd]{revtex4}
\usepackage{epsfig}
\usepackage{wasysym}

\def\al{\alpha}
\def\be{\beta}
\def\ga{\gamma}
\def\de{\delta}
\def\ep{\epsilon}
\def\ve{\varepsilon}
\def\ze{\zeta}
\def\et{\eta}
\def\th{\theta}

\def\ka{\kappa}
\def\la{\lambda}

\def\rh{\rho}

\def\si{\sigma}

\def\ta{\tau}

\def\ph{\phi}

\def\ch{\chi}
\def\ps{\psi}
\def\om{\omega}
\def\Ga{\Gamma}
\def\De{\Delta}

\def\La{\Lambda}
\def\Si{\Sigma}
\def\Up{\Upsilon}
\def\Ph{\Phi}

\def\Om{\Omega}
\def\cA{{\cal A}}
\def\cB{{\cal B}}
\def\cC{{\cal C}}

\def\cF{{\cal F}}
\def\cl{{\cal L}}
\def\cL{{\cal L}}

\def\cS{\Si}

\def\mn{{\mu\nu}}

\def\fr#1#2{{{#1} \over {#2}}}
\def\half{{\textstyle{1\over 2}}}

\def\frac#1#2{{\textstyle{{#1}\over {#2}}}}

\def\lsim{\mathrel{\rlap{\lower4pt\hbox{\hskip1pt$\sim$}}
    \raise1pt\hbox{$<$}}}
\def\gsim{\mathrel{\rlap{\lower4pt\hbox{\hskip1pt$\sim$}}
    \raise1pt\hbox{$>$}}}
\def\sqr#1#2{{\vcenter{\vbox{\hrule height.#2pt
         \hbox{\vrule width.#2pt height#1pt \kern#1pt
         \vrule width.#2pt}
         \hrule height.#2pt}}}}

\def\prt{\partial}
\def\lrpartial{\raise 1pt\hbox{$\stackrel\leftrightarrow\partial$}}

\def\lrvec#1{ \stackrel{\leftrightarrow}{#1} }
\def\part2{\partial_\alpha \partial^\alpha}

\def\etal{{\it et al.}}

\def\pt#1{\phantom{#1}}

\def\sss{s^{\mu\nu}}
\def\ttt{t^{\ka\la\mu\nu}}

\def\sb{\overline{s}}
\def\tb{\overline{t}}
\def\ub{\overline{u}}

\def\stw{\tilde{s}}
\def\ttw{\tilde{t}}
\def\utw{\tilde{u}}
\def\Btw{\tilde{B}}

\def\hsy{h_{\mu\nu}}  
\def\nsy{\et_{\mu\nu}}

\def\he{h^E}

\def\xx'{|\vec x -\vec x'|}

\def\vb#1#2{e_{#1}^{{\pt{#1}}#2}}

\def\b2{b^\al b_\al}
\def\ff{\ve}

\newcommand{\beq}{\begin{equation}}
\newcommand{\eeq}{\end{equation}}
\newcommand{\bea}{\begin{eqnarray}}
\newcommand{\eea}{\end{eqnarray}}
\newcommand{\bit}{\begin{itemize}}
\newcommand{\eit}{\end{itemize}}
\newcommand{\rf}[1]{(\ref{#1})}

\begin{document}

\title{Signals for Lorentz Violation in Post-Newtonian Gravity}

\author{Quentin G.\ Bailey and V.\ Alan Kosteleck\'y}

\affiliation{Physics Department, Indiana University, 
Bloomington, IN 47405, U.S.A.}

\date{IUHET 489, March 2006; Physical Review D, in press}

\begin{abstract}

The pure-gravity sector of the minimal Standard-Model Extension 
is studied in the limit of Riemann spacetime. 
A method is developed to extract the modified Einstein field equations 
in the limit of small metric fluctuations about the Minkowski vacuum,
while allowing for the dynamics 
of the 20 independent coefficients for Lorentz violation.
The linearized effective equations 
are solved to obtain the post-newtonian metric.
The corresponding post-newtonian behavior of a perfect fluid 
is studied and applied to the gravitating many-body system.
Illustrative examples of the methodology are provided
using bumblebee models.
The implications of the general theoretical results are studied 
for a variety of existing and proposed gravitational experiments,
including lunar and satellite laser ranging,
laboratory experiments with gravimeters and torsion pendula,
measurements of the spin precession of orbiting gyroscopes,
timing studies of signals from binary pulsars,
and the classic tests involving the perihelion precession
and the time delay of light. 
For each type of experiment considered,
estimates of the attainable sensitivities are provided.
Numerous effects of local Lorentz violation
can be studied in existing or near-future experiments 
at sensitivities ranging from parts in $10^4$
down to parts in $10^{15}$.

\end{abstract}

\maketitle

\section{Introduction}
\label{Introduction}

At the classical level,
gravitational phenomena 
are well described by general relativity,
which has now survived nine decades of experimental 
and theoretical scrutiny.
In the quantum domain,
the Standard Model of particle physics
offers an accurate description
of matter and nongravitational forces.
These two field theories 
provide a comprehensive and successful description of nature.
However,
it remains an elusive challenge to find 
a consistent quantum theory of gravity 
that would merge them into a single underlying unified theory 
at the Planck scale.

Since direct measurements at the Planck scale are infeasible,
experimental clues about this underlying theory are scant. 
One practical approach is to search for properties 
of the underlying theory that could be manifest 
as suppressed new physics effects,
detectable in sensitive experiments 
at attainable energy scales.
Promising candidate signals of this type include ones
arising from minuscule violations of Lorentz symmetry
\cite{cpt04,reviews}. 

Effective field theory is a useful tool for describing
observable signals of Lorentz violation
\cite{kp}.
Any realistic effective field theory 
must contain the Lagrange densities 
for both general relativity and the Standard Model,
possibly along with suppressed operators of higher mass dimension.
Adding also all terms that involve operators for Lorentz violation 
and that are scalars under coordinate transformations
results in an effective field theory 
called the Standard-Model Extension (SME).
The leading terms in this theory include 
those of general relativity
and of the minimally coupled Standard Model,
along with possible Lorentz-violating terms 
constructed from gravitational and Standard-Model fields.

Since the SME is founded on well-established physics
and constructed from general operators,
it offers an approach to describing Lorentz violation
that is largely independent of the underlying theory.
Experimental predictions of realistic theories 
involving relativity modifications
are therefore expressible in terms of the SME
by specifying the SME coefficient values.  
In fact,
the explicit form of all dominant Lorentz-violating terms 
in the SME is known 
\cite{akgrav}.
These terms consist of Lorentz-violating operators
of mass dimension three or four,
coupled to coefficients with Lorentz indices
controlling the degree of Lorentz violation.
The subset of the theory containing 
these dominant Lorentz-violating terms
is called the minimal SME.

Since Lorentz symmetry underlies 
both general relativity and the Standard Model,
experimental searches for violations
can take advantage either of 
gravitational or of nongravitational forces,
or of both.
In the present work,
we initiate an SME-based study of gravitational experiments
searching for violations of local Lorentz invariance.
To restrict the scope of the work to a reasonable size
while maintaining a good degree of generality,
we limit attention here 
to the pure-gravity sector of the minimal SME in Riemann spacetime. 
This neglects possible complexities associated with
matter-sector effects 
and with Riemann-Cartan
\cite{hhkn}
or other generalized spacetimes,
but it includes all dominant Lorentz-violating signals 
in effective action-based metric theories of gravity.

The Minkowski-spacetime limit of the minimal SME
\cite{ck}
has been the focus of various experimental studies,
including ones
with photons \cite{photonexpt,photonth1,photonth2},
electrons \cite{eexpt,eexpt2,eexpt3},
protons and neutrons \cite{ccexpt,spaceexpt,bnsyn},
mesons \cite{hadronexpt},
muons \cite{muexpt},
neutrinos \cite{nuexpt},
and the Higgs \cite{higgs}.
To date,
no compelling evidence for nonzero coefficients 
for Lorentz violation has been found,
but only about a third of the possible signals
involving light and ordinary matter 
(electrons, protons, and neutrons)
have been experimentally investigated,
while some of the other sectors remain virtually unexplored.
Our goal here is to provide the theoretical basis 
required to extend the experimental studies
into the post-newtonian regime,
where asymptotically Minkowski spacetime
replaces the special-relativistic limit.

Nonzero SME coefficients for Lorentz violation 
could arise via several mechanisms.
It is convenient to distinguish two possibilities,
spontaneous Lorentz violation and explicit Lorentz violation.
If the Lorentz violation is spontaneous
\cite{ksp},
the SME coefficients arise from underlying dynamics,
and so they must be regarded as fields
contributing to the dynamics through
the variation of the action. 
In contrast,
if the Lorentz violation is explicit,
the SME coefficients for Lorentz violation 
originate as prescribed spacetime functions
that play no dynamical role.
Remarkably,
the geometry of Riemann-Cartan spacetimes,
including the usual Riemann limit,
is inconsistent with explicit Lorentz violation
\cite{akgrav}.
In principle,
a more general non-Riemann geometry 
such as a Finsler geometry might allow for explicit violation
\cite{akgrav,gyb},
but this possibility remains an open issue at present. 
We therefore limit attention 
to spontaneous Lorentz violation in this work.
Various scenarios for the underlying theory
are compatible with a spontaneous origin for Lorentz violation,
including ones based on
string theory
\cite{ksp},
noncommutative field theories
\cite{ncqed},
spacetime-varying fields
\cite{spacetimevarying},
quantum gravity
\cite{qg},
random-dynamics models
\cite{fn},
multiverses
\cite{bj},
and brane-world scenarios
\cite{brane}.

Within the assumption of spontaneous Lorentz breaking,
the primary challenge to extracting the post-newtonian physics 
of the pure-gravity minimal SME
lies in accounting correctly 
for the fluctuations around the vacuum values
of the coefficients for Lorentz violation,
including the massless Nambu-Goldstone modes
\cite{bkgrav}.
Addressing this challenge is the subject of 
Sec.\ \ref{theory} of this work.
The theoretical essentials for the analysis are presented
in Sec.\ \ref{basics},
while Sec.\ \ref{linearization} 
describes our methodology for obtaining 
the effective linearized field equations
for the metric fluctuations in a general scenario.
The post-newtonian metric of the pure-gravity minimal SME
is obtained in Sec.\ \ref{metric},
and it is used to discuss modifications 
to perfect-fluid and many-body dynamics 
in Sec.\ \ref{dynamics}. 
In recent decades,
a substantial effort has been invested 
in analysing weak-field tests of general relativity 
in the context of post-newtonian expansions of an arbitrary metric,
following the pioneering theoretical works of Nordtvedt and Will
\cite{ppn,cmw}.
Some standard and widely used forms of these expansions
are compared and contrasted to our results 
in Sec.\ \ref{othermetrics}.
The theoretical part of this work concludes
in Sec.\ \ref{bbmodel},
where the key ideas for our general methodology 
are illustrated in the context
of a class of bumblebee models.

The bulk of the present paper concerns the implications
for gravitational experiments
of the post-newtonian metric
for the pure-gravity sector of the minimal SME.
This issue is addressed in Sec.\ \ref{experimental applications}.
To keep the scope reasonable,
we limit attention to leading and subleading effects.
We identify experiments 
that have the potential to measure SME coefficients
for Lorentz violation,
and we provide estimates of the attainable sensitivities.
The analysis begins 
in Sec.\ \ref{general considerations} 
with a description of some general considerations 
that apply across a variety of experiments,
along with a discussion of existing bounds.
In Sec.\ \ref{llr},
we focus on experiments involving laser ranging,
including ranging to the Moon and to artificial satellites.
Section \ref{Earth laboratory experiments}
studies some promising laboratory experiments
on the Earth,
including gravimeter measurements of vertical acceleration
and torsion-pendulum tests of horizontal accelerations.
The subject of Sec.\ \ref{gyroscope experiment}
is the precession of an orbiting gyroscope,
while signals from binary pulsars 
are investigated in Sec.\ \ref{binary pulsars}.
The role of the classic tests of general relativity 
is discussed in Sec.\ \ref{classic}.
To conclude the paper,
a summary of the main results
is provided in Sec.\ \ref{summary},
along with a tabulation of the estimated attainable experimental
sensitivities for the SME coefficients for Lorentz violation. 
Some details of the orbital analysis
required for our considerations of laser ranging 
are relegated to Appendix A.
Throughout this work,
we adopt the notation and conventions of
Ref.\ \cite{akgrav}.

\section{Theory}
\label{theory}

\subsection{Basics}
\label{basics}

The SME action with gravitational couplings
is presented in Ref.\ \cite{akgrav}.
In the general case,
the geometric framework assumed is a Riemann-Cartan spacetime,
which allows for nonzero torsion.
The pure gravitational part of the SME Lagrange density 
in Riemann-Cartan spacetime
can be viewed as the sum of two pieces,
one Lorentz invariant and the other Lorentz violating:
\beq
{\cl}_{\rm gravity} =
{\cl}_{\rm LI} +{\cl}_{\rm LV}.
\label{gravlag}
\eeq
All terms in this Lagrange density are invariant
under {\it observer} transformations,
in which all fields and backgrounds transform.
These include observer local Lorentz transformations
and observer diffeomorphisms
or general coordinate transformations.
The piece ${\cl}_{\rm LI}$ also remains invariant 
under {\it particle} transformations,
in which the localized fields and particles 
transform but the backgrounds remain fixed.
These include particle local Lorentz transformations
and particle diffeomorphisms. 
However,
for vanishing fluctuations of the coefficients for Lorentz violation,
the piece ${\cl}_{\rm LV}$ 
changes under particle transformations 
and thereby breaks Lorentz invariance.

The Lorentz-invariant piece ${\cl}_{\rm LI}$
is a series in powers of 
the curvature $R_{\ka\la\mu\nu}$, 
the torsion $T_{\la\mu\nu}$,
the covariant derivatives $D_\ka$,
and possibly other dynamical fields
determining the pure-gravity properties of the theory.
The leading terms in ${\cl}_{\rm LI}$
are usually taken as the  
Einstein-Hilbert and cosmological terms
in Riemann-Cartan spacetime.
The Lorentz-violating piece ${\cl}_{\rm LV}$
is constructed by combining
coefficients for Lorentz violation
with gravitational field operators to produce individual terms 
that are observer invariant under both local Lorentz 
and general coordinate transformations.
The explicit form of this second piece can also be written
as a series in the curvature, torsion,
covariant derivative, and possibly other fields:
\bea
{\cl}_{\rm LV}
&=&
e (k_T)^{\la\mu\nu} T_{\la\mu\nu}
+ e (k_R)^{\ka\la\mu\nu} R_{\ka\la\mu\nu}
\nonumber\\ &&
+ e (k_{TT})^{\al\be\ga\la\mu\nu} T_{\al\be\ga} T_{\la\mu\nu}
\nonumber\\ &&
+ e (k_{DT})^{\ka\la\mu\nu} D_{\ka} T_{\la\mu\nu}
+\ldots ,
\label{lvlag}
\eea
where $e$ is the determinant of the vierbein $\vb \mu a$.
The coefficients for Lorentz violation
$(k_T)^{\la\mu\nu}$,
$(k_R)^{\ka\la\mu\nu}$,
$(k_{TT})^{\al\be\ga\la\mu\nu}$,
$(k_{DT})^{\ka\la\mu\nu}$
can vary with spacetime position.
Since particle local Lorentz violation is always accompanied
by particle diffeomorphism violation
\cite{bkgrav},
the coefficients for Lorentz violation
also control diffeomorphism violation in the theory.

In the present work,
we focus on the Riemann-spacetime limit of the SME,
so the torsion is taken to vanish.
We suppose that the Lorentz-invariant piece 
of the theory is the Einstein-Hilbert action,
and we also restrict attention 
to the leading-order Lorentz-violating terms. 
The gravitational terms that remain in this limit
form part of the minimal SME.
The basic features of the resulting theory
are discussed in Ref.\ \cite{akgrav},
and those relevant for our purposes
are summarized in this subsection.

The effective action of the minimal SME 
in this limit can be written as
\beq
S = S_{\rm EH} + S_{\rm LV} + S^\prime.
\label{act}
\eeq
The first term in \rf{act} is the Einstein-Hilbert action
of general relativity.  
It is given by
\beq
S_{\rm EH} = \fr {1}{2\ka} \int d^4x e(R- 2 \La),
\label{leh}
\eeq
where $R$ is the Ricci scalar,
$\La$ is the cosmological constant,
and $\ka = 8 \pi G$.
As usual,
in the present context of a Riemann spacetime,
the independent degrees of freedom of the gravitational field
are contained in the metric $g_\mn$.
Since we are ultimately focusing on 
the post-newtonian limit of \rf{act}, 
in which the effects of $\La$ are known to be negligible,
we set $\La=0$ for the remainder of this work.

The second term in Eq.\ \rf{act} contains
the leading Lorentz-violating gravitational couplings.
They can be written as 
\beq
S_{\rm LV} = \fr {1}{2\ka} \int d^4x e (-u R 
+\sss R^T_\mn + \ttt C_{\ka\la\mu\nu}).
\label{llv}
\eeq
In this equation,
$R^T_\mn$ is the trace-free Ricci tensor 
and $C_{\ka\la\mu\nu}$ is the Weyl conformal tensor.  
The coefficients for Lorentz violation 
$s^\mn$ and $t^{\ka\la\mu\nu}$
inherit the symmetries
of the Ricci tensor and the Riemann curvature tensor,
respectively.  
The structure of Eq.\ \rf{llv} 
implies that $s^\mn$ can be taken traceless 
and that the various traces of $t^{\ka\la\mu\nu}$ 
can all be taken to vanish.  
It follows that Eq.\ \rf{llv} 
contains 20 independent coefficients,
of which one is in $u$,
9 are in $s^\mn$, 
and 10 are in $t^{\ka\la\mu\nu}$.

The coefficients $u$, $s^\mn$, and $t^{\ka\la\mu\nu}$
typically depend on spacetime position.
Their nature depends in part on the origin
of the Lorentz violation.
As mentioned in the introduction,
explicit Lorentz violation is incompatible with Riemann spacetime
\cite{akgrav}.
We therefore limit attention 
in this work to spontaneous Lorentz violation
in Riemann spacetime,
for which the coefficients $u$, $s^\mn$, $t^{\ka\la\mu\nu}$
are dynamical fields.
Note that spontaneous local Lorentz violation 
is accompanied by spontaneous diffeomorphism violation,
so as many as 10 symmetry generators can be broken 
through the dynamics,
with a variety of interesting attendant phenomena
\cite{bkgrav}.
Note also that $u$, $s^\mn$, $t^{\ka\la\mu\nu}$
may be composites of fields in the underlying theory.
Examples for this situation are discussed in Sec.\ \ref{bbmodel}.

The third term in Eq.\ \rf{act} is the general matter action
$S^\prime$.
In addition to determining the dynamics of ordinary matter, 
it includes contributions from the coefficients 
$u$, $s^\mn$, $t^{\ka\la\mu\nu}$,
which for our purposes must be considered in some detail.
The action $S^\prime $ could also be taken to include
the SME terms describing Lorentz violation in the matter sector.
These terms, 
given in Ref.\ \cite{akgrav},
include Lorentz-violating matter-gravity couplings
with potentially observable consequences,
but addressing these effects lies beyond the scope 
of the present work.
Here,
we focus instead on effects 
from the gravitational and matter couplings 
of the coefficients 
$u$, $s^\mn$, $t^{\ka\la\mu\nu}$
in Eq.\ \rf{llv}.

Variation with respect to the metric $g_{\mu\nu}$
while holding 
$u$, $s^\mn$, and $t^{\ka\la\mu\nu}$ 
fixed yields the field equations
\beq
G^\mn
-(T^{Rstu})^\mn
= \ka (T_g)^\mn.
\label{meq}
\eeq
In this expression,
\bea
(T^{Rstu})^\mn 
\hskip -5 pt
&\equiv &
-\half D^\mu D^\nu u-\half D^\nu D^\mu u
+ g^\mn D^2 u + u G^\mn
\nonumber\\
&&
+ \half s^{\al\be} R_{\al\be} g^\mn
+ \half D_\al D^\mu s^{\al\nu}
+ \half D_\al D^\nu s^{\al\mu}
\nonumber\\
&&
- \half D^2 s^\mn
-\half g^\mn D_\al D_\be s^{\al\be}
+ \half t^{\al\be\ga\mu} R_{\al\be\ga}^{\pt{\al\be\ga}\nu}
\nonumber\\
&&
+ \half t^{\al\be\ga\nu} R_{\al\be\ga}^{\pt{\al\be\ga}\mu}
+ \half t^{\al\be\ga\de} R_{\al\be\ga\de}g^\mn
\nonumber\\
&&
- D_\al D_\be t^{\mu\al\nu\be}
- D_\al D_\be t^{\nu\al\mu\be},
\label{trst}
\eea
while the general matter energy-momentum tensor 
is defined as usual by
\beq
\half e (T_g)^\mn \equiv \de {\cl}^\prime / \de g_\mn, 
\label{Tg}
\eeq
where $\cl^\prime$ is the Lagrange density 
of the general matter action $S^\prime$. 

\subsection{Linearization}
\label{linearization}

One of the central goals of this work is to 
use the SME to
obtain the newtonian and leading post-newtonian corrections
to general relativity induced by Lorentz violations.
For this purpose,
it suffices to work at linear order in metric fluctuations
about a Minkowski background.
We can therefore adopt the usual asymptotically inertial coordinates
and write
\beq
g_\mn = \nsy + \hsy.
\label{metricexp}
\eeq
In this subsection,
we derive the effective linearized field equations for $\hsy$
in the presence of Lorentz violation. 

\subsubsection{Primary linearization}

A key issue is the treatment of the 
dynamics of the coefficient fields 
$u$, $s^\mn$, and $t^{\ka\la\mu\nu}$.
As described above,
these are assumed to induce spontaneous violation 
of local Lorentz invariance
and thereby acquire vacuum expectation values
$\ub$, $\sb^{\mu\nu}$, and $\tb^{\ka\la\mu\nu}$,
respectively.
Denoting the field fluctuations 
about these vacuum solutions 
as $\utw$, $\stw^\mn$, and $\ttw^{\ka\la\mu\nu}$,
we can write 
\bea
u &=& \ub + \utw,
\nonumber\\
\sss &=& \sb^\mn + \stw^\mn,
\nonumber\\
\ttt &=& \tb^{\ka\la\mu\nu} + \ttw^{\ka\la\mu\nu}.
\label{texp}
\eea
Note that the fluctuations include as a subset
the Nambu-Goldstone modes 
for local Lorentz and diffeomorphism violation,
which are described in Ref.\ \cite{bkgrav}.
For present purposes
it suffices to work at linear order in the fluctuations,
so in what follows nonlinear terms at 
$O(h^2)$, $O(h\utw)$, $O(\stw^2)$, etc.,
are disregarded,
and we adopt the standard practice
of raising and lowering indices on linear quantities 
with the Minkowski metric $\nsy$.

Deriving the effective linearized field equations for $\hsy$
involves applying the expressions \rf{metricexp}-\rf{texp} 
in the asymptotically inertial frame.
It also requires developing methods to account for effects
on $\hsy$ due to the fluctuations
$\utw$, $\stw^\mn$, $\ttw^{\ka\la\mu\nu}$.
In fact, 
as we show below,
five key assumptions about the properties of these fluctuations 
suffice for this purpose.

The first assumption concerns the vacuum expectation values.
We assume
(i): 
{\it 
the vacuum values 
$\ub$, $\sb^{\mu\nu}$, $\tb^{\ka\la\mu\nu}$
are constant in asymptotically inertial cartesian coordinates.}
Explicitly, we take
\bea
\prt_\al \ub &=& 0,
\nonumber\\
\prt_\al \sb^\mn &=& 0,
\nonumber\\
\prt_\al \tb^{\ka\la\mu\nu} &=& 0.
\label{conds}
\eea
More general conditions could be adopted, 
but assumption (i) ensures that translation invariance 
and hence energy-momentum conservation are preserved 
in the asymptotically Minkowski regime.
The reader is cautioned that this assumption is typically 
different from the requirement of covariant constancy.

The second assumption ensures small Lorentz-violating effects.
We suppose (ii): 
{\it the dominant effects are linear in the vacuum values 
$\ub$, $\sb^{\mu\nu}$, $\tb^{\ka\la\mu\nu}$.} 
This assumption has been widely adopted in 
Lorentz-violation phenomenology.
The basic reasoning is that any Lorentz violation in nature 
must be small,
and hence linearization typically suffices.
Note, however, that in the present context 
this assumption applies only to the vacuum values 
of the coefficients for Lorentz violation in the SME action \rf{act}.
Since these coefficients are related 
via undetermined coupling constants
to the vacuum values of the dynamical fields 
in the underlying theory,
assumption (ii) provides no direct information
about the sizes of the latter.

With these first two assumptions,
we can extract the linearized version 
of the field equations \rf{trst}
in terms of $\hsy$ and the fluctuations 
$\utw$, $\stw^\mn$, and $\ttw^{\ka\la\mu\nu}$.
Some calculation shows that
the linearized trace-reversed equations 
can be written in the form
\beq
R_\mn = \ka (S_g)_\mn + \cA_\mn + \cB_\mn.
\label{leq}
\eeq
In this expression,
$(S_g)_\mn$ is the trace-reversed energy-momentum tensor
defined by
\beq
(S_g)_\mn = (T_g)_\mn - \half \nsy (T_g)^\al_{\pt{\al}\al},
\label{trev}
\eeq
where $(T_g)_\mn$ is the linearized energy-momentum tensor 
containing the energy-momentum density of
both conventional matter and the fluctuations 
$\utw$, $\stw^\mn$, and $\ttw^{\ka\la\mu\nu}$.
Also,
the terms $\cA_\mn$ and $\cB_\mn$ 
in Eq.\ \rf{leq}
are given by
\bea
\cA_\mn &=& 
-\prt_\mu \prt_\nu \utw - \frac 12 \nsy \Box \utw
+ \prt_\al \prt_{(\mu} \stw^{\al}_{\pt{\al}\nu)}
-\frac 12 \Box \stw_\mn 
\nonumber\\
&&
+ \frac 14 \nsy \Box \stw^\al_{\pt{\al}\al}
-2 \prt_\al \prt_\be \ttw^{\pt\mu\al\pt\nu\be}_{\mu\pt\al\nu}
+\nsy \prt_\al \prt_\be \ttw^{\al\ga\be}_{\pt{\al\ga\be}\ga}
\nonumber\\
&&
+\sb^{\al}_{\pt{\al}(\mu } \prt_\al \Ga^{\be}_{\pt{\be}\nu)\be} 
+\sb^{\al\be} \prt_\al \Ga_{(\mu\nu)\be}  
-\sb^{\al}_{\pt{\al}(\mu} \prt^\be \Ga_{\nu)\be\al} 
\nonumber\\
&&
+\frac 12 \nsy \sb^{\al}_{\pt{\al}\be} \prt^\ga \Ga^{\be}_{\pt{\be}\ga\al}
-4\tb^{\pt{(\mu}\al\pt{\nu)}\be}_{(\mu\pt{\al}\nu)} 
\prt_\al \Ga^{\ga}_{\pt{\ga}\ga\be}, 
\label{cA}
\eea
and 
\bea
\cB_\mn &=& \ub R_\mn
- \frac 12 \nsy \sb^{\al\be} R_{\al\be} + \sb^{\al}_{\pt{\al}(\mu} R_{\nu)\al}
\nonumber\\
&&
+ 2 \tb^{\al\be\ga}_{\pt{\al\be\ga}(\mu} R_{\al\be\ga\nu)}
- \frac 32 \nsy \tb^{\al\be\ga\de}R_{\al\be\ga\de}
\nonumber\\
&&
- 2 \tb^{\pt{(\mu}\al\pt{\nu)}\be}_{(\mu\pt{\al}\nu)} R_{\al\be}.
\label{cB}
\eea
The connection coefficients appearing in Eq.\ \rf{cA} 
are linearized Christoffel symbols,
where indices are lowered with the Minkowski metric as needed.
The terms $R$, 
$R_{\al\be}$, and $R_{\al\be\ga\de}$ appearing in 
Eqs.\ \rf{leq}-\rf{cB} and elsewhere below 
are understood to be the linearized Ricci scalar, 
Ricci tensor, and Riemann curvature tensor, respectively. 

\subsubsection{Treatment of the energy-momentum tensor}

To generate the effective equation of motion for $\hsy$ alone,
the contributions from the fluctuations 
$\utw$, $\stw^\mn$, $\ttw^{\ka\la\mu\nu}$
must be expressed in terms of 
$\hsy$, its derivatives, and the vacuum values
$\ub$, $\sb^{\mu\nu}$, $\tb^{\ka\la\mu\nu}$.
In general,
this is a challenging task.
We adopt here a third assumption 
that simplifies the treatment of the dynamics
sufficiently to permit a solution
while keeping most interesting cases.
We assume (iii):
{\it the fluctuations 
$\utw$, $\stw^\mn$, $\ttw^{\ka\la\mu\nu}$
have no relevant couplings to conventional matter.}
This assumption is standard in alternate theories of gravity,
where it is desired to introduce new fields to modify gravity 
while maintaining suitable matter properties.
It follows that $(S_g)_\mn$ can be split as
\beq
(S_g)_\mn = (S_M)_\mn + (S_{stu})_\mn,
\label{etsplt}
\eeq
where $(S_M)_\mn$ is the trace-reversed energy-momentum tensor
for conventional matter.

To gain insight about assumption (iii),
consider the contributions of the fluctuations 
$\utw$, $\stw^\mn$, $\ttw^{\ka\la\mu\nu}$
to the trace-reversed energy-momentum tensor $(S_g)_\mn$.
\it A priori, \rm
the fields 
$u$, $s^\mn$, $t^{\ka\la\mu\nu}$
can have couplings to matter currents 
in the action $S^\prime$,
which would affect the energy-momentum contribution of 
$\utw$, $\stw^\mn$, $\ttw^{\ka\la\mu\nu}$.
However,
through the vacuum values 
$\ub$, $\sb^{\mu\nu}$, $\tb^{\ka\la\mu\nu}$,
these couplings would also induce 
coefficients for Lorentz violation
in the SME matter sector, 
which are generically known to be small 
from nongravitational experiments.
The couplings of the fluctuations 
$\utw$, $\stw^\mn$, $\ttw^{\ka\la\mu\nu}$
to the matter currents
are therefore also generically expected to be small,
and so it is reasonable to take the
the matter as decoupled from the fluctuations
for present purposes.

Possible exceptions to the decoupling of the fluctuations
and matter can arise.
For example,
in the case of certain bumblebee models, 
including those described in Sec.\ \ref{bbmodel},
the fluctuations $\stw^\mn$ can be written in terms
of a vacuum value and a propagating vector field.
This vector field can be identified with the photon 
in an axial gauge \cite{bkgrav}.
A sufficiently large charge current $j^\mu$ 
could then generate significant fluctuations $\stw^\mn$,
thereby competing with gravitational effects.
However, 
for the various systems considered in this work,
the gravitational fields dominate over the
electrodynamic fields by a considerable amount,
so it is again reasonable to adopt assumption (iii).

Given the decomposition \rf{etsplt},
the task of decoupling the fluctuations
$\utw$, $\stw^\mn$, $\ttw^{\ka\la\mu\nu}$ in $(S_g)_\mn$
reduces to expressing the partial energy-momentum tensor 
$(S_{stu})_\mn$
in terms of $\hsy$, its derivatives, and the vacuum values
$\ub$, $\sb^{\mu\nu}$, $\tb^{\ka\la\mu\nu}$.
For this purpose,
we can apply a set of four identities,
derived from the traced Bianchi identities,
that are always satisfied by the
gravitational energy-momentum tensor \cite{akgrav}.
The linearized versions of these conditions suffice here.
They read
\beq
\ka \prt^\mu (T_{g})_\mn = 
- \sb^{\al\be} \prt_\be R_{\al\nu}
-2 \tb^{\al\be\ga\de} \prt_\de R_{\al\be\ga\nu}.
\label{bi}
\eeq
Using the fact that $(T_M)_\mn$
is separately conserved,
one can show that the four conditions \rf{bi}
are satisfied by
\bea
\ka (S_{stu})_\mn &=& 
-2 \sb^{\al}_{\pt{\al}(\mu} R_{\nu)\al}
+ \half \sb_\mn R 
+ \nsy \sb^{\al\be} R_{\al\be}
\nonumber\\
&&
- 4 \tb^{\al\be\ga}_{\pt{\al\be\ga}(\mu} R_{\al\be\ga\nu)}
+ 2 \nsy \tb^{\al\be\ga\de} R_{\al\be\ga\de}
\nonumber\\
&&
+ 4 \tb^{\al\pt\mu\be}_{\pt\al\mu\pt\be\nu} R_{\al\be}
+ \cS_\mn .
\label{set2}
\eea
In this equation,
the term $\cS_\mn$ obeys 
\beq
\prt^\mu (\cS_\mn - \frac 12 \nsy \cS^{\al}_{\pt{\al}\al}) = 0.
\label{cT}
\eeq
It represents an independently conserved piece
of the linearized energy-momentum tensor
that is undetermined by the conditions \rf{bi}.

For calculational purposes,
it is convenient in what follows to adopt assumption (iv): 
{\it the independently conserved piece of the 
trace-reversed energy momentum tensor vanishes,
$\cS_\mn=0$}.
Since $\cS_\mn$ is independent 
of other contributions to $(S_{stu})_\mn$,
one might perhaps suspect that it vanishes in most theories,
at least to linear order.
However,
theories with nonzero $\cS_\mn$ do exist and may even be generic.
Some simple examples are discussed in Sec.\ \ref{bbmodel}.
Nonetheless,
it turns out that assumption (iv) suffices for meaningful progress
because in many models 
the nonzero contributions from $\cS_\mn$ 
merely act to scale the effective linearized Einstein equations
relative to those obtained in the zero-$\cS_\mn$ limit.
Models of this type are said to violate assumption (iv) weakly,
and their linearized behavior is closely related to 
that of the zero-$\cS_\mn$ limit.
In contrast,
a different behavior is exhibited by  
certain models with ghost kinetic terms for the basic fields.
These have nonzero contributions to $\cS_\mn$ 
that qualitatively change 
the behavior of the effective linearized Einstein equations
relative to the zero-$\cS_\mn$ limit.
Models with this feature are said to violate assumption (iv) strongly. 
It is reasonable to conjecture 
that for propagating modes a strongly nonzero $\cS_\mn$ 
is associated with ghost fields, 
but establishing this remains an open issue 
lying outside the scope of the present work.

\subsubsection{Decoupling of fluctuations}

At this stage, 
the trace-reversed energy-momentum tensor $(S_g)_\mn$ 
in Eq.\ \rf{leq}
has been linearized and expressed 
in terms of $\hsy$, its derivatives, and the vacuum values
$\ub$, $\sb^{\mu\nu}$, $\tb^{\ka\la\mu\nu}$.
The term $\cB_\mn$ in Eq.\ \rf{leq} already has the
desired form,
so it remains to determine $\cA_\mn$.
The latter explicitly contains the fluctuations
$\stw^\mn$, $\ttw^{\ka\la\mu\nu}$, and $\utw$.
By assumption (iii),
the fluctuations couple only to gravity,
so it is possible in principle
to obtain them as functions of $\hsy$ alone
by solving their equations of motion. 
It follows that $\cA_\mn$ can be expressed
in terms of derivatives of these functions.
Since the leading-order dynamics is controlled by
second-order derivatives, 
the leading-order result for $\cA_\mn$ is also
expected to be second order in derivatives. 
We therefore adopt assumption (v):
{\it the undetermined terms in $\cA_\mn$ are constructed from
linear combinations of two partial spacetime derivatives
of $\hsy$ and the vacuum values
$\nsy$, $\ub$, $\sb^\mn$, and $\tb^{\ka\la\mu\nu}$.}  
More explicitly,
the undetermined terms take the generic form 
$M_\mn^{\pt{\mn}\al\be\ga\de} \prt_\al \prt_\be h_{\ga\de}$.
This assumption ensures 
a smooth match to conventional general relativity 
in the limit of vanishing coefficients 
$M_\mn^{\pt{\mn}\al\be\ga\de}$.

To constrain the form of $\cA_\mn$,
we combine assumption (v) with invariance properties
of the action, 
notably those of diffeomorphism symmetry.
Since we are considering spontaneous symmetry breaking,
which maintains the symmetry of the full equations of motion,
the original particle diffeomorphism invariance can be applied.
In particular,
it turns out that
the linearized particle diffeomorphism transformations
leaves the linearized equations of motion 
\rf{leq} invariant.
To proceed,
note first that 
the vacuum values 
$\ub$, $\sb^\mn$, $\tb^{\ka\la\mu\nu}$
and $\nsy$ are invariant
under a particle diffeomorphism parametrized by $\xi^\mu$
\cite{bkgrav},
while the metric fluctuation transforms as
\beq
\hsy \rightarrow 
\hsy - \prt_\mu \xi_\nu - \prt_\nu \xi_\mu .
\eeq
The form of Eq.\ \rf{set2}
then implies that $(S_g)_\mn$ is invariant at linear order,
and from Eq.\ \rf{cB} we find $\cB_\mn$ is also invariant.
It then follows from Eq.\ \rf{leq} 
that $\cA_\mn$ must be particle diffeomorphism invariant also. 

The invariance of $\cA_\mn$ can also be checked directly. 
Under a particle diffeomorphism,
the induced transformations on the fluctuations 
$\stw^\mn$, $\ttw^{\ka\la\mu\nu}$, $\utw$
can be derived from the original transformation of the fields
$u$, $s^\mn$, and $t^{\ka\la\mu\nu}$
in the same manner that the induced transformation of $h_\mn$ 
is obtained from the transformation for $g_\mn$.
We find
\bea
\ttw^{\ka\la\mu\nu} &\rightarrow&
\ttw^{\ka\la\mu\nu} + \tb^{\al\la\mu\nu} \prt_\al \xi^{\ka}
+\tb^{\ka\al\mu\nu} \prt_\al \xi^{\la}
\nonumber\\
&&
+\tb^{\ka\la\al\nu} \prt_\al \xi^{\mu}
+\tb^{\ka\la\mu\al} \prt_\al \xi^{\nu},
\nonumber\\
\stw^\mn &\rightarrow& \stw^\mn
+ \sb^{\mu\al} \prt_\al \xi^\nu
+\sb^{\al\nu}\prt_\al \xi^\mu,
\nonumber\\
\utw &\rightarrow& \utw.
\label{utrans}
\eea
Using these transformations,
one can verify explicitly that $\cA_\mn$ is invariant 
at leading order under particle diffeomorphisms.

The combination of assumption (v) and 
the imposition of particle diffeomorphism invariance
suffices to extract a covariant form for $\cA_\mn$
involving only the metric fluctuation $\hsy$.
After some calculation,
we find $\cA_\mn$ takes the form
\bea
\cA_\mn &=& -2 a \ub R_\mn 
+ \sb^{\al\be} R_{\al (\mu \nu) \be}
- \sb^\al_{ \pt{\al} (\mu } R_{\nu ) \al}
\nonumber\\
&&
- b\tb^{\al\pt{\mu}\be}_{\pt{\al}\mu \pt{\be}\nu} R_{\al\be}
- \frac 14 b \nsy \tb^{\al\be\ga\de} R_{\al\be\ga\de}
\nonumber\\
&&
- b \tb^{\al\be\ga}_{\pt{\al\be\ga}(\mu} R_{\nu)\ga\al\be}.
\label{cAf}
\eea
Some freedom still remains in the structure of $\cA_\mn$,  
as evidenced in Eq.\ \rf{cAf} through the presence
of arbitrary scaling factors $a$ and $b$.
These can take different values in distinct explicit theories.

We note in passing that,
although the above derivation makes use of
particle diffeomorphism invariance,
an alternative possibility exists. 
One can instead apply observer diffeomorphism invariance,
which is equivalent to invariance
under general coordinate transformations. 
This invariance is unaffected by any stage 
of the spontaneous symmetry breaking,
so it can be adapted to a version of the above reasoning.
Some care is required in this procedure.
For example,
under observer diffeomorphisms the vacuum values 
$\ub$, $\sb^\mn$, $\tb^{\ka\la\mu\nu}$
transform nontrivially,
and the effects of this transformation on
Eqs.\ \rf{conds}
must be taken into account.
In contrast,
the fluctuations
$\stw^\mn$, $\ttw^{\ka\la\mu\nu}$, $\utw$
are unchanged at leading order under observer diffeomorphisms.
In any case,
the result \rf{cAf} provides the necessary structure of $\cA_\mn$ 
when the assumptions (i) to (v) are adopted. 

\subsubsection{Effective linearized Einstein equations}
\label{elee}

The final effective Einstein equations 
for the metric fluctuation $\hsy$ are obtained 
upon inserting 
Eqs.\ \rf{cB}, \rf{set2}, and \rf{cAf} 
into Eq.\ \rf{leq}.
We can arrange the equations in the form
\beq
R_\mn = \ka (S_M)_\mn + \Ph^{\ub}_\mn + \Ph^{\sb}_\mn
+ \Ph^{\tb}_\mn.
\label{fleq}
\eeq
The quantities on the right-hand side are given by
\bea
\Ph^{\ub}_\mn &=& \ub R_\mn,
\nonumber
\\
\Ph^{\sb}_\mn &=& \half \nsy \sb^{\al\be} R_{\al\be}
-2 \sb^\al_{\pt{\al}(\mu} R_{\al\nu)} + \half \sb_\mn R
\nonumber\\
&&
+ \sb^{\al\be} R_{\al\mu\nu\be},
\nonumber
\\
\Ph^{\tb}_\mn &=& 
2\tb^{\al\be\ga}_{\pt{\al\be\ga}(\mu} R_{\nu)\ga\al\be}
+ 2\tb^{\pt{\mu}\al\pt{\nu}\be}_{\mu\pt{\al}\nu} R_{\al\be}
\nonumber\\
&&
+ \frac 12 \nsy \tb^{\al\be\ga\de} R_{\al\be\ga\de}
\nonumber\\
&=& 0.
\label{tphi}
\eea
Each of these quantities is independently conserved, 
as required by the linearized Bianchi identities.
Each is also invariant under particle diffeomorphisms,
as can be verified by direct calculation.
The vanishing of $\Ph^{\tb}_\mn$ at the linearized level
is a consequence of these conditions and
of the index structure of the coefficient $\tb^{\al\be\ga\de}$,
which implies the identity
\cite{eh}
$\tb_{[\al\be}^{\pt{[\al\be}[\ga\de}\de^{\mu]}_{\pt{\mu]}\nu]} = 0$.

In Eq.\ \rf{fleq},
the coefficients of 
$\Ph^{\ub}_\mn$, 
$\Ph^{\sb}_\mn$, 
$\Ph^{\tb}_\mn$
are taken to be unity by convention.
In fact,
scalings can arise from the terms in Eq.\ \rf{cAf} or, 
in the case of models weakly violating assumption (iv),
from a nonzero $\cS_\mn$ term in Eq.\ \rf{set2}.
However,
these scalings can always be absorbed into 
the definitions of the vacuum values 
$\ub$, $\sb^\mn$, $\tb^{\ka\la\mu\nu}$.
In writing Eq.\ \rf{fleq} 
we have implemented this rescaling of vacuum values,
since it is convenient for the calculations to follow.
However,
the reader is warned that 
as a result the vacuum values 
of the fields $u$, $s^\mn$, $t^{\ka\la\mu\nu}$
appearing in Eq.\ \rf{llv}
differ from the vacuum values  
in Eqs.\ \rf{tphi} by a possible scaling
that varies with the specific theory being considered.

Since Eq.\ \rf{fleq} is linearized 
both in $\hsy$ and in the vacuum values,
the solution for $\hsy$ can be split into 
the sum of two pieces,
\beq
\hsy = \he_\mn + \tilde h_\mn,
\label{hexp}
\eeq
one conventional
and one depending on the vacuum values. 
The first term $\he_\mn$ is defined
by the requirement that it satisfy  
\beq
R_\mn = \ka (S_M)_\mn,
\label{eq}
\eeq
which are the standard linearized Einstein
equations of general relativity.
The second piece controls the deviations due to Lorentz violation.
Denoting by $\widetilde R_\mn$
the Ricci tensor constructed with $\tilde h_\mn$,
it follows that this second piece is determined by
the expression
\beq
\widetilde R_\mn = 
\Ph^{\ub}_\mn + \Ph^{\sb}_\mn ,
\label{Rp}
\eeq
where it is understood that 
the terms on the right-hand side
are those of Eqs.\ \rf{tphi} 
in the limit $\hsy \rightarrow \he_\mn$,
so that only terms at linear order in Lorentz violation are kept.

Equation \rf{Rp} is the desired end product
of the linearization process.
It determines the leading corrections to general relativity 
arising from Lorentz violation
in a broad class of theories.
This includes any modified theory of gravity 
that has an action with leading-order contributions 
expressible in the form \rf{llv} 
and satisfying assumptions (i)-(v).
Note that the fields $u$, $s^\mn$, $t^{\ka\la\mu\nu}$ 
can be composite,
as occurs in the bumblebee examples
discussed in Sec.\ \ref{bbmodel}.
Understanding the implications of Eq.\ \rf{Rp}
for the post-newtonian metric
and for gravitational experiments
is the focus of the remainder of this work. 

\section{Post-newtonian Expansion}
\label{post-newtonian expansion}

This section performs a post-newtonian analysis
of the linearized effective Einstein equations \rf{Rp}
for the pure-gravity sector of the minimal SME.
We first present the post-newtonian metric 
that solves the equations.
Next,
the equations of motion for a perfect fluid in this metric 
are obtained.
Applying them to a system of massive self-gravitating bodies
yields the leading-order acceleration 
and the lagrangian in the point-mass limit.
Finally,
a comparison of the post-newtonian metric for the SME
with some other known post-newtonian metrics is provided. 

\subsection{Metric}
\label{metric}

Following standard techniques
\cite{sw},
we expand the linearized effective Einstein equations \rf{eq}
and leading-order corrections \rf{Rp}
in a post-newtonian series.
The relevant expansion parameter is the 
typical small velocity $\overline v$
of a body within the dynamical system, 
which is taken to be $O(1)$.
The dominant contribution to the metric fluctuation $h_\mn$ 
is the newtonian gravitational potential $U$.
It is second order,
$O(2) = \overline v^2 
\approx G \overline M / \overline r$,
where $\overline M$ is the typical body mass 
and $\overline r$ is the typical system distance.
The source of the gravitational field 
is taken to be a perfect fluid,
and its energy-momentum tensor
is also expanded in a post-newtonian series.
The dominant term is the
mass density $\rh$.
The expansion for $h_\mn$ begins at $O(2)$
because the leading-order gravitational equation 
is the Poisson equation 
$\vec \nabla^2 U = -4 \pi G \rh$.

The focus of the present work
is the dominant Lorentz-violating effects.
We therefore restrict attention to the newtonian $O(2)$
and sub-newtonian $O(3)$ corrections
induced by Lorentz violation.
For certain experimental applications,
the $O(4)$ metric fluctuation $h_{00}$ 
would in principle be of interest,
but deriving it requires solving the sub-linearized
theory of Sec.\ \ref{basics} and lies beyond the
scope of the present work.

As might be expected from the form of
$\Ph^{\ub}_\mn$ in Eq.\ \rf{tphi},
which involves a factor $\ub$ multiplying the Ricci tensor,
a nonzero $\ub$ acts merely to scale
the post-newtonian metric derived below.
Moreover,
since the vacuum value $\ub$ is a scalar
under particle Lorentz transformations
and is also constant in asymptotically inertial coordinates,
it plays no direct role in considerations of Lorentz violation.
For convenience and simplicity,
we therefore set $\ub = 0$ in what follows. 
However,
no assumptions are made about the sizes 
of the coefficients for Lorentz violation,
other than assuming they are sufficiently small 
to validate the perturbation techniques adopted in 
Sec.\ \ref{linearization}.
In terms of the post-newtonian bookkeeping,
we treat the coefficients for Lorentz violation as $O(0)$.
This ensures that we keep all possible
Lorentz-violating corrections 
implied by the linearized field equations \rf{fleq} 
at each post-newtonian order considered.

The choice of observer frame of reference
affects the coefficients for Lorentz violation
and must therefore be specified
in discussions of physical effects.
For immediate purposes,
it suffices
to assume that the reference frame chosen for the analysis
is approximately asymptotically inertial
on the time scales relevant for any experiments.
In practice,
this implies adopting
a reference frame that is comoving
with respect to the dynamical system under consideration.
The issue of specifying the observer frame of reference
is revisited in more detail as needed in subsequent sections.

As usual,
the development of the post-newtonian series for the metric
involves the introduction of certain potentials 
for the perfect fluid
\cite{cmw}.
For the pure-gravity sector
of the minimal SME
taken at $O(3)$,
we find that the following potentials are required:
\bea
U &=& G \int d^3x' \fr {\rh (\vec x', t)} {R} ,
\nonumber\\
U^{jk} &=& G \int d^3x' \fr {\rh ( \vec x', t) R^j R^k}{R^3},
\nonumber\\
V^j &=& G \int d^3x' \fr {\rh ( \vec x', t) v^j(\vec x',t)}{R},
\nonumber\\
W^j &=& G \int d^3x' \fr {\rh ( \vec x', t) 
v^k ( \vec x',t) R^k R^j} {R^3},
\nonumber\\
X^{jkl} &=& G \int d^3x' \fr {\rh ( \vec x', t) v^j (\vec x',t) 
R^k R^l} {R^3},
\nonumber\\
Y^{jkl} &=& G \int d^3x' \fr { \rh ( \vec x', t) v^m (\vec x',t) 
R^m R^j R^k R^l} {R^5} ,
\label{pots}
\eea
where $R^j = x^j - x'^j$ and $R= |\vec x - \vec x'|$ with
the euclidean norm.

The potential $U$ is the usual newtonian gravitational potential.
In typical gauges,
the potentials $V^j$, $W^j$ occur in the post-newtonian expansion
of general relativity,
where they control various gravitomagnetic effects.
In these gauges,
the potentials $U^{jk}$, $X^{jkl}$, and $Y^{jkl}$
lie beyond general relativity.
To our knowledge the latter two,
$X^{jkl}$ and $Y^{jkl}$,
have not previously been considered in the literature.
However,
they are needed to construct the contributions 
to the $O(3)$ metric 
arising from general leading-order Lorentz violation.

A `superpotential' $\ch$ defined by
\beq
\ch = -G \int d^3 x' \rh (\vec x', t) R
\label{chi}
\eeq
is also used in the literature
\cite{cmw}.
It obeys the identities
\bea
\prt_j \prt_k \ch &=& U^{jk} - \de^{jk} U, 
\nonumber\\
\prt_0 \prt_j \ch &=& V^j - W^j.
\label{chiident}
\eea
In the present context,
it is convenient to introduce two additional superpotentials.
We define
\bea
\ch^j &=& G \int d^3 x' \rh (\vec x',t) v^j (\vec x', t) R
\label{chij}
\eea
and
\bea
\ps &=& G \int d^3 x' \rh (\vec x',t) v^j (\vec x', t)
R^j R.
\label{psi}
\eea
These obey several useful identities including,
for example,
\bea
\prt_j \prt_k \ch^l &=& \de^{jk} V^l - X^{ljk},
\nonumber \\
\prt_j \prt_k \prt_l \ps &=& 3 \de^{(jk} V^{l)}-3\de^{(jk} W^{l)}
-3X^{(jkl)} + 3Y^{jkl} .
\nonumber
\\
\label{chijident}
\eea
In the latter equation,
the parentheses denote total symmetrization with a factor of $1/3$. 

In presenting the post-newtonian metric,
it is necessary to fix the gauge.
In our context,
it turns out that calculations can be substantially simplified
by imposing the following gauge conditions:
\bea
\prt_j g_{0j} &=& \fr 12 \prt_0 g_{jj},
\nonumber\\
\prt_j g_{jk} &=& \fr 12 \prt_k
(g_{jj} - g_{00}).
\label{gauge2}
\eea
It is understood that these conditions apply to $O(3)$.
Although the conditions \rf{gauge2} 
appear superficially similar to those 
of the standard harmonic gauge
\cite{sw},
the reader is warned that in fact they differ at $O(3)$.

With these considerations in place,
direct calculation now yields the post-newtonian metric at $O(3)$
in the chosen gauge.
The procedure is to break the effective linearized equations \rf{Rp}
into temporal and spatial components,
and then to use the usual Einstein equations
to eliminate the pieces $h^E_{\mu\nu}$
of the metric on the right-hand side 
in favor of the potentials \rf{pots} in the chosen gauge,
keeping appropriate track of the post-newtonian orders.
The resulting second-order differential equations 
for $\tilde h_{\mu\nu}$
can be solved in terms of the potentials \rf{Rp}.

After some work,
we find that the metric 
satisfying Eqs.\ \rf{eq} and \rf{Rp} 
can be written at this order as 
\bea
g_{00} &=& -1 + 2U + 3 \sb^{00} U 
+\sb^{jk} U^{jk} - 4 \sb^{0j} V^j + O(4), 
\nonumber\\
\label{g00}
\\
g_{0j} &=& -\sb^{0j}U - \sb^{0k} U^{jk} 
- \frac 72 (1 + \frac {1}{28} \sb^{00})V^j 
+\frac 34 \sb^{jk} V^k 
\nonumber\\
&&
- \frac 12 (1+\frac {15}{4} \sb^{00})W^j
+\frac 54 \sb^{jk} W^k
\nonumber\\
&&
+\frac 94 \sb^{kl} X^{klj}
-\frac {15}{8} \sb^{kl} X^{jkl}
-\frac 38 \sb^{kl} Y^{klj},
\label{g0j}
\\
g_{jk} &=& \de^{jk} 
+ (2 - \sb^{00})\de^{jk} U
\nonumber\\
&& 
+ ( \sb^{lm} \de^{jk} 
- \sb^{jl} \de^{mk}
-\sb^{kl} \de^{jm}
+ 2\sb^{00} \de^{jl} \de^{km} ) U^{lm}.
\nonumber\\
\label{gjk}
\eea
Although they are unnecessary 
for a consistent $O(3)$ expansion,
the $O(3)$ terms for $g_{0j}$ 
and the $O(2)$ terms for $g_{jk}$
are displayed explicitly
because they are useful for part of the analysis to follow.
The $O(4)$ symbol in the expression for $g_{00}$
serves as a reminder of the terms missing
for a complete expansion at $O(4)$.
Note that the metric potentials 
for general relativity in the chosen gauge
are recovered upon setting
all coefficients for Lorentz violation to zero,
as expected. 
Note also that a nonzero $\bar u$ would merely act
to scale the potentials in the above equations
by an unobservable factor $(1+\bar u)$.

The properties of this metric under spacetime transformations 
are induced from those of the SME action.
As described in Sec.\ \ref{basics},
two different kinds of spacetime transformation
can be considered:
observer transformations and particle transformations.
The SME is invariant under observer transformations,
while the coefficients for Lorentz violation
determine both the particle local Lorentz violation
and the particle diffeomorphism violation
in the theory.
Since the SME includes all observer-invariant sources 
of Lorentz violation,
the $O(3)$ post-newtonian metric of the minimal SME 
given in Eqs.\ \rf{g00}-\rf{gjk} must have 
the same observer symmetries 
as the $O(3)$ post-newtonian metric of general relativity.

One relevant set of transformations 
under which the metric of general relativity is covariant 
are the post-galilean transformations 
\cite{pg}.
These generalize the galilean transformations 
under which newtonian gravity is covariant.
They correspond to Lorentz transformations 
in the asymptotically Minkowski regime.
A post-galilean transformation 
can be regarded as the post-newtonian product
of a global Lorentz transformation 
and a possible gauge transformation applied 
to preserve the chosen post-newtonian gauge. 
Explicit calculation verifies that 
the $O(3)$ metric of the minimal SME is unchanged 
by an observer global Lorentz transformation,
up to an overall gauge transformation
and possible effects from $O(4)$.
This suggests that the post-newtonian metric 
of the minimal SME indeed takes the same form 
\rf{g00}-\rf{gjk}
in all observer frames related by post-galilean transformations,
as expected.

In contrast, 
covariance of the minimal-SME metric \rf{g00}-\rf{gjk}
fails under a particle post-galilean transformation,
despite the freedom to perform gauge transformations.
This behavior can be traced to the invariance 
(as opposed to covariance)
of the coefficients for Lorentz violation
appearing in Eqs.\ \rf{g00}-\rf{gjk} 
under a particle post-galilean transformation,
which is a standard feature of vacuum values arising 
from spontaneous Lorentz violation.
The metric of the minimal SME 
therefore breaks the particle post-galilean symmetry 
of ordinary general relativity.

\subsection{Dynamics}
\label{dynamics}

Many analyses of experimental tests
involve the equations of motion of the gravitating sources.
In particular,
the many-body equations of motion
for a system of massive bodies
in the presence of Lorentz violation
are necessary for the tests we consider 
in this work.
Here,
we outline the description of massive bodies as perfect fluids
and obtain the equations of motion and action
for the many-body dynamics
in the presence of Lorentz violation.

\subsubsection{The post-newtonian perfect fluid}

Consider first the description of each massive body.
Adopting standard assumptions
\cite{cmw},
we suppose the basic properties of each body 
are adequately described
by the usual energy-momentum tensor 
$(T_M)^\mn$ for a perfect fluid.
Given the fluid element four-velocity $u^\mu$, 
the mass density $\rh$, 
the internal energy $\Pi$,
and the pressure $p$,
the energy-momentum tensor is
\beq
(T_M)^\mn = (\rh + \rh \Pi + p)u^\mu u^\nu + p g^\mn.
\label{tfl}
\eeq
The four equations of motion for the perfect fluid are 
\beq
D_\mu (T_M)^\mn = 0.
\label{fleqns}
\eeq
Note that the construction of the linearized effective
Einstein equations \rf{fleq} in Sec.\ \ref{linearization} 
ensures that this equation is satisfied in our context.

To proceed,
we separate the temporal and spatial components
of Eq.\ \rf{fleqns}
and expand the results in a post-newtonian series
using the metric of the minimal SME 
given in Eqs.\ \rf{g00}-\rf{gjk},
together with the associated Christoffel symbols.
As usual,
it is convenient 
to define a special fluid density $\rh^*$ 
that satisfies the continuity equation
\beq
\prt_0 \rh^* + \prt_j (\rh^* v^j)=0.
\label{rho*}
\eeq
Explicitly,
we have 
\beq
\rh^* = \rh (-g)^{1/2} u^0,
\label{rhodef}
\eeq
where $g$ is the determinant of the post-newtonian metric.

For the temporal component $\nu=0$, 
we find 
\bea
0 &=& \prt_0 [\rh^* (1 + \Pi + \frac 12 v^j v^j + U)]
\nonumber\\
&&
+\prt_j [\rh^* (1 + \Pi + \frac 12 v^k v^k + U)v^j + p v^j]
\nonumber\\
&&
-\rh^* \prt_0 U - 2 \rh^* v^j \prt_j U
\nonumber\\
&&
+\frac 12 \prt_0 [\rh^* (3\sb^{00} U + \sb^{jk} U^{jk})]
\nonumber\\
&&
+\frac 12 \prt_j [\rh^* (3 \sb^{00} U + \sb^{kl} U^{kl})v^j]
\nonumber\\
&&
-\frac 12 \rh^* \prt_0 (3 \sb^{00} U + \sb^{jk} U^{jk}) 
\nonumber\\
&&
-\rh^* v^j \prt_j (3 \sb^{00} U + \sb^{kl} U^{kl}).
\label{euler}
\eea
The first four terms of this equation
reproduce the usual generalized Euler equations 
as expressed in the post-newtonian approximation.
The terms involving $\sb^\mn$
represent the leading corrections due to Lorentz violation.

The key effects on the perfect fluid 
due to Lorentz violation
arise in the spatial components $\nu=j$ of Eq.\ \rf{fleqns}.
These three equations
can be rewritten in terms of $\rh^*$
and then simplified by using the continuity equation \rf{rho*}.
The result is an expression describing 
the acceleration of the fluid elements.
It can be written in the form 
\beq
\rh^* \fr {d v^j}{dt} +A^j_E +A^j_{LV}=0,
\label{accdens}
\eeq
where 
$A^j_E$ is the usual set of terms arising in general relativity 
and 
$A^j_{LV}$ contains terms that violate Lorentz symmetry.

For completeness,
we keep terms in $A^j_E$ to $O(4)$ 
and terms in $A^j_{LV}$ to $O(3)$ 
in the post-newtonian expansion.
This choice preserves the dominant corrections
to the perfect fluid equations of motion due to Lorentz violation.
Explicitly,
we find that $A^j_E$ is given by 
\bea
A^j_E &=& - \rh^* \prt_j U + \prt_j [p (1+3U)]
\nonumber\\
&&
-(\prt_j p) (\Pi + \frac 12 v^k v^k + p/\rh^*)
\nonumber\\
&&
-v^j (\rh^* \prt_0 U - \prt_0 p)
\nonumber\\
&&
+4 \rh^* d(U v^j - V^j)/dt 
\nonumber\\
&&
+ \frac 12 \rh^* \prt_0 (V^j - W^j) 
\nonumber\\
&&
+ 4 \rh^* v^k \prt_j V^k
- \rh^* \prt_j \Ph
\nonumber\\
&&
-\rh^* (\prt_j U)( v^k v^k + 3p/\rh^*),
\label{AE}
\eea
where $\Ph$ is a perfect-fluid potential at $O(4)$ given by
\bea
\Ph &=& G \int d^3x' \fr {\rh (\vec x', t)}{R}
[2 v^k v^k (\vec x', t) + 2 U(\vec x', t) 
\nonumber\\
&& 
\pt{\int d^3x' \rh}
+ \Pi(\vec x', t)
+ 3 p (\vec x', t)/\rh (\vec x', t)].
\label{potO4}
\eea
Similarly,
the Lorentz-violating piece $A^j_{LV}$ is found to be 
\bea
A^j_{LV} &=& - \frac 12 \rh^* \prt_j (3\sb^{00} U + \sb^{kl} U^{kl})
\nonumber\\
&&
- \rh^* \prt_0 ( \sb^{j0} U + \sb^{0k} U^{jk})
\nonumber\\
&&
- \rh^* v^k \prt_k ( \sb^{j0} U + \sb^{0l} U^{jl})
\nonumber\\
&&
+ \rh^* v^k \prt_j ( \sb^{0k} U + \sb^{0l} U^{kl})
\nonumber\\
&&
+ 2 \rh^* \prt_j \sb^{0k} V^k.
\label{ALV}
\eea

We note in passing that Eqs.\ \rf{rho*} and \rf{accdens}
can be used to demonstrate that the standard formula 
\beq
\fr{d\Pi}{dt} = -p \prt_k v^k
\eeq
remains valid in the presence of Lorentz violation. 

\subsubsection{Many-body dynamics}

As a specific relevant application of these results,
we seek the equations of motion for 
a system of massive bodies described as a perfect fluid.
Adopting the general techniques of Ref.\ \cite{cmw},
the perfect fluid can be separated into 
distinct self-gravitating clumps of matter,
and the equations \rf{accdens} 
can then be appropriately integrated 
to yield the coordinate acceleration of each body.

The explicit calculation of the acceleration
requires the introduction
of various kinematical quantities for each body.
For our purposes,
it suffices to define the conserved mass of the $a$th body as
\beq
m_a = \int d^3x' \rh^* (t, \vec x'),
\label{ma} 
\eeq
where $\rh^*$ is specified by Eq.\ \rf{rhodef}.
Applying the continuity properties of $\rh^*$,
we can introduce the position $x^j_a$, 
the velocity $v^j_a$, 
and the acceleration $a^j_a$ of the $a$th body as 
\bea
x^j_a &=& m^{-1}_a \int d^3x \rh^* (t, \vec x) x^j,
\label{xa}
\\
v^j_a &=& \fr {dx^j_a}{dt} 
= m^{-1}_a \int d^3x \rh^* (t, \vec x) \fr {dx^j}{dt},
\label{va}
\\
a^j_a &=& \fr {d^2 x^j_a}{dt^2}
= m^{-1}_a \int d^3x \rh^* (t, \vec x) \fr {d^2 x^j}{dt^2}.
\label{aa}
\eea
The task is then to insert Eq.\ \rf{accdens} into Eq.\ \rf{aa}
and integrate. 

To perform the integration,
the metric potentials are split
into separate contributions from each body
using the definitions \rf{xa} and \rf{va}.
We also expand each potential in a multipole series. 
Some of the resulting terms involve tidal contributions 
from the finite size of each body,
while others involve integrals over each individual body.
In conventional general relativity,
the latter vanish if equilibrium conditions
are imposed for each body.
In the context of general metric expansions
it is known that some parameters may introduce
self accelerations
\cite{cmw},
which would represent a violation 
of the gravitational weak equivalence principle.
However,
for the gravitational sector of the minimal SME,
we find that no self contributions arise
to the acceleration $a^j_a$ at post-newtonian order $O(3)$
once standard equilibrium conditions for each body
are imposed.

For many applications in subsequent sections,
it suffices to disregard
both the tidal forces across each body 
and any higher multipole moments.
This corresponds to taking a point-particle limit.
In this limit, 
some calculation reveals that
the coordinate acceleration of the $a$th body
due to $N$ other bodies is given by
\bea
\fr {d^2x^j_a}{dt^2} &=& -(1 + \frac 32 \sb^{00})
\sum_{b \neq a} \fr { G m_b r^j_{ab}}{r^3_{ab}}
\nonumber\\
&&
+\sb^{jk} \sum_{b \neq a} \fr {G m_b r^k_{ab}} {r^3_{ab}}
\nonumber\\
&&
- \frac 32 \sb^{kl} \sum_{b \neq a} 
\fr {G m_b r^j_{ab} r^k_{ab} r^l_{ab}} {r^5_{ab}}
\nonumber\\
&&
+ \sb^{0j} \sum_{b \neq a} \fr {G m_b (v^k_b r^k_{ab}
- 2 v^k_a r^k_{ab}) } {r^3_{ab}}
\nonumber\\
&&
+\sb^{0k} \sum_{b \neq a} \fr {G m_b ( 2 v^k_a r^j_{ab}
- v^j_b r^k_{ab}+ v^k_b r^j_{ab})}{r^3_{ab}}
\nonumber\\
&&
+3 \sb^{0k} \sum_{b \neq a} \fr {G m_b v^l_b r^l_{ab} 
r^j_{ab} r^k_{ab}} {r^5_{ab}} 
\nonumber\\
&&
+ \ldots . 
\label{nbody}
\eea
The ellipses represent additional acceleration terms
arising at $O(4)$ in the post-newtonian expansion. 
These higher-order terms include 
both corrections from 
general relativity arising via Eq.\ \rf{AE}
and $O(4)$ effects controlled 
by SME coefficients for Lorentz violation.

Note that these point-mass equations
can be generalized to include tidal forces 
and multipole moments using standard techniques.
As an additional check,
we verified the above result is also obtained
by assuming that each body travels along a geodesic
determined by the presence of all other bodies.
The geodesic equation for the $a$th body then yields
Eq.\ \rf{nbody}.

The coordinate acceleration \rf{nbody}
can also be derived as the equation of motion 
from an effective nonrelativistic action principle
for the $N$ bodies,
\beq
S=\int dt L.
\eeq
A calculation reveals that the associated lagrangian $L$
can be written explicitly as
\bea
L 
\hskip -4pt
&=&
\frac 12 \sum_{a} m_a \vec v^2_a 
+ \frac 12 \sum_{ab} \fr {G m_a m_b} {r_{ab}} (1+\frac 32 \sb^{00} 
+ \frac 12 \sb^{jk} \hat r^j_{ab} \hat r^k_{ab} )
\nonumber\\
&&
-\frac 12 \sum_{ab} \fr {G m_a m_b }{r_{ab}}( 3 \sb^{0j} v^j_a 
+ \sb^{0j} \hat r^j_{ab} v^k_a \hat r^k_{ab})
\nonumber\\
&&
+ \ldots ,
\label{lpp}
\eea
where it is understood that the summations omit $b=a$.
Note that the $O(2)$ potential has a form
similar to that arising in Lorentz-violating electrostatics
\cite{km,bak}.

Using standard techniques,
one can show that there are
conserved energy and momenta for this system of point masses.  
This follows from the temporal and spatial translational invariance
of the lagrangian $L$,
which in turn is a consequence of the choice 
$\prt_\mu \sb^{\nu\la} = 0$ 
in Eq.\ \rf{conds}.

As an illustrative example,
consider the simple case of two point masses $m_1$ and $m_2$
in the limit where only $O(3)$ terms are considered.
The conserved hamiltonian can be written
\bea
H &=& \fr {\vec P^2_1}{2 m_1} + \fr {\vec P^2_2}{2 m_2}
- \fr {G m_1 m_2}{r} (1 + \frac 12 \sb^{jk} \hat r^j \hat r^k )
\nonumber\\
&&
+ \fr {3 G \sb^{0j} ( m_2 P^j_1 + m_1 P^j_2)} {2r}
\nonumber\\
&&
+ \fr {G \sb^{0j} r^j ( m_2 P^k_1 + m_1 P^k_2) r^k}{2r^3}.
\label{ham}
\eea
In this expression,
$\vec r = \vec r_1 - \vec r_2$ is the relative separation 
of the two masses,
and the canonical momenta are
given by $\vec P_n = \prt L/ \prt \vec r_n$ for each mass.
It turns out that the total conserved momentum is the sum of 
the individual canonical momenta.
Explicitly,
we find 
\bea
P^j &=& P^j_N - \fr {3 G m_1 m_2 \sb^{0j}}{r}
- \fr {G m_1 m_2 \sb^{0k} r^k r^j}{r^3},
\label{mom1}
\eea
where the first term is the usual newtonian center-of-mass momentum 
$\vec P_N = m_1 \vec v_1 + m_2 \vec v_2$.
Defining $M = m_1 + m_2$, 
we can adopt $\vec V = \vec P_N/M$ as
the net velocity of the bound system.

Further insight can be gained by 
considering the time average of the total conserved momentum 
for the case where the two masses 
are executing periodic bound motion.
If we keep only results 
at leading order in coefficients for Lorentz violation,
the trajectory $\vec r$ for an elliptical orbit
can be used in Eq.\ \rf{mom1}. 
Averaging over one orbit then gives
the mean condition 
\beq
<\vec P> = M <\vec V> + \vec a , 
\label{mom2}
\eeq
where
\beq
a^j = -\fr {G m_1 m_2}{a e^2} 
[3e^2 \sb^{0j} + \ve (e)\sb_P P^j + (1-e^2)^{1/2} \ve (e)\sb_Q Q^j)]
\label{aeff}
\eeq
is a constant vector.
In the latter equation,
$a$ is the semimajor axis of the ellipse,
$\ve (e)$ is a function of its eccentricity $e$,
$\vec P$ points towards the periastron,
while $\vec Q$ is a perpendicular vector determining
the plane of the orbit.
The explicit forms of $\ve (e)$, $\vec P$, and $\vec Q$
are irrelevant for present purposes,
but the interested reader can find them in
Eqs.\ \rf{k} and \rf{eccfns} of Sec.\ \ref{binary pulsars}.
Here,
we remark that 
Eq.\ \rf{mom2} has a parallel in the fermion sector 
of the minimal SME.
For a single fermion with only a nonzero coefficient
for Lorentz violation of the $a^\mu$ type,
the nonrelativistic limit of the motion 
yields a momentum having the same form as that in Eq.\ \rf{mom2} 
(see Eq.\ (32) of the first of Refs.\ \cite{ck}).
Note also that the conservation of $\vec P$ 
and the constancy of $\vec a$ 
imply that the measured mean net velocity $<\vec V>$ 
of the two-body system is constant,  
a result consistent 
with the gravitational weak equivalence principle. 

\subsection{Connection to other post-newtonian metrics}
\label{othermetrics}

This subsection examines the relationship
between the metric of the pure-gravity sector of the minimal SME,
obtained under assumptions (i)-(v) of Sec.\ \ref{linearization},
and some existing post-newtonian metrics.
We focus here on two popular cases,
the parametrized post-newtonian (PPN) formalism
\cite{ppn}
and the anisotropic-universe model 
\cite{nanis}.

The philosophies of the SME
and these two cases are different. 
The SME begins with an observer-invariant action
constructed to incorporate all known physics and fields
through the Standard Model and general relativity.
It categorizes particle local Lorentz violations
according to operator structure.
Within this approach,
the dominant Lorentz-violating effects 
in the pure-gravity sector are controlled
by 20 independent components 
in the traceless coefficients 
$\ub$, $\sb^\mn$, $\tb^{\ka\la\mu\nu}$.
No assumptions about post-newtonian physics are required
\it a priori. \rm
In contrast,
the PPN formalism is based directly on the expansion 
of the metric in a post-newtonian series.
It assumes isotropy in a special frame,
and the primary terms are chosen based on simplicity
of the source potentials.
The corresponding effects involve 10 PPN parameters.
The anisotropic-universe model is again different:
it develops an effective 
$N$-body classical point-particle lagrangian,
with leading-order effects controlled by 11 parameters.
These differences in philosophy and methodology
mean that a complete match cannot be expected
between the corresponding post-newtonian metrics.
However,
partial matches exist,
as discussed below.

\subsubsection{PPN formalism}
\label{ppn}

Since the $O(4)$ terms in $g_{00}$
in the minimal-SME metric \rf{g00} remain unknown,
while many of the 10 PPN parameters appear 
only at $O(4)$ in $g_{00}$,
a detailed comparison of the 
pure-gravity sector of the minimal SME and the PPN formalism
is infeasible at present.
However,
the basic relationship can be extracted  
via careful matching of the known metrics in different frames,
as we demonstrate next.

We first consider the situation in an observer frame
that is comoving with the expansion of the universe,
often called the universe rest frame.
In this frame,
the PPN metric has a special form.
Adopting the gauge \rf{gauge2} 
and keeping terms at $O(3)$,
the PPN metric is found to be 
\bea
g_{00} &=& -1 + 2U + O(4),
\nonumber\\
g_{0j} &=& - \frac 12 (6 \ga + 1 + \frac 12 \al_1) V^j 
-\frac 12 (3 - 2\ga + \frac 12 \al_1) W^j, 
\nonumber\\
g_{jk} &=& \de^{jk} [1 + (3\ga-1) U] - (\ga - 1) U^{jk}.
\label{ppnurf}
\eea
Of the 10 parameters in the PPN formalism,
only two appear at $O(3)$ in this gauge and this frame.
The reader is cautioned that 
the commonly used form of the PPN metric 
in the standard PPN gauge differs from that in Eq.\ \rf{ppnurf}
by virtue of the gauge choice \rf{gauge2}.

The PPN formalism assumes that physics is isotropic
in the universe rest frame.
In contrast,
the SME allows for anisotropies in this frame.
To compare the corresponding post-newtonian metrics in this frame 
therefore requires restricting the SME to an isotropic limit. 
In fact,
the combination of isotropy
and assumption (ii) of Sec.\ \ref{linearization},
which restricts attention to effects
that are linear in the SME coefficients,
imposes a severe restriction:
of the 19 independent SME coefficients contained in
$\sb^\mn$ and $\tb^{\ka\la\mu\nu}$,
only one combination is observer invariant
under spatial rotations and hence isotropic.
Within this assumption,
to have any hope of matching to the PPN formalism,
the SME coefficients must therefore be restricted 
to the isotropic limiting form
\bea
\sb^\mn &=& 
\left(
\begin{array}{cccc} 
\sb^{00} & 0 & 0 & 0 \\
0 & \frac 13 \sb^{00} & 0 & 0 \\
0 & 0 & \frac 13 \sb^{00} & 0 \\
0 & 0 & 0 & \frac 13 \sb^{00} \\
\end{array} 
\right),
\nonumber\\
\tb^{\ka\la\mu\nu} &=& 0.
\label{fc}
\eea
As discussed in Sec.\ \ref{metric},
the coefficient $\ub$ is unobservable in the present context 
and can be set to zero.
In the special limit \rf{fc},
the minimal-SME metric
in Eqs.\ \rf{g00}, \rf{g0j}, and \rf{gjk}
reduces to
\bea
g_{00} &=& -1 + (2 + \frac {10}{3} \sb^{00})U + O(4),
\nonumber\\
g_{0j} &=& - \frac 12 (7 + \sb^{00}) V^j
-\frac 12 (1 + \frac 53 \sb^{00}) W^j,
\nonumber\\
g_{jk} &=& \de^{jk} [1 + (2 - \frac 23 \sb^{00})U]
+\frac 43 \sb^{00} U^{jk}.
\label{smeurf}
\eea

With these restrictions and in the universe rest frame,
a match between the minimal-SME and the PPN metrics 
becomes possible.
The gravitational constant in the restricted minimal-SME metric 
\rf{smeurf} must be redefined as
\beq
G_{\rm new} = G (1 + \frac 53 \sb^{00}).
\label{g}
\eeq
The match is then
\bea
\al_1 &=& -\frac {16} 3 \sb^{00},
\nonumber\\
\ga &=& 1 - \frac 4 3 \sb^{00}.
\label{urfmatch}
\eea

From this match,
we can conclude that 
{\it the pure-gravity sector of the minimal SME
describes many effects that lie outside
the PPN formalism,}
since 18 SME coefficients cannot be matched
in this frame.
Moreover,
the converse is also true.
In principle,
a similar match in the universe rest frame could be made at $O(4)$,
where the minimal-SME metric in the isotropic limit 
can still be written in terms of just one coefficient $\sb^{00}$,
while the PPN formalism requires 10 parameters.
It follows that 
{\it the PPN formalism in turn describes many effects
that lie outside the pure-gravity sector of the minimal SME,}
since 9 PPN parameters cannot be matched
in this frame.
The mismatch between the two 
is a consequence of the differing philosophies:
effects that dominate at the level
of a pure-gravity realistic action
(gravitational sector of the minimal SME) 
evidently differ from those selected by requirements of simplicity
at the level of the post-newtonian metric (PPN).

Further insight about the relationship between
the gravitational sector of the minimal SME and the PPN formalism
can be gained by transforming to 
a Sun-centered frame comoving with the solar system.
This frame is of direct relevance for experimental tests.
More important in the present context,
however,
is that the transformation between the universe rest frame
and the Sun-centered frame mixes terms 
at different post-newtonian order,
which yields additional matching information.

Suppose the Sun-centered frame is moving with 
a velocity $\vec w$ of magnitude $|\vec w| \ll 1$ 
relative to the universe rest frame.
Conversion of a metric from one frame to the other
can be accomplished 
with an observer post-galilean transformation.
A complete transformation would require
using the transformation law of the metric, 
expressing the potentials in the new coordinates,
and transforming also the SME coefficients for Lorentz violation 
or the PPN parameters,
all to the appropriate post-newtonian order.
For the minimal-SME metric \rf{smeurf}
in the isotropic limit,
this procedure would include transforming $\sb^{00}$,
including also the change in 
the new gravitational coupling $G_{\rm new}$ 
via its dependence on $\sb^{00}$.
However,
for some purposes it is convenient
to perform only the first two of these steps.
Indeed,
in the context of the PPN formalism,
this two-step procedure 
represents the standard choice adopted in the literature 
\cite{cmw}.
In effect, 
this means the PPN parameters appearing in the PPN metric
for the Sun-centered frame 
remain expressed in the universe rest frame.
For comparative purposes,
we therefore adopt in this subsection
a similar procedure for the isotropic limit
of the minimal-SME metric \rf{smeurf}.

Explicitly,
we find that
the PPN metric in the Sun-centered frame 
and in the gauge \rf{gauge2} is given by
\bea
g_{00} &=& -1 + 2U -(\al_1-\al_2-\al_3)w^2 U 
\nonumber\\
&&
-\al_2 w^j w^k U^{jk} +(2\al_3-\al_1)w^j V^j 
\nonumber\\
&&
+(2\al_2 - \frac 12 \al_1) w^j (V^j-W^j) + O(4),
\nonumber\\
g_{0j} &=& -\frac 14 \al_1 w^j U -\frac 14 \al_1 w^k U^{jk}
\nonumber\\
&&
- \frac 12 (6 \ga + 1 + \frac 12 \al_1) V^j 
-\frac 12 (3 - 2\ga + \frac 12 \al_1) W^j, 
\nonumber\\
g_{jk} &=& \de^{jk} [1 + (3\ga-1) U] - (\ga - 1) U^{jk}.
\label{ppnscf}
\eea
This expression depends on four of the 10 PPN parameters.
It includes all terms at $O(3)$ 
and also all terms in $g_{00}$ dependent on $\vec w$.
Since $\vec w$ is $O(1)$,
some of the latter are at $O(4)$,
including those involving the two parameters 
$\al_2$ and $\al_3$ that
are absent from the PPN metric \rf{ppnurf}
in the universe rest frame.
The dependence on $\vec w$ is a key feature
that permits a partial comparison of these $O(4)$ terms
with the minimal SME
and hence provides matching information
that is unavailable in the universe rest frame.
 
For the minimal SME,
the isotropic limit of the pure-gravity sector
in the Sun-centered frame and the gauge \rf{gauge2} 
is found to be 
\bea
g_{00} &=& -1 + (2 + \frac {10}{3} \sb^{00})U 
+ 4 w^2 \sb^{00}U 
\nonumber\\
&&
+ \frac 43 w^j w^k \sb^{00} U^{jk} 
+\frac {16}{3} \sb^{00} w^j V^j +O(4),
\nonumber\\
g_{0j} &=& \frac 43 \sb^{00} w^j U + \frac 43 \sb^{00} w^k U^{jk}
\nonumber\\
&&
-\frac 12 (7 + \sb^{00}) V^j
-\frac 12 (1 + \frac 53 \sb^{00}) W^j,
\nonumber\\
g_{jk} &=& \de^{jk} [1 + (2 - \frac 23 \sb^{00})U]
+\frac 43 \sb^{00} U^{jk},
\label{smescf}
\eea
where the coefficient $\sb^{00}$ remains defined
in the universe rest frame.
Since the general results \rf{g00}-\rf{gjk}
for the minimal-SME metric 
take the same form in any post-galilean observer frame,
Eq.\ \rf{smescf} can be derived
by transforming the coefficient $\sb^{00}$ 
from the universe rest frame
to the corresponding coefficients $\sb^{\mu\nu}$
in the Sun-centered frame
and then substituting the results into
Eqs.\ \rf{g00}-\rf{gjk}.
Like the PPN metric \rf{ppnscf},
the expression \rf{smescf} contains all terms at $O(3)$
along with some explicit $O(4)$ terms that depend on $\vec w$. 

By rescaling the gravitational constant as in Eq.\ \rf{g}
and comparing the two post-newtonian metrics
\rf{ppnscf} and \rf{smescf},
we recover the previous matching results \rf{urfmatch}
and obtain two additional relationships: 
\bea
\al_1 &=& -\frac {16} 3 \sb^{00},
\nonumber\\
\al_2 &=& -\frac {4} 3 \sb^{00},
\nonumber\\
\al_3 &=& 0,
\nonumber\\
\ga &=& 1 - \frac 4 3 \sb^{00}.
\label{aaag}
\eea
The vanishing of $\al_3$ is unsurprising.
This parameter is always zero in semiconservative theories
\cite{cmw,lln},
while assumption (i) of Sec.\ \ref{linearization}
and Eq.\ \rf{conds}
imply a constant asymptotic value for $\sb^{00}$,
which in turn ensures global energy-momentum conservation.
A more interesting issue is the generality of the condition 
\beq
\al_1=4\al_2
\label{al12cond}
\eeq
implied by Eq.\ \rf{aaag}.
It turns out that this condition depends 
on assumption (iv) of Sec.\ \ref{linearization},
which imposes the vanishing of the independent
energy-momentum contribution $\Si^{\mu\nu}$. 
The relationship between the conditions $\Si^{\mu\nu} \neq 0$ 
and $\al_1 \neq 4\al_2$ 
is considered further in Sec.\ \ref{bbmodel} below.

We emphasize that
all quantities in Eq.\ \rf{aaag}
are defined in the universe rest frame.
For experimental tests of the SME,
however,
it is conventional
to report measurements of the coefficients for Lorentz violation
in the Sun-centered frame. 
The conversion between the two 
takes a simple form 
for the special isotropic limit involved here.
It can be shown that 
the minimal-SME coefficients $\sb^{\mu\nu}_{\rm S}$
in the Sun-centered frame
and the isotropic coefficient $\sb^{00}_{\rm U}$
in the universe rest frame 
are related by 
\bea
\sb^{00}_{\rm S}
&=&
(1 + \frac 4 3 w^j w^j)\sb^{00}_{\rm U},
\nonumber\\
\sb^{0j}_{\rm S}
&=&
-\frac 4 3 w^j \sb^{00}_{\rm U},
\nonumber\\
\sb^{jk}_{\rm S}
&=&
\frac 1 3 (\de^{jk} + 4 w^j w^k)\sb^{00}_{\rm U}.
\label{sssu}
\eea
These equations can be used to relate 
the results in this subsection 
to ones expressed in terms of the coefficients $\sb^{\mu\nu}$
in the Sun-centered frame. 

\begin{figure}
\centerline{\psfig{figure=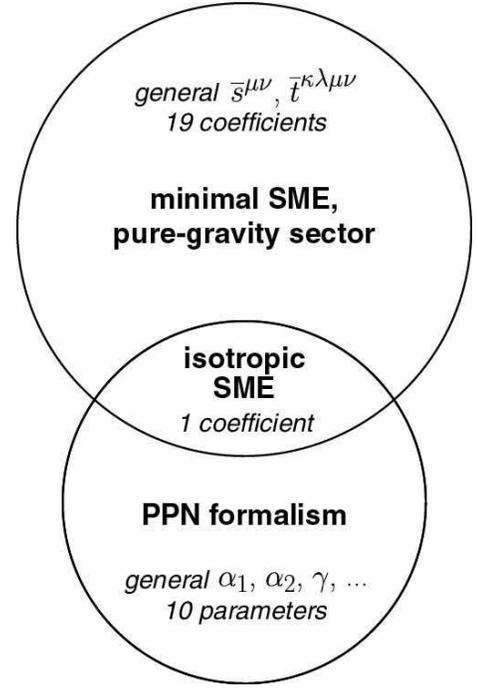,width=0.7\hsize}}
\caption{\label{Venn_v8} 
Schematic relationship 
between the pure-gravity sector of the minimal SME 
and the PPN formalism.}
\end{figure}

The relationship between the pure-gravity sector
of the minimal SME and the PPN formalism 
can be represented as the Venn diagram in Fig.\ \ref{Venn_v8}.
The overlap region,
which corresponds to the isotropic limit,
is a one-parameter region in the universe rest frame.
This overlap region encompasses only a small portion
of the effects governed by 
the pure-gravity sector of the minimal SME 
and by the PPN formalism. 
Much experimental work has been done to explore the PPN parameters.
However,
the figure illustrates that
a large portion of coefficient space 
associated with dominant effects in a realistic action
(SME) remains open for experimental exploration.
We initiate the theoretical investigation 
of the various possible experimental searches for these effects
in Sec.\ \ref{experimental applications}.

We note in passing that the above matching considerations
are derived for the pure-gravity sector of the minimal SME.
However,
even in the isotropic limit,
the matter sector of the minimal SME contains
numerous additional coefficients for Lorentz violation.
Attempting a match between the isotropic limit
of the minimal SME with matter
and the PPN formalism would also be of interest
but lies beyond our present scope.

\subsubsection{Anisotropic-universe model}
\label{aumodel}

An approach adopting a somewhat different philosophy
is the anisotropic-universe model 
\cite{nanis}.
This model is conservative and is based on the formulation 
of an effective classical point-particle lagrangian.
Possible anisotropies in a preferred frame
are parametrized via a set of 
spatial two-tensors and spatial vectors,
and the post-newtonian metric is constructed.
The model includes velocity-dependent terms 
and three-body interactions up to $O(4)$.
The connection to the PPN formalism
is discussed in Ref.\ \cite{nanis}.

A detailed match between
the anisotropic-universe model 
and the classical point-particle limit of the
pure-gravity sector of the minimal SME 
is impractical at present,
since the $O(4)$ terms for the latter are undetermined.
However,
some suggestive features of the relationship can be obtained.
For this purpose,
it suffices to consider the restriction 
of the anisotropic-universe model
to two-body terms at $O(3)$.
In this limit, 
the lagrangian is
\bea
L_{\rm AU} &=& \frac 12 \sum_{a} m_a \vec v^2_a 
+ \frac 12 \sum_{ab} \fr {G m_a m_b} {r_{ab}} 
(1 + 2 \Om^{jk} \hat r^j_{ab} \hat r^k_{ab} )
\nonumber\\
&&
+ \sum_{ab} \fr {G m_a m_b }{r_{ab}}( \Ph^{j} v^j_a 
+ \Up^{j} \hat r^j_{ab} v^k_a \cdot \hat r^k_{ab}).
\label{au}
\eea
The anisotropic properties of the model
at this order are controlled by 11 parameters,
collected into one symmetric spatial two-tensor 
$\Om^{jk}$
and two three-vectors 
$\Ph^{j}$, $\Up^{j}$.

To gain insight into the relationship,
we can compare Eq.\ \rf{au}
with the $O(3)$ point-particle lagrangian \rf{lpp}
obtained in Sec.\ \ref{dynamics}.
Consider the match in the 
preferred frame of the anisotropic-universe model.
Comparison of Eqs.\ \rf{au} and \rf{lpp}
gives the correspondence 
\bea
\Om^{jk} &=& \frac 1 4\sb^{jk} - \frac 1{12}\sb^{00}\de^{jk}, 
\nonumber\\
\Ph^{j} &=& -\frac 32 \sb^{0j},
\nonumber\\
\Up^{j} &=& -\half \sb^{0j}.
\label{smatch}
\eea
The five parameters $\Om^{jk}$
are determined by the SME coefficients $\sb^{jk}$ and $\sb^{00}$,
while the two three-vectors $\Ph^{j}$, $\Up^{j}$
are determined by $\sb^{0j}$.

The SME is a complete effective field theory 
and so contains effects beyond any point-particle description,
including that of the anisotropic-universe model.
Even in the point-particle limit,
it is plausible that the pure-gravity sector of the minimal SME
describes effects outside the anisotropic-universe model
because the latter is based on two-tensors and vectors
while the SME contains the four-tensor $\tb^{\ka\la\mu\nu}$.
Nonetheless,
in this limit the converse is also plausible:
the anisotropic-universe model is likely 
to contain effects outside the pure-gravity sector
of the minimal SME.
The point is that 
the match \rf{smatch} implies 
the 11 parameters $\Om^{jk}$, $\Ph^{j}$, $\Up^{j}$
are fixed by only 9 independent SME coefficients $\sb^{\mu\nu}$.
This suggests that two extra degrees of freedom appear
in the anisotropic-universe model already at $O(3)$.
Some caution with this interpretation may be advisable,
as the condition $\Ph^{j}= 3\Up^{j}$
implied by Eq.\ \rf{smatch}
in the context of the minimal SME
arises from the requirement that the form of the lagrangian
be observer invariant under post-galilean transformations.
To our knowledge
the observer transformation properties 
of the parameters in the anisotropic-universe model
remain unexplored in the literature,
and adding this requirement
may remove these two extra degrees of freedom. 
At $O(4)$,
however,
the full anisotropic-universe model contains additional two-tensors 
for which there are no matching additional coefficients 
in the pure-gravity sector of the minimal SME.
It therefore appears likely that the correspondence between 
the two approaches is again one of partial overlap.
It is conceivable that effects from the 
SME matter sector or from nonminimal SME terms 
could produce a more complete correspondence.

\section{Illustration: bumblebee models}
\label{bbmodel}

The analysis presented 
in Secs.\ \ref{theory} and \ref{post-newtonian expansion}
applies to all theories with terms for Lorentz violation 
that can be matched to the general form \rf{llv}.
An exploration of its implications for experiments
is undertaken in Sec.\ \ref{experimental applications}.
Here,
we first take a short detour 
to provide a practical illustration of the general methodology 
and to illuminate the role of the five assumptions adopted
in the linearization procedure of Sec.\ \ref{linearization}.
For this purpose,
we consider a specific class of theories,
the bumblebee models.
Note,
however,
that the material in this section is inessential for
the subsequent analysis of experiments,
which is independent of specific models.
The reader can therefore proceed directly
to Sec.\ \ref{experimental applications} at this stage
if desired.
 
Bumblebee models are theories involving a vector field $B^\mu$
that have spontaneous Lorentz violation
induced by a potential $V(B^\mu)$.
The action relevant for our present purposes 
can be written as 
\bea
S_{B} &=&
\int d^4 x
\Big[ \fr 1 {2\ka} (eR + \xi eB^\mu B^\nu R_{\mu\nu})
\nonumber\\
&&
\hskip -10pt
\pt {\int d^4 x }
- \frac 14 \al e B^{\mu\nu} B_{\mu\nu}
+ \frac 12 \be e (D_\mu B_\nu)(D^\mu B^\nu)
\nonumber\\
&&
\hskip -10pt 
\pt {\int d^4 x }
- eV(B^\mu)
+ \cL_{\rm M} 
\Big],
\label{bb}
\eea
where $\al$ and $\be$ are real.
In Minkowski spacetime and with vanishing potential $V$,
a nonzero value of $\be$ introduces a St\"uckelberg ghost. 
However,
the ghost term with $\be$ potentially nonzero
is kept in the action here 
to illustrate some features of the assumptions made 
in Sec.\ \ref{linearization}. 
In Eq.\ \rf{bb},
the potential $V$ is taken to have the functional form 
\beq
V(B^\mu) = V(B^\mu B_\mu \pm b^2),
\eeq 
where $b^2$ is a real number.
This potential induces a nonzero vacuum value $B^\mu = b^\mu$
obeying $b^\mu b_\mu = \mp b^2$.
The theory \rf{bb} is understood to be taken 
in the limit of Riemann spacetime,
where the field-strength tensor can be written as 
$B_\mn= \prt_\mu B_\nu-\prt_\nu B_\mu$.
Bumblebee models involving nonzero torsion
in the more general context of Riemann-Cartan spacetime
are investigated in Refs.\ \cite{akgrav,bkgrav}.

Theories coupling gravity to a vector field 
with a vacuum value
have a substantial history in the literature.
A vacuum value as a gauge choice
for the photon was discussed by Nambu 
\cite{nambu}.
Models of the form \rf{bb} without a potential term
but with a vacuum value for $B^\mu$ 
and nonzero values of $\al$ and $\be$ 
were considered by 
Will and Nordtvedt
\cite{wn}
and by 
Hellings and Nordtvedt
\cite{hn}.
The idea of using a potential $V$ 
to break Lorentz symmetry spontaneously
and hence to enforce a nonzero vacuum value for $B^\mu$
was introduced by
Kosteleck\'y and Samuel
\cite{ks2},
who studied both the smooth `quadratic' case
$V = - \la e(B^\mu B_\mu \pm b^2)^2/2$ 
and the limiting Lagrange-multiplier case 
$V = - \la e(B^\mu B_\mu \pm b^2)$
for $\be = 0$
and in $N$ dimensions.
The spontaneous Lorentz breaking 
is accompanied by qualitatively new features,
including 
a Nambu-Goldstone sector of massless modes
\cite{bkgrav},
the necessary breaking of U(1) gauge invariance
\cite{akgrav},
and implications for the behavior of the matter sector
\cite{kleh},
the photon 
\cite{bkgrav},
and the graviton 
\cite{akgrav,kpssb}.
More general potentials have also been investigated,
and some special cases with hypergeometric $V$ 
turn out to be renormalizable in Minkowski spacetime
\cite{aknonpoly}.
The situation with a Lagrange-multiplier potential
for a timelike $b_\mu$
and both $\al$ and $\be$ nonzero
has been explored by Jacobson and collaborators 
\cite{jm}.
Related analyses have been performed in 
Refs.\ \cite{bb1,bb2}.

In what follows,
we consider the linearized and post-newtonian limits
of the theory \rf{bb} for some special cases.
The action for the theory contains 
a pure-gravity Einstein piece,
a gravity-bumblebee coupling term controlled by $\xi$,
several terms determining the bumblebee dynamics,
and a matter Lagrange density. 
Comparison of Eq.\ \rf{bb} with the action \rf{llv}
suggests a correspondence between the gravity-bumblebee
coupling and the fields $u$ and $s^\mn$,
with the latter being related to the traceless part of
the product $B^\mu B^\nu$.
It is therefore reasonable to expect the linearization analysis
of the previous sections can be applied,
at least for actions with appropriate bumblebee dynamics.

\subsection{Cases with $\be = 0$}
\label{ghostfree}

Consider first the situation without a ghost term, $\be = 0$.
Suppose for definiteness that $\al = 1$.
The modified Einstein and bumblebee field equations 
for this case are given in Ref.\ \cite{akgrav}.
To apply the general formalism developed 
in the previous sections, 
we first relate the bumblebee action \rf{bb}
to the action \rf{act}
of the pure-gravity sector of the minimal SME,
and then we consider the linearized version 
of the equations of motion.

The match between the bumblebee and SME actions
involves identifying $u$ and $s^\mn$ as composite fields
of the underlying bumblebee field and the metric,
given by
\bea
u &=& \frac 1 4 \xi B^\al B^\be g_{\al\be},
\nonumber\\ 
s^\mn &=& \xi B^\mu B^\nu 
- \frac 1 4 \xi B^\al B^\be g_{\al\be} g^{\mn}.
\label{usmatch}
\eea
This is a nonlinear relationship
between the basic fields of a given model and the SME,
a possibility noted in Sec.\ \ref{basics}.

Following the notation 
of Sec.\ \ref{linearization},
the bumblebee field $B^\mu$ can be expanded around
its vacuum value $b^\mu$ as
\beq
B^\mu = b^\mu + \Btw^\mu.
\label{bvac}
\eeq
In an asympotically cartesian coordinate system,
the vacuum value is taken to obey 
\beq
\prt_\nu b^\mu=0.
\label{bcond}
\eeq
This ensures that assumption (i)
of Sec.\ \ref{linearization} holds.

The expansion of the SME fields $u$, $s^\mn$, $t^{\ka\la\mn}$
about their vacuum values is given in Eq.\ \rf{texp}. 
At leading order,
the $\xi$-dependent term in the bumblebee action \rf{bb}
is reproduced by the Lorentz-violating piece \rf{llv}
of the SME action by making the identifications
\bea
\ub &=& \frac 14 \xi \b2 ,
\nonumber\\
\utw &=& \half \xi (b_\al \Btw^\al + \half b^\al b^\be h_{\al\be}),
\nonumber\\
\sb^\mn &=& \xi ( b^\mu b^\nu - \frac 14 \et^\mn \b2 ),
\nonumber\\
\stw^\mn &=& \xi (b^\mu \Btw^\nu + b^\nu \Btw^\mu 
+ \frac 14 h^\mn \b2 
\nonumber\\
&&
\pt{\xi}
-\frac 12 \et^\mn b_\al \Btw^\al
-\frac 14 \et^\mn b^\al b^\be h_{\al\be}),
\nonumber\\
\tb^{\ka\la\mu\nu} &=& 0,
\nonumber\\
\ttw^{\ka\la\mu\nu} &=& 0.
\label{bexps}
\eea
We see that assumption (ii) of Sec.\ \ref{linearization}
holds if the combination $\xi b^\mu b^\nu$ is small.

The next step is to examine the field equations.
The nonlinear nature of the match \rf{usmatch}
and its dependence on the metric imply that 
the direct derivation of the linearized effective Einstein equations
from the bumblebee theory \rf{bb}
differs in detail from the derivation
of the same equations in terms of $s^\mn$ and $u$
presented in Sec.\ \ref{linearization}.
For illustrative purposes and to confirm the results
obtained via the general linearization process,
we pursue the direct route here.

Linearizing in both the metric fluctuation $\hsy$ and  
the bumblebee fluctuation $\Btw^\mu$
but keeping all powers of the vacuum value $b^\mu$,
the linearized Einstein equations can be extracted
from the results of Ref.\ \cite{akgrav}. 
Similarly,
the linearized equations of motion for the bumblebee fluctuations 
can be obtained.
For simplicity,
we disregard the possibility of direct couplings
between the bumblebee fields and the matter sector,
which means that 
assumption (iii) of Sec.\ \ref{linearization} holds.
The bumblebee equations of motion then take the form 
\beq
\prt^\mu B_\mn = 2V'b_\nu - \frac {\xi}{\ka} b^\mu R_{\mu\nu}, 
\label{bbfe1}
\eeq
where the prime denotes the derivative with respect to the argument.
Once the potential $V$ is specified,
Eq.\ \rf{bbfe1} for the bumblebee fluctuations
can be inverted using Fourier decomposition in momentum space.

Consider,
for example,
the situation for a smooth potential 
$V = - \la e(B^\mu B_\mu \pm b^2)^2/2$.
For this case,
Eq.\ \rf{bbfe1} can be written as
\bea
\lefteqn{(\et_\mn\Box  - \prt_\mu \prt_\nu 
- 4 \la b_\mu b_\nu ) \Btw^\mu = }
\nonumber\\
&&
- b^\al \prt^\mu (\prt_\mu h_{\nu\al}
- \prt_\nu h_{\mu\al})
+ 2 \la b_\nu b^\al b^\be h_{\al\be}
- \frac {\xi}{\ka} b^\al R_{\al\nu}.
\nonumber\\
\label{bbfe2}
\eea
Converting to momentum space,
the propagator for the $\Btw^\mu$ field 
for both timelike and spacelike $b^\mu$
is found to be
\bea
G^\mn (p) &=& -\fr {\et^\mn}{p^2} 
+ \fr {(b^\mu p^\nu + b^\nu p^\mu)}{p^2 b_\al p^\al}
\nonumber\\
&&
\qquad
\qquad
- \fr {(4\la \b2 + p^2) p^\nu p^\mu}{4\la p^2 (b_\al p^\al)^2 }.
\label{prop}
\eea
In this equation,
$p^\mu$ is the four momentum and $p^2 \equiv p^\mu p_\mu$. 
This propagator matches results from other analyses 
\cite{aknonpoly,bb1}. 

The propagator \rf{prop} can be used to solve for the
harmonic bumblebee fluctuation $\Btw^\mu (p)$ in momentum space
in terms of the metric.
We find
\bea
\Btw^\mu (p) &=& -h^{\mu\al} b_\al 
+ \fr {p^\mu b^\al b^\be h_{\al\be}}{2 b^\al p_\al}
\nonumber\\
&&
- \fr {\xi b^\mu R}{2 \ka p^2} 
+ \fr {\xi p^\mu R}{8 \ka \la b^\al p_\al} 
+ \fr {\xi p^\mu b^\al b_\al R}{2 \ka p^2 b^\al p_\al}
\nonumber\\  
&&
+ \fr {\xi b_\al R^{\al\mu}}{\ka p^2} 
- \fr {\xi p^\mu b^\al b^\be R_{\al\be}}{\ka p^2 b^\al p_\al} .
\label{btilde}
\eea
This solution can be reconverted to position space and substituted 
into the linearized gravitational field equations 
to generate the effective Einstein equations for $\hsy$.
At leading order in the coupling $\xi$,
these take the expected general form \rf{fleq}
with the identifications in Eq.\ \rf{bexps},
except that the coefficient of $\Ph^{\ub}_\mn$
appears as $-3$ rather than unity.
To match the normalization conventions adopted
in Eq.\ \rf{fleq},
a rescaling of the type discussed in Sec.\ \ref{linearization}
must be performed,
setting $\ub \rightarrow -\ub/3$.    
The calculation shows that
assumption (v) in Sec.\ \ref{linearization} holds.
Moreover,
the independently conserved piece $\cS_\mn$
of the energy-momentum tensor $(S_{stu})_\mn$ 
contains only a trace term generating 
the rescaling of $\Ph^{\ub}_\mn$.
This bumblebee theory therefore provides an explicit illustration
of a model that weakly violates assumption (iv) 
of Sec.\ \ref{linearization}.

A similar analysis can be performed for the 
Lagrange-multiplier potential
$V = - \la e(B^\mu B_\mu \pm b^2)$,
for both timelike and spacelike $b^\mu$.
We find that the linearized limits of these models
are also correctly described 
by the general formalism developed in the previous sections,
including the five assumptions of Sec.\ \ref{linearization}.

It follows that the match \rf{aaag} to the PPN formalism
holds for all these bumblebee models
in the context of the isotropic limit,
for which $b^0$ is the only nonzero coefficient
in the universe rest frame.
Note that the condition \rf{al12cond} is valid 
both for the Lagrange-multiplier potential
and for the smooth potential.
In fact,
the $\be = 0$ model with zero potential term $V=0$
but a nonzero isotropic vacuum value for $B^\mu$ 
also satisfies the condition \rf{al12cond}
at leading order 
\cite{hn,cmw}.

We note in passing that 
the limit of zero coupling $\xi$ implies the vanishing 
of all the coefficients and fluctuations in Eq.\ \rf{bexps}.
In the pure gravity-bumblebee sector,
any Lorentz-violating effects in the effective linearized theory 
must then ultimately be associated with the bumblebee potential $V$.
Within the linearization assumptions we have made above,
this result is compatible with that
obtained in Ref.\ \cite{bkgrav}
for the effective action 
of the bumblebee Nambu-Goldstone fluctuations,
for which the Einstein-Maxwell equations are recovered
in the same limit.
Further insight can be obtained
by examining the bumblebee trace-reversed energy-momentum tensor 
$(S_B)_\mn$,
obtained by varying the minimally coupled parts of
the bumblebee action \rf{bb} with respect to the metric.
This variation gives
\bea
(S_B)_\mn &=& b_\mu \prt^\al B_{\al\nu} + b_\nu \prt^\al B_{\al\mu}
-\et_\mn b^\be \prt^\al B_{\al\be} 
\nonumber\\
&&
- 2 (b_\mu b_\nu-\frac 12 \et_\mn b^\al b_\al)V' .
\label{bbem}
\eea
Note that the composite nature of $u$ and $s^\mn$
means that this expression cannot be readily identified
with any of the various pieces of the energy-momentum tensor
introduced in Sec.\ \ref{linearization}.
Using the bumblebee field equations,
$(S_B)_\mn$ can be expressed entirely in terms of $V'$ 
and terms proportional to $\xi$,
whatever the chosen potential.
For the smooth quadratic potential,
insertion of the bumblebee modes $\tilde B^\mu$ 
obtained in Eq.\ \rf{btilde} yields
\bea
V' &= & \la (2 b^\al \tilde B_\al + b^\al b^\be h_{\al\be})
= \fr {\xi R}{4 \ka}.
\label{vprime}
\eea
This shows that all terms in $(S_B)_\mn$ are proportional to $\xi$,
thereby confirming that these modes contribute 
no Lorentz-violating effects to the behavior of $h_{\mu\nu}$ 
in the limit of vanishing $\xi$.
Appropriate calculations for the analogous modes of $\tilde B^\mu$
in the case of the Lagrange-multiplier potential
gives the same result.

\subsection{Cases with $\be \neq 0$}
\label{ghost theories}

As an example of a model that lies 
within the minimal SME 
but outside the mild assumptions of Sec.\ \ref{linearization},
consider the limit $\al = 0$ and $\be = 1$
of the theory \rf{bb}.
The kinetic term of this theory 
is expected to include negative-energy contributions
from the ghost term.

Following the path adopted in Sec.\ \ref{ghostfree},
we expand the bumblebee field about its vacuum value
as in Eq.\ \rf{bvac},
impose the condition \rf{bcond} of
asymptotic cartesian constancy,
make the identifications \rf{bexps},
and disregard bumblebee couplings to the matter sector.
It then follows that the limit $\al = 0$, $\be = 1$
also satisfies assumptions (i), (ii) and (iii)
of Sec.\ \ref{linearization}.
In this limit,
the bumblebee equations of motion become 
\beq
D^\mu D_\mu B_\nu = 2V'b_\nu - \frac {\xi}{\ka} b^\mu R_{\mu\nu}.
\label{bbfe3}
\eeq

For specific potentials,
the linearized form of Eq.\ \rf{bbfe3} 
can be inverted by Fourier decomposition
and the propagator obtained.
Here,
we consider for definiteness the potential 
$V = - \la e(B^\mu B_\mu \pm b^2)^2/2$.
However,
most of the results that follow
also hold for the Lagrange-multiplier potential.

Inverting and substituting into the 
linearized modified Einstein equations
produces the effective equations for $\hsy$.
For our purposes,
it suffices to study the case $\xi=0$.
In the harmonic gauge,
we find the perturbation $\hsy$ is determined by 
the linearized effective equations
\bea
R_\mn &=& \ka [\frac 12 b^\al b_{(\mu} \Box h_{\nu ) \al}
-\frac 12 b^\al b^\be \prt_\al \prt_{(\mu} h_{\nu ) \be}
\nonumber\\
&&
+\frac 12 (b^\al \prt_\al)^2 \hsy 
-\frac 14 b^\al b_{(\mu} \prt_{\nu )} \prt_\al h^{\ga}_{\pt{\ga}\ga}
\nonumber\\
&&
-\frac 14 \nsy b^\al b^\be \Box h_{\al\be}
+(S_M)_\mn ].
\label{gheq}
\eea
The right-hand side of this equation
fails to match the generic form \rf{fleq}
of the linearized equations 
derived in Sec.\ \ref{linearization}. 
In fact,
the non-matter part can be regarded as an effective contribution 
to the independently conserved piece $\cS_\mn$
of the energy-momentum tensor $(S_{stu})_\mn$.
We can therefore conclude that 
the ghost model with $\al = 0$, $\be = 1$
strongly violates assumption (iv) of Sec.\ \ref{linearization}.

The ghost nature of the bumblebee modes 
in this model implies that an exploration 
of propagating solutions of Eq.\ \rf{gheq}
can be expected to encounter problems with negative energies.
Nonetheless,
a post-newtonian expansion for the metric
can be performed.
For definiteness,
we take the effective coefficients 
for Lorentz violation $\ka b^\mu b^\nu$ to be small,
and we adopt the isotropic-limit assumption
that in the universe rest frame
only the coefficient $b^0$ is nonzero.
A calculation then reveals that 
the post-newtonian metric at $O(3)$
in the Sun-centered frame 
and in the gauge \rf{gauge2} 
is given by
\bea
g_{00} &=& -1 + 2U - \ka (b^0)^2 w^2 U + \ka (b^0)^2 w^j w^k U^{jk}
\nonumber\\
&&
- 2 \ka (b^0)^2 w^j (V^j - W^j) + O(4),
\nonumber\\
g_{0j} &=& -\frac 72 V^j - \frac 12 W^j,
\nonumber\\
g_{jk} &=& \de^{jk} (1+2U),
\label{ghmet} 
\eea
where $G$ has been appropriately rescaled.
Comparison with the PPN metric \rf{ppnurf} 
in the same gauge yields
the following parameter values
in this ghost model: 
\bea
\al_1 &=& 0,
\nonumber\\
\al_2 &=& - \ka (b^0)^2,
\nonumber\\
\al_3 &=& 0,
\nonumber\\
\ga &=& 1 .
\label{aaag2}
\eea
This model therefore fails to obey the condition \rf{al12cond}.
The violation of assumption (iv)
evidently affects the general structure 
of the post-newtonian metric.

In light of the results 
obtained above,
we conjecture that 
quadratic ghost kinetic terms are associated with 
a nonzero value of $\cS_\mn$
that strongly violates assumption (iv) of Sec.\ \ref{linearization}
and that violates the condition \rf{al12cond}
at linear order in the isotropic limit.
This conjecture is consistent with the results
obtained above with the Lagrange-multiplier potential
and with the smooth potential,
both for the case $\be = 0$ and for the case $\be \neq 0$.
Moreover,
studies in the context of the PPN formalism
of various models with $V=0$
but a nonzero vacuum value for $B^\mu$
also suggest that ghost terms are associated with
violations of the condition \rf{al12cond}.
For example,
this is true of the post-newtonian limit of the ghost model
with vanishing $V = 0$ and $\xi = 0$
\cite{wn}.
Similarly,
inspection of the general case
with $V=0$ but $\al\neq 0$ and $\be\neq 0$
\cite{cmw} 
shows that the condition \rf{al12cond}
is satisfied at linear order
in the PPN parameters when $\be$ vanishes.
A proof of the conjecture in the general context
appears challenging to obtain but would be of definite interest.

\section{Experimental Applications}
\label{experimental applications}

The remainder of this paper
investigates various gravitational experiments
to determine signals for nonzero coefficients for Lorentz violation
and to estimate the attainable sensitivities.
The dominant effects of Lorentz violation in these experiments
are associated with newtonian $O(2)$ 
and post-newtonian $O(3)$ terms in $g_{00}$
and with post-newtonian $O(2)$ terms in $g_{0j}$.
They are controlled by combinations 
of the 9 coefficients $\sb^\mn$.
Effects from $O(4)$ terms in $g_{00}$
lie beyond the scope of the present analysis.
However,
some effects involving 
$O(3)$ terms in $g_{0j}$
or $O(2)$ terms in $g_{jk}$ 
are accessible by focusing on specific measurable signals.
For example,
experiments involving
the classic time-delay effect 
on a photon passing near a massive body
or the spin precession of a gyroscope in curved spacetime
can achieve sensitivity to terms at these orders.

We begin in Sec.\ \ref{general considerations}
with a discussion of our frame conventions
and transformation properties,
which are applicable to many of the experimental scenarios
considered below.
The types of experimental constraints 
that might be deduced from prior experiments are also summarized.
The remaining subsections treat
distinct categories of experiments.
Section \ref{llr} examines 
measurements obtained from lunar and satellite laser ranging. 
Section \ref{Earth laboratory experiments}
considers terrestrial experiments 
involving gravimeter and laboratory tests.
Orbiting gyroscopes provide another source of information,
as described in Sec.\ \ref{gyroscope experiment}.
The implications for Lorentz violation of
observations of binary-pulsar systems
are discussed in Sec.\ \ref{binary pulsars}.
The sensitivies of the classic tests,
in particular the perihelion shift and the time-delay effect,
are considered in Sec.\ \ref{classic}.
More speculative applications of the present theory,
for example
to the properties of dark matter or to the Pioneer anomaly,
are also of interest
but their details lie beyond the scope of this work
and will be considered elsewhere.

\subsection{General Considerations}
\label{general considerations}

\subsubsection{Frame conventions and transformations}
\label{framesetc}

\begin{figure}
\centerline{\psfig{figure=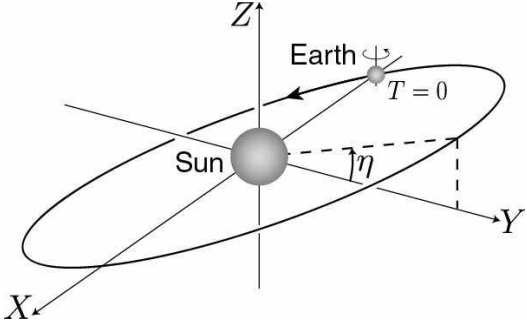,width=0.8\hsize}}
\caption{\label{fig1}Sun-centered celestial-equatorial frame.}
\end{figure}

The comparative analysis of signals for Lorentz violation
from different experiments is facilitated 
by adopting a standard inertial frame.
The canonical reference frame for SME experimental studies
in Minkowski spacetime
is a Sun-centered celestial-equatorial frame
\cite{km}. 
In the present context of post-newtonian gravity,
the standard inertial frame is chosen as 
an asymptotically inertial frame
that is comoving with the rest frame of the solar system
and that coincides with the canonical Sun-centered frame.
The cartesian coordinates in the Sun-centered frame 
are denoted by 
\beq
x^{\Xi}=(T, X^J)=(T,X,Y,Z)
\eeq
and are labeled with capital Greek letters.
By definition,
the $Z$ axis is aligned with the rotation axis of the Earth,
while the $X$ axis points along the direction 
from the center of the Earth to the Sun at the vernal equinox. 
The inclination of the Earth's orbit is denoted $\et$.
The origin of the coordinate time $T$ 
is understood to be the time when the Earth
crosses the Sun-centered $X$ axis at the vernal equinox.
This standard coordinate system is depicted in Fig.\ \ref{fig1}.
The corresponding coordinate basis vectors are denoted 
\beq
{\bf e}_{\Xi} = ({\bf e}_{T}, {\bf e}_{J})
= (\prt_T, \prt_J).
\eeq 

In the Sun-centered frame,
as in any other inertial frame,
the post-newtonian spacetime metric 
takes the form given in Eqs.\ \rf{g00}-\rf{gjk}.
This metric,
along with the point-mass equations of motion \rf{nbody},
forms the basis of our experimental studies to follow.
The corresponding line element can be written in the form
\bea
ds^2 &=& -[ 1 - h_{TT} (T, \vec X) + O(4)] dT^2
\nonumber\\
&&
+2 h_{TJ} (T, \vec X) dT dX^J 
\nonumber\\
&&
+[\de_{JK} + h_{JK} (T, \vec X)] dX^J dX^K,
\label{scfmetric}
\eea
where $h_{TT}$ is taken to $O(3)$, 
$h_{TJ}$ is taken to $O(3)$
and $h_{JK}$ is taken to $O(2)$.
For the purposes of this work,
it typically suffices to include contributions 
to the metric fluctuations from the Sun and the Earth.

Of particular interest for later applications are
various sets of orthonormal basis vectors 
that can be defined in the
context of the line element \rf{scfmetric}.
One useful set is appropriate for an observer at rest,
$dX^J/dT=0$,
at a given point $(T, \vec X)$ in the Sun-centered frame.
Denoting the four elements of this basis set by ${\bf e}_{\mu}$
with $\mu \equiv (t, j)$,
we can write 
\bea
{\bf e}_{t} &=& 
\de_{\pt{T}t}^{T} [ 1 + \frac 12 h_{TT} (T, \vec X) 
+ O(4)] {\bf e}_{T} ,
\nonumber\\
{\bf e}_{j} &=& 
\de^J_{\pt{J}j} [ {\bf e}_{J}
-\frac 12 h_{JK} (T, \vec X) {\bf e}_{K} ] 
+ \de^J_{\pt{J}j} h_{TJ} (T, \vec X) {\bf e}_{T} .
\nonumber\\
\label{ortho1}
\eea 
Direct calculation shows that this basis satisfies 
\beq
ds^2 = -{\bf e}_{t}^2 + {\bf e}_{j}^2
\eeq
to post-newtonian order.

Another useful set of basis vectors,
appropriate for an observer in arbitrary motion,
can be obtained from the basis set \rf{ortho1} 
by applying a local Lorentz transformation.
Denoting this new set of vectors by ${\bf e}_{\hat \mu}$,
we have
\beq
{\bf e}_{\hat \mu} = 
\La^{\nu}_{\pt{\nu}\hat \mu} (\tau) {\bf e}_{\nu}.
\label{ortho2}
\eeq
It is understood that all quantities 
on the right-hand side of this equation 
are to be evaluated along the observer's worldline,
which is parametrized by proper time $\tau$.

For the experimental applications in the present work,
it suffices to expand the local Lorentz transformation
in Eq.\ \rf{ortho2} in a post-newtonian series.
This gives
\bea
{\bf e}_{\hat t} &=& 
\de^{t}_{\pt{t}\hat t} (1 + \frac 12 v^2) {\bf e}_t 
+ v^j {\bf e}_j,
\nonumber\\
{\bf e}_{\hat j} &=& 
\de^j_{\pt{j}\hat j} v^k R^{kj} {\bf e}_t 
+ \de^j_{\pt{j}\hat j} 
(\de^{kl} + \frac 12 v^k v^l ) R^{lj} {\bf e}_k.
\label{ortho3}  
\eea
The components $({\bf e}_{\hat t})^\Xi$ 
coincide with the observer's four-velocity $u^\Xi$
in the Sun-centered frame.
In this expression,
$v^j$ is the coordinate velocity
of the observer as measured in the frame \rf{ortho1},
and $R^{jk}$ is an arbitrary rotation.
The reader is cautioned that the coordinate velocities 
$v^j$ and $v^J$ typically differ at the post-newtonian level.
The explicit relationship can be 
derived from \rf{ortho1}
and is found to be
\beq
v^j = 
\de_{J}^{\pt{J}j} v^J (1+ \frac 12 h_{TT})
+ \frac 12 \de_{J}^{\pt{J}j} h_{JK} v^K + O(4).
\label{velocity}
\eeq

\subsubsection{Current bounds}
\label{current bounds}

In most modern tests of local Lorentz symmetry 
with gravitational experiments,
the data have been analyzed 
in the context of the PPN formalism.
The discussion in Sec.\ \ref{othermetrics}
shows there is a correspondence between
the PPN formalism and the pure-gravity sector
of the minimal SME in a special limit,
so it is conceivable \it a priori \rm that 
existing data analyses could directly yield
partial information about SME coefficients.
Moreover,
a given experiment might in fact have sensitivity
to one or more SME coefficients 
even if the existing data analysis fails to identify it,
so the possibility arises that 
new information can be extracted
from available data by adopting the SME context. 
The specific SME coefficients 
that could be measured by the reanalysis of existing experiments
depend on details of the experimental procedure.
Nonetheless,
a general argument can be given that provides
a crude estimate of the potential sensitivities
and the SME coefficients that might be constrained.

The standard experimental analysis for Lorentz violation
in gravitational experiments is based on the assumed existence 
of a preferred-frame vector $\vec w$.
Usually, 
this is taken to have
a definite magnitude and orientation for the solar system
and is identified with the velocity of the solar system 
with respect to the rest frame 
of the cosmic microwave background radiation
\cite{cmw}.
The strength of the coupling of the vector to gravity 
is then determined by the two PPN parameters
$\al_1$ and $\al_2$.

Performing a data analysis of this type 
under the assumption that the vector $\vec w$
is the source of isotropy violations
is equivalent in the SME context to supposing 
that only parallel projections 
of the coefficients $\sb^{\mu\nu}$
along the unit vector $\hat w$
contribute to any signals.
This holds regardless of the choice of $\vec w$.
For the coefficients $\sb^{\mu\nu}$,
there are two possible parallel projections,
given by
\bea
\sb^{0j}_{\parallel} &=& \hat w^j \hat w^k \sb^{0k},
\nonumber \\
\sb^{jk}_{\parallel} &=& \hat w^j \hat w^k \hat w^l \hat w^m \sb^{lm}.
\label{sjkp}
\eea
Also,
the coefficients $\tb^{\ka\la\mu\nu}$ 
play no role in the analysis at this order,
in accordance with the discussion in Sec.\ \ref{elee}.
It therefore follows 
that sensitivity to at most 2 of the 19 SME coefficients 
$\sb^{\mu\nu}$ and $\tb^{\ka\la\mu\nu}$
could be extracted from these types of analyses
of experimental data.
Together with the 10 coefficients $\tb^{\ka\la\mu\nu}$,
the 7 perpendicular projections given by
\bea
\sb^{0j}_{\perp} &=& \sb^{0k}-\hat w^j \hat w^k \sb^{0k},
\nonumber\\
\sb^{jk}_{\perp} &=& \sb^{jk}-
\hat w^j \hat w^k \hat w^l \hat w^m \sb^{lm}
\label{sjkperp}
\eea
remain unexplored in experimental analyses to date.
 
The two independent measurements of parallel projections
of SME coefficients could be obtained, 
for example,
from a reanalysis of existing data from lunar laser ranging 
\cite{al1llr,al2llr}.
A crude estimate of the sensitivity to be expected 
can be obtained by 
assuming that the bounds attained on a particular term 
of the PPN metric in the Sun-centered frame
roughly parallel those that might be achieved
for the corresponding term in the pure-gravity minimal-SME metric.
This procedure gives
$\sb^{0j}_{\parallel} \lsim 10^{-8}$ 
and $\sb^{jk}_{\parallel} \lsim 10^{-11}$.
However, 
these estimates are untrustworthy 
because the substantial differences between the two metrics
imply unknown effects on the data analysis,
so these constraints are no more than a guide
to what might be achieved.  

A more interesting constraint can be obtained 
by considering implications of the observed close alignment 
of the Sun's spin axis 
with the angular-momentum axis of the planetary orbits.
The key point is that the metric term $\sb^{jk} U^{jk}$
violates angular-momentum conservation.
As a result,
a spinning massive body experiences a self-torque
controlled by the coefficients $\sb^{jk}$.
For the Sun,
this effect induces a precession of the spin axis,
which over the lifetime of the solar system
would produce a misalignment
of the spin axis relative to the ecliptic plane.
This idea was introduced by Nordtvedt and used 
to bound the PPN parameter $\al_2$ 
\cite{nsol}.
In the context of the pure-gravity sector of the minimal SME,
the analysis is similar and so is not presented here. 
It turns out that the final result 
involves a particular combination 
of SME coefficients in the Sun-centered frame given by
\bea
\sb_{\rm SSP} &=& 
\sqrt{\sb_{\rm t}^{JK}\sb_{\rm t}^{JL} \hat S^K \hat S^L 
- (\sb_{\rm t}^{JK} \hat S^J \hat S^K)^2}.
\label{ssp}
\eea
In this expression,
$\hat S^J$ are the Sun-frame components of the
unit vector pointing in the direction of the present solar spin.
The coefficients $\sb_{\rm t}^{JK}$ are the traceless 
components of $\sb^{JK}$,
given by 
\beq
\sb_{\rm t}^{JK} = \sb^{JK} - \frac 1 3 \de^{JK} \sb^{TT}.
\eeq 
Under the assumption that the coefficients $\sb_{\rm t}^{JK}$ 
are small enough for perturbation theory to be valid,
we find an approximate bound of
$\sb_{\rm SSP} \lsim 10^{-13}$.

\subsection{Lunar and Satellite Ranging}
\label{llr}

Lunar laser ranging is among the most sensitive tests
of gravitational physics within the solar system to date.
Dominant orbital perturbations
in the context of the PPN metric
were obtained in Ref.\ \cite{nllr},  
while oscillations arising from a subset 
of the parameters for the anisotropic-universe model 
were calculated in Ref.\ \cite{nanisllr}.
Here,
we obtain the dominant effects on the motion of 
a satellite orbiting the Earth 
that arise from nonzero coefficients $\sb^\mn$
for Lorentz violation.
Our results are applicable to the Earth-Moon system 
as well as to artificial satellites.

\begin{figure}
\centerline{\psfig{figure=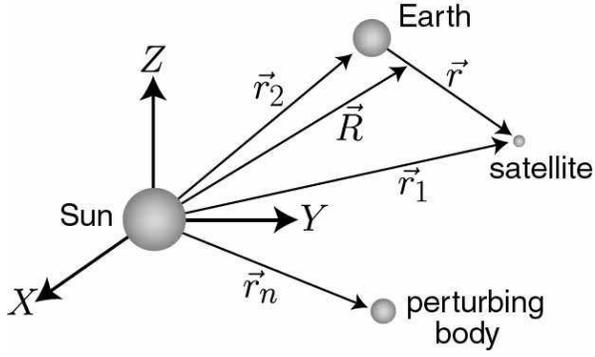,width=0.9\hsize}}
\caption{\label{fig2}The Earth-satellite system 
in the Sun-centered frame.}
\end{figure}

The analysis is performed in the Sun-centered frame.
The satellite motion is affected by the Earth,
the Sun, and other perturbing bodies.
For the analysis,
it is useful to introduce quantities 
to characterize the masses,
positions, sidereal frequencies, etc., 
relevant for the motion.
These are listed in Table 1.
The various position vectors 
are depicted in Fig.\ \ref{fig2}.

This subsection begins by presenting
the coordinate acceleration of the Earth-satellite separation,
which is the primary observable in laser-ranging experiments.
To gain insight into the content of the resulting expressions,
we next perform a perturbative analysis
that extracts the dominant frequencies 
and corresponding amplitudes 
for oscillations driven by Lorentz violation.
This analysis is somewhat lengthy,
and it is largely relegated to Appendix
\ref{dlso},
with only the primary results presented in the main text.
Finally,
we estimate the likely sensitivities attainable
in experimental analyses
of ranging to the Moon
and to various artificial satellites. 

\begin{widetext}
\begin{center}
\begin{tabular} {ll} 
\hline
Quantity & Definition \\
\hline
$m_1$ & satellite mass \\
$m_2$ & Earth mass \\
$M=m_1+m_2$ & total Earth-satellite mass \\
$\de m = m_2 - m_1$ & Earth-satellite mass difference \\
$m_n$ & mass of $n$th perturbing body\\
$M_{\odot}$ & Sun mass \\
$r^J_1$ & satellite position \\ 
$r_0$ & mean satellite orbital distance to Earth\\ 
$r^J_2$ & Earth position \\
$R$ & mean Earth orbital distance to Sun\\
$R_{\oplus}$ & Earth radius \\
$r^J_n$ & position of $n$th perturbing body\\
$r^J = r^J_1 - r^J_2 = (x,y,z)$ & Earth-satellite separation,
with magnitude $r = |\vec r_1 - \vec r_2|$\\
$\de r$ & deviation of Earth-satellite distance $r$ 
from the mean $r_0$ \\
$R^J = (m_1 r^J_1 + m_2 r^J_2)/M$ 
\qquad\qquad 
& position of Earth-satellite newtonian center of mass \\
$R^J_n = R^J - r^J_n$ 
& separation of newtonian center of mass 
and $n$th perturbing body \\
$\om$ & mean satellite frequency \\
$\om_0$ & anomalistic satellite frequency \\
$\om_n$ & $n$th harmonic of satellite frequency \\
$\Om_{\oplus}= \sqrt{G M_{\odot}/R^3} $ & mean Earth orbital frequency\\
$v_0$ & mean satellite orbital velocity \\
$V_{\oplus} = \Om_{\oplus} R$ & mean Earth orbital velocity \\
$v^J = v^J_1 - v^J_2 = dr^J/dT$ & relative Earth-satellite velocity \\ 
$V^J = (m_1 v^J_1 + m_2 v^J_2)/M$ 
\qquad\qquad 
& velocity of Earth-satellite newtonian center of mass \\
$Q=J_2 GM R^2_{\oplus}$ 
& Earth quadrupole moment, with $J_2 \approx 0.001$ \\
\hline
\end{tabular}
\end{center}
\begin{center}
Table 1.\ Definitions for analysis of laser ranging 
to the Moon and to artificial satellites. 
\label{llrdefns}
\end{center}
\end{widetext}

\subsubsection{Earth-satellite dynamics}
\label{esdynamics}

The coordinate acceleration 
$\al^J_{\rm ES}$
of the relative Earth-satellite separation
can be obtained from Eq.\ \rf{nbody}.
Allowing for perturbative effects from the Sun 
and other bodies,
we can write 
\bea
\al^J_{\rm ES} &\equiv& \fr {d^2 r^J}{dT^2}
\nonumber\\
&=& \al^J_{\rm N} + \al^J_{\rm T} + \al^J_{\rm Q} 
+ \al^J_{\rm LV} + \ldots 
\nonumber\\
\label{esacc}
\eea
The first three terms in this expression
are newtonian effects that are independent
of Lorentz violation.
The first is the newtonian acceleration for a point mass
in the newtonian gravitational field of the Earth-satellite system.
It is given by
\beq
\al^J_{\rm N} = - \fr {GM}{r^3}r^J.
\label{nacc}
\eeq
In this expression and what follows,
$G$ has been rescaled by a factor of $(1 - 3\sb^{TT}/2)$,
which is unobservable in this context. 
The second term is the newtonian tidal quadrupole term,
which takes the form
\beq
\al^J_{\rm T} = \sum_{n \neq 1,2} \fr {G m_n (3\hat R^J_n 
\hat R^K _n r^K - r^J)}{R^3_n}.
\label{T}
\eeq
This expression basically represents the leading effects from 
external bodies at the newtonian level of approximation.
The third term is 
\bea
\al^J_{\rm Q} = -\fr {3 Q z \de^{ZJ}}{r^5} 
- \fr {3 Q (5z^2+r^2) r^J}{2 r^7}.
\label{Q}
\eea
It represents the acceleration due to the Earth's quadrupole moment.

The term $\al^J_{\rm LV}$ in Eq.\ \rf{esacc}
contains the leading Lorentz-violating effects.
It can be split into two pieces,
\beq
\al^J_{\rm LV} = \al^J_{\rm LV,ES} + \al^J_{\rm LV, tidal} .
\label{lv}
\eeq
The first of these is given by
\bea
\al^J_{\rm LV,ES} &=& \fr {GM \sb^{JK} r^K}{r^3}
- \fr {3G M \sb^{KL} r^K r^L r^J}{2 r^5} 
\nonumber\\
&&
+ \fr {(3GM V^K \sb^{TK} + 2G\de m v^K \sb^{TK})r^J}{r^3}
\nonumber\\
&&
- \fr {(GM V^K r^K \sb^{TJ} + 2G\de m v^K r^K \sb^{TJ})}{r^3}  
\nonumber\\
&&
- \fr {GM V^J \sb^{TK} r^K}{r^3} 
+ \fr {3GM V^K \sb^{TL} r^K r^L r^J}{r^5}.
\nonumber\\
\label{lvES}
\eea
It contains the Lorentz-violating accelerations
arising from the Earth-satellite system alone.
The term $\al^J_{\rm LV, tidal}$ represents 
the Lorentz-violating tidal accelerations 
due to the presence of other bodies. 
For the Earth-Moon case,
the dominant Lorentz-violating tidal accelerations
are due to the Sun,
and we find the leading contributions are 
\bea
\al^J_{\rm LV, tidal} &=& \Om^2_{\oplus} \sb^{JK} 
(r^K - 3 \hat R^K \hat R^L r^L)
\nonumber\\
&&
-\frac 32 \Om^2_{\oplus} \sb^{KL} (r^J \hat R^K \hat R^L 
+ 2 r^K \hat R^L \hat R^J 
\nonumber\\
&&
\pt{-\frac 32 \Om^2_{\oplus} \sb^{KL}}
- 5 \hat R^M r^M \hat R^K \hat R^L \hat R^J) 
\nonumber\\
&&
+2 \Om^2_{\oplus} \sb^{TK}(r^J V^K 
+ v^K R^J + \frac {\de m}{M} r^J v^K
\nonumber\\
&&
\pt{+2 \Om^2_{\oplus}}
-3 \hat R^L r^L \hat R^J V^K 
-3 \frac {\de m}{M} \hat R^L r^L \hat R^J v^K)
\nonumber\\
&&
- 2 \Om^2_{\oplus} \sb^{TJ}(r^K V^K + v^K R^K 
+ \frac {\de m}{M} r^K v^K
\nonumber\\
&&
\pt{ -2 \Om^2_{\oplus}}
-3 \hat R^L r^L \hat R^K V^K 
-3 \frac {\de m}{M} \hat R^L r^L \hat R^K v^K).
\nonumber\\
\label{lvtidal}
\eea
In contrast,
for artificial satellite orbits 
the corresponding Lorentz-violating 
tidal effects from the Moon and Sun
are suppressed by about six orders of magnitude 
and can safely be ignored.

Finally,
the ellipses in Eq.\ \rf{esacc} represent
higher-order tidal corrections from perturbing bodies, 
higher multipole corrections from the Earth's potential,
and general relativistic corrections entering 
at the $O(4)$ post-newtonian level.
These smaller effects are ignored in the analysis that follows.

\subsubsection{Perturbative expansion}
\label{pertexp}

To fit laser-ranging data, 
Eq.\ \rf{esacc} could be entered
into an appropriate computer code,
along with modeling data for other systematics 
of the satellite orbit 
\cite{al1llr}.
However, 
to get a sense of the types of sensitivities attainable, 
it is useful to study analytically the acceleration \rf{esacc} 
in a perturbation scheme.
Here,
we focus on oscillatory corrections to the
observable relative Earth-satellite separation
$r \equiv |\vec r_1 - \vec r_2|$
in the Sun-centered frame.
Note that the boost between the Sun-centered frame and the Earth,
where observations are in fact performed,
introduces corrections to $r$ only at $O(2)$
and hence corrections to Eq.\ \rf{esacc} at $O(4)$,
as can be verified by determining 
the proper distance measured by an Earth observer.
Similarly,
modifications to the deviation $\de r$ of $r$ from 
the mean Earth-satellite distance $r_0$
also enter only at $O(2)$
and can therefore be neglected for our analysis.
See, for example, Ref.\ \cite{nllr}.

\begin{figure}
\centerline{\psfig{figure=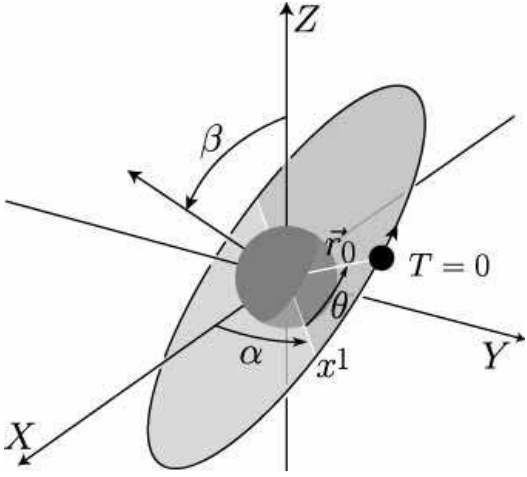,width=0.8\hsize}}
\caption{\label{fig3}
Satellite orbital parameters in the Sun-centered frame.
For simplicity,
the Earth is shown as if it were translated
to the origin of the Sun-centered coordinates.}
\end{figure}

Due to its length, 
the explicit calculation of the Lorentz-violating
corrections to $r$ is relegated to Appendix \ref{dlso}.
We focus here on the main results.
In presenting them,
some definitions associated with  
the satellite orbit are needed. 
These are depicted in Fig.\ \ref{fig3}.
The orientation of the orbit
is described by two angles,
the longitude of the node $\al$
and the inclination of the orbit $\be$.
It is also useful to introduce the circular orbit phase $\th$,
which represents the angle at time $T=0$
subtended between the position vector $\vec r_0$
and the line of ascending nodes in the plane
of the unperturbed orbit.
Note that corrections to all these angles
arising from the boost factor between 
the Sun-centered frame and the Earth 
are negligible in the present context.

The oscillatory radial corrections $\de r$ 
arising from the Lorentz-violating terms in Eq.\ \rf{esacc} 
take the generic form
\beq
\de r = \sum_n [A_n \cos (\om_n T + \phi_n) 
+ B_n \sin (\om_n T + \phi_n)].
\label{delr}
\eeq
The dominant amplitudes $A_n$ and $B_n$ 
and the values of the corresponding phases $\ph_n$
are listed in Table 2.
The amplitudes are taken directly from Eqs.\ \rf{2om-om0_rad}.
The phases $\ph_n$ are related as indicated
to the circular-orbit phase $\th$ shown in Fig.\ \ref{fig3}.

\begin{center}
\begin{tabular}{lc}
\hline
Amplitude & Phase \\ 
\hline
\hline
$A_{2\om} = -\fr {1}{12}(\sb^{11}-\sb^{22}) r_0 $ & $2\th$ \\
$B_{2\om} = -\fr 16 \sb^{12} r_0 $ & $ 2\th$  \\
$A_{2\om-\om_0} = -\om e r_0 (\sb^{11}-\sb^{22})/
16(\om-\om_0) $ & $2\th$ \\
$B_{2\om-\om_0} 
= - \om e r_0 \sb^{12} /8(\om-\om_0)$  & $2\th$ \\
$A_{\om} = - \om (\de m) v_0 r_0 \sb^{02}/M(\om-\om_0)$ & $\th$ \\
$B_{\om} = \om (\de m) v_0 r_0 \sb^{01}/M(\om-\om_0)$ & $\th$ \\
$A_{\Om_{\oplus}} =  V_{\oplus} 
r_0 (b_1/b_2) \sb_{\Om_{\oplus} c} $ & $0$ \\
$B_{\Om_{\oplus}} =  V_{\oplus} 
r_0 (b_1/b_2) \sb_{\Om_{\oplus} s} $ & $0$ \\
\hline
\end{tabular}
\end{center}
\begin{center}
Table 2.\ Dominant Earth-satellite range oscillations.
\label{rangeosc}
\end{center}

The mass, position, frequency, and velocity variables
appearing in Table 2 are defined in Table 1.
The eccentricity of the orbit is denoted $e$,
and for the analysis it is assumed to be much less than unity.
The quantities $b_1$ and $b_2$ are defined 
in Eq.\ \rf{b2} of Appendix \ref{dlso}.
The anomalistic frequency $\om_0$ 
is the frequency of the natural eccentric oscillations.
Note that $\om_0$ differs from $\om$ 
in the presence of perturbing bodies and quadrupole moments.
In the lunar case,
for example, 
$2 \pi/ (\om_0-\om) \approx 8.9$ years
\cite{anomecc}.

Some of the combinations of SME coefficients
appearing in Table 2 are labeled 
with indices 1 and 2,
which refer to projections on the orbital plane. 
The label 1 represents projection onto the line of ascending nodes,
while the label 2 represents projection
onto the perpendicular direction in the orbital plane. 
Explicitly,
the combinations are given in terms of 
the basic Sun-centered coefficients for Lorentz violation as
\bea
\sb^{11}-\sb^{22} &=& (\cos^2 \al - \sin^2 \al \cos^2 \be) \, \sb^{XX}
\nonumber\\
&&
+(\sin^2 \al - \cos^2 \al \cos^2 \be) \, \sb^{YY}
- \sin^2 \be \, \sb^{ZZ}  
\nonumber\\
&&
+ 2 \sin \al \cos \al (1+ \cos^2 \be) \, \sb^{XY} 
\nonumber\\
&&
+ 2 \sin \be \cos \be \sin \al \, \sb^{XZ}
\nonumber\\
&&
- 2 \sin \be \cos \be \cos \al \, \sb^{YZ}, 
\nonumber\\
\sb^{12} &=& -\sin \al \cos \al \cos \be \, \sb^{XX}
\nonumber\\
&&
+ \cos \al \sin \al \cos \be \, \sb^{YY}
\nonumber\\
&&
+(\cos^2 \al - \sin^2 \al) \cos \be \, \sb^{XY}
\nonumber\\
&&
 +\cos \al \sin \be \, \sb^{XZ}
 +\sin \al \sin \be \, \sb^{YZ},
\nonumber\\
\sb^{01} &=& \cos \al \, \sb^{TX} + \sin \al \, \sb^{TY},
\nonumber\\
\sb^{02} &=& -\sin \al \cos \be \, \sb^{TX} 
+ \cos \al \cos \be \, \sb^{TY}
\nonumber\\
&&
+ \sin \be \, \sb^{TZ}.
\label{sb02}
\eea

Different combinations of coefficients for Lorentz violation
also appear in Table 2,
associated with the amplitudes 
of the oscillations at frequency $\Om_{\oplus}$.
The relevant combinations of these coefficients 
in the Sun-centered frame are explicitly given by 
\bea
\sb_{\Om_{\oplus} c} &=& 
\frac 12(\sin \al \sin \et \sin \be \cos \be 
\nonumber\\
&&
\qquad
+ \sin \al \cos \al \cos \et \cos^2 \be 
\nonumber\\
&&
\qquad
- \sin \al \cos \et \cos \al) \sb^{TX} 
\nonumber\\
&&
- (3 \cos \et + \frac 12 
\cos \et \sin^2 \al 
\nonumber\\
&&
\qquad
+ \frac 12 \cos \al \sin \et \sin \be \cos \be
\nonumber\\
&&
\qquad
+\frac 12 \cos \et \cos^2 \al \cos^2 \be ) \sb^{TY}
\nonumber\\
&&
-(3\sin \et + \frac 12 \cos \al \cos \et \sin \be \cos \be 
\nonumber\\
&&
\qquad
+ \frac 12 \sin^2 \be \sin \et)\sb^{TZ} ,
\nonumber\\
\sb_{\Om_{\oplus} s} &=& 
-(3+ \frac 12 \sin^2 \al \cos^2 \be 
+ \frac 12 \cos^2 \al)\sb^{TX} 
\nonumber\\
&&
\qquad
-\frac 12 \cos \al \sin \al \sin^2 \be \sb^{TY} 
\nonumber\\
&&
\qquad
+\frac 12 \sin \al \sin \be \cos \be \sb^{TZ}.
\label{ucos} 
\eea

\subsubsection{Experiment}
\label{llrexpt}

A feature of particular potential interest for experiments 
is the dependence of the coefficients in Table 2
on the orientation of the orbital plane.
To illustrate this,
consider first an equatorial satellite,
for which $\be=0$. 
The observable combinations of coefficients for Lorentz violation
then reduce to
\bea
\sb^{11}-\sb^{22} &=& \cos 2 \al (\sb^{XX} - \sb^{YY})
+ 2 \sin 2 \al \, \sb^{XY},
\nonumber\\    
\sb^{12} &=& -\frac 12 \sin 2 \al (\sb^{XX} - \sb^{YY}) 
\nonumber\\
&&
+ \cos 2 \al \, \sb^{XY}.
\label{beta0}
\eea
This implies that
the $2\om$ and $2\om - \om_0$ frequency bands of this orbit 
have sensitivity to the coefficients 
$\sb^{XX}$, $\sb^{YY}$, and $\sb^{XY}$.
If instead a polar satellite is considered, 
for which $\be = \pi /2$,
the same frequency bands have sensitivity to all six 
coefficients in $\sb^{JK}$:
\bea
\sb^{11}-\sb^{22} &=& \frac 12 (\sb^{XX} + \sb^{YY}) 
+  \frac 12 \cos 2 \al (\sb^{XX} - \sb^{YY})
\nonumber\\
&&
+ \sin 2 \al \, \sb^{XY} - \sb^{ZZ},
\nonumber\\    
\sb^{12} &=& \cos \al \, \sb^{XZ} + \sin \al \, \sb^{YZ}.
\label{beta90}
\eea
These examples show that satellites with different orbits 
can place independent bounds on coefficients for Lorentz violation.
A similar line of reasoning shows sensitivity to the 
$\sb^{TJ}$ coefficients can be acquired as well.

For lunar laser ranging,
the even-parity coefficients $\sb^{JK}$ can be extracted from
Eqs.\ \rf{sb02} using the appropriate values
of the longitude of the node and the inclination.
For definiteness and to acquire insight,
we adopt the values
$\al \approx 125^{\circ}$ and $\be \approx 23.5^{\circ}$.
However,
these angles vary for the Moon 
due to comparatively large newtonian perturbations,
so some caution is needed in using the equations that follow.
In any case,
with these values
we find the even-parity coefficients are given by
\bea
(\sb^{11}-\sb^{22})_{\leftmoon} 
&=& 0.08 (\sb^{XX} + \sb^{YY} - 2 \sb^{ZZ})
\nonumber\\
&&
-0.31 (\sb^{XX} - \sb^{YY}) 
- 1.7 \sb^{XY} 
\nonumber\\
&&
+0.60 \sb^{XZ} + 0.42 \sb^{YZ},
\nonumber\\
(\sb^{12})_{\leftmoon} &=& 0.43 (\sb^{XX} - \sb^{YY}) 
- 0.31 \sb^{XY} 
\nonumber\\
&&
- 0.23 \sb^{XZ} - 0.33 \sb^{YZ} .
\label{m2}
\eea
Evidently,
lunar laser ranging measures two independent combinations
of the $\sb^{JK}$ coefficients.

For the odd-parity coefficients $\sb^{TJ}$,
adopting the same values for $\al$ and $\be$
shows that lunar laser ranging 
offers sensitivity to the combinations 
\bea
(\sb^{01})_{\leftmoon} &=& -0.60 \sb^{TX} + 0.82 \sb^{TY},
\nonumber\\
(\sb^{02})_{\leftmoon} &=& - 0.53 \sb^{TY} -0.75 \sb^{TX} +0.40 \sb^{TZ},
\nonumber\\
(\sb_{\Om_{\oplus} c})_{\leftmoon} &=& -3.1 \sb^{TY} - 1.1 \sb^{TZ} 
+0.094 \sb^{TX},
\nonumber\\
(\sb_{\Om_{\oplus} s})_{\leftmoon} &=& -3.4 \sb^{TX}  + 0.037 \sb^{TY} 
+0.15 \sb^{TZ}.
\label{m6}
\eea
This reveals that
measuring the amplitudes at frequency $\om$ and phase $\th$
provides sensitivity to 2 combinations 
of the 3 independent coefficients $\sb^{TJ}$.
Provided one can also independently measure the amplitudes 
at frequency $\Om$ and zero phase, 
the third independent coefficient in $\sb^{TJ}$
can also be accessed.  

Together the above considerations imply that 
{\it lunar laser ranging can in principle 
measure at least 5 independent combinations 
of coefficients for Lorentz violation.}
In particular,
the odd-parity coefficients $\sb^{TJ}$ can be 
completely determined.

To estimate the experimental sensitivity attainable
in lunar laser ranging, 
we assume ranging precision at the centimeter level,
which has already been achieved
\cite{onecm}.
With the standard lunar values 
$r_0 = 3.8 \times 10^8$ m,  
$V_{\oplus} = 1.0 \times 10^{-4}$, 
$\om/(\om-\om_0) \approx 120$,
and $e=0.055$,
we find that lunar laser ranging can attain
the following sensitivities:
parts in $10^{11}$ 
on the coefficient 
$(\sb^{12})_{\leftmoon}$,
parts in $10^{10}$ 
on the combination
$(\sb^{11}-\sb^{22})_{\leftmoon}$,
and parts in $10^{7}$ on the coefficients 
$\sb^{01}$, $\sb^{02}$,
$\sb_{\Om_{\oplus} s}$,
$\sb_{\Om_{\oplus} c}$.
These sensitivities may be substantially improved with the 
new Apache Point Observatory Lunar Laser-ranging Operation 
(APOLLO),
which is currently under development in New Mexico 
with the goal of millimeter-ranging precision 
\cite{apollo}.

Next,
we consider ranging to artificial satellites.
In practice,
the realistic modeling of 
satellite orbits is more challenging than the Moon
due to substantial perturbative effects.
However,
we show here that ranging to artificial satellites 
has the potential to be particularly useful for measuring 
independent combinations of the $\sb^{JK}$ coefficients 
that are less readily accessible to lunar laser ranging 
due to the approximately fixed orientation of the 
plane of the lunar orbit.

The relevant coefficients for a generic satellite orbit
are given in Eqs.\ \rf{sb02} in terms
of the right ascension $\al$ and inclination $\be$.
For the odd-parity coefficients,
a similar argument applies as for the lunar case, 
so any one satellite can 
in principle measure all 3 coefficients $\sb^{TJ}$.
Using reasoning similar to that for lunar laser ranging,
we find that any one satellite can make
only two independent measurements of 
combinations of the even-parity coefficients $\sb^{JK}$.
It follows that
{\it any 2 satellites at different orientations 
can in principle measure 3 of the 4 combinations 
of components of $\sb^{JK}$ 
to which lunar laser ranging is insensitive.}

The reason that only 3 of the 4 can be measured 
is that the combination 
$\sb^{XX}+\sb^{YY}+\sb^{ZZ}=\sb^{TT}$ 
is absent from Eqs.\ \rf{sb02}
and hence only 5 combinations of the 6 coefficients
in $\sb^{JK}$ 
are measurable via the amplitudes in Table 2
for any set of orbit orientations.
However,
higher harmonics in $\om$, $\Om$, and other frequencies 
are likely to arise from the eccentric motion 
of the Moon or satellite.
Although the corresponding amplitudes  
would be suppressed compared to those in Table 2,
they could include terms from which the combination
$\sb^{XX}+\sb^{YY}+\sb^{ZZ}=\sb^{TT}$ might be measured.
The numerical code used to fit the data
would include these amplitudes.
Note that the strength of some signals
might be enhanced by a suitable orientation 
of the satellite orbit,
since the amplitudes in Table 2 depend
on the orientation angles $\al$ and $\be$
through the quantities $b_1$ and $b_2$ 
defined in Eq.\ \rf{b2}.

\begin{center}
\begin{tabular}{lllllc}
\hline
Satellite & No.\ & \quad $r_0$(m) 
& $\be$ & Ref.\ \\
\hline
\hline 
ETALON & $2$ & $2.5 \times 10^7$ & $65^{\circ}$ & \cite{ilrs} \\
GALILEO & $30$ & $3.0 \times 10^7$ & $56^{\circ}$ & \cite{galileo} \\
GLONASS & $3$ & $2.5 \times 10^7$ & $65^{\circ}$ & \cite{ilrs} \\
GPS & $2$ & $2.6 \times 10^7$ & $55^{\circ}$ & \cite{ilrs} \\
LAGEOS I & $1$ & $ 1.2 \times 10^7$ & $110^{\circ}$ & \cite{ilrs} \\
LAGEOS II 
\qquad 
\qquad 
& $1$ 
\qquad 
\qquad 
& $1.2 \times 10^7$ 
\qquad
\qquad
& $53^{\circ}$ 
\qquad 
\qquad 
& \cite{ilrs} \\
\hline
\end{tabular}
\end{center}
\begin{center}
Table 3.\ Some high-orbit satellites.
\label{sats}
\end{center}

Some promising high-orbit satellite missions,
both current and future, 
are listed in Table 3.  
Although beyond our present scope,
it would be interesting to determine 
whether data from ranging to any of these satellites 
could be adapted to search for the signals in Table 2.
As an estimate of attainable sensitivities,
suppose centimeter-ranging sensitivity is feasible 
and take
$r_0 \approx 10^7$ m, 
$\om/(\om-\om_0) \approx 2 \times 10^3$,
and $e = 0.010$.
These values produce the following estimated
sensitivities from laser ranging to artificial satellites:
parts in $10^{9}$ on $\sb^{11}-\sb^{22}$,
parts in $10^{10}$ on $\sb^{12}$,
parts in $10^{8}$ on $\sb^{01}$, $\sb^{02}$,
and 
parts in $10^{5}$ on $\sb_{\Om_{\oplus} s}$,
$\sb_{\Om_{\oplus} c}$. 

Interesting possibilities may also exist 
for observing secular changes to the orbits
of near-Earth satellites.
Sensitivities comparable to some of those mentioned above
may be attainable for certain coefficients 
\cite{nesat}.
A detailed analysis of this situation would require incorporating
also contributions to 
the Earth-satellite acceleration $\al^J_{\rm ES}$
proportional to the spherical moment of inertia of the Earth.
These contributions,
which can arise from Lorentz-violating effects proportional to 
the potential $U^{JK}$ in Eq.\ \rf{pots},
are small for high-orbit satellites or the Moon
and so have been neglected in the analysis above, 
but they can be substantial for near-Earth orbits.

\subsection{Laboratory experiments}
\label{Earth laboratory experiments}

This section considers some sensitive laboratory experiments
on the Earth.
Among ones already performed are gravimeter tests,
analyzed in Refs.\ \cite{gcmw,grvmet}.
For this case,
the basic idea is to measure the force on a test mass 
held fixed above the Earth's surface,
seeking apparent variations in the locally measured value 
of Newton's gravitational constant $G$.
The coefficient space of the pure-gravity sector of the minimal SME
to which this experiment is sensitive is explored below.
Other sensitive existing laboratory experiments use 
various torsion pendula,
with some already being applied to probe Lorentz violation 
in the fermion sector 
\cite{eexpt2}.
This section also discusses some torsion-pendulum tests 
that could perform sensitive measurements 
of Lorentz symmetry in the gravitational sector.

\subsubsection{Theory}
\label{Earth laboratory theory}

To explore these ideas,
we must first determine the dynamics of a test particle 
and establish the local acceleration at a point on
the Earth's surface
in the presence of Lorentz violation controlled by
the coefficients $\sb^{\mu\nu}$.
The results are required in 
a suitable observer coordinate system for experiments.

An observer on the surface of the Earth
is accelerating and rotating with respect to 
the asymptotic inertial space.
The appropriate coordinate system is therefore 
the proper reference frame 
of an accelerated and rotated observer 
\cite{nz,mtw}.
At second order in the coordinate distance $x^{\hat j}$,
the metric for this coordinate system 
can be written in the form
\bea
ds^2 &=& -(dx^{\hat 0})^2 \big[ 1 + 2 a^{\hat j} x^{\hat j} 
+ (a^{\hat j} x^{\hat j})^2 + (\om^{\hat j} x^{\hat j})^2 
\nonumber\\
&&
- \om^2 x^{\hat j} x^{\hat j} + R_{\hat 0 \hat j \hat 0 \hat k} 
x^{\hat j} x^{\hat k} \big]
\nonumber\\
&& 
+ 2 dx^{\hat 0} dx^{\hat j} ( \ep^{\hat j \hat k \hat l} 
\om^{\hat k} x^{\hat l} - \frac 23 R_{\hat 0 \hat k \hat j 
\hat l} x^{\hat k} x^{\hat l})
\nonumber\\
&&
+ dx^{\hat j} dx^{\hat k} (\de_{\hat j \hat k} 
- \frac 13 R_{\hat j \hat l \hat k \hat m} x^{\hat l} x^{\hat m}).
\label{amet}
\eea  
Here,
we have introduced $x^{\hat 0}$,
which coincides along the worldline with
the observer's proper time.

From this metric,
an effective Lagrange density can be established
for test particle dynamics in the proper reference frame
of an accelerated and rotated observer.
The action for a test particle of mass $m$ takes the form 
\bea
S &=& - m \int dx^{\hat 0} \sqrt{\fr {-ds^2}{(dx^{\hat 0})^2} }
\nonumber\\
&=& \int dx^{\hat 0} L.
\label{effact}
\eea
For present purposes,
it suffices to express the lagrangian 
as a post-newtonian series.
Denote the coordinate position of the test particle by $x^{\hat j}$
and its coordinate velocity by $\dot x^{\hat j}$,
where the dot signifies derivative with respect to $x^{\hat 0}$.
This velocity and the rotation velocity 
of the Earth's surface are taken to be $O(1)$.
The lagrangian then becomes
\bea
m^{-1}L &=& \frac 12 (\dot{\vec x})^2 
 - \vec a \cdot \vec x 
+ \fr 12 (\vec \om \times \vec x)^2 
\nonumber\\
&&
+ \dot{\vec x} \cdot (\vec \om \times \vec x)
- \frac 12 R_{\hat 0 \hat j \hat 0 \hat k} x^{\hat j} x^{\hat k} 
+ \ldots . 
\label{leff}
\eea
In this expression,
$\vec a$ is the observer acceleration,
$\vec \om$ is the observer angular velocity,
and the various curvature components are taken to contain 
only the dominant contributions 
proportional to $\vec\nabla^2 g_{00}$.
The ellipses represent terms that contribute at $O(4)$
via post-newtonian quantities and sub-dominant curvature terms.

The dominant Lorentz-violating effects arise through $\vec a$. 
To establish the local acceleration in the proper reference frame
in the presence of Lorentz violation,
consider first the observer's four acceleration 
in an arbitrary post-newtonian coordinate frame. 
This is given by the standard formula
\beq
a^\mu = \fr {du^\mu}{d\tau} + \Ga^{\mu}_{\pt{\mu}\nu\ka} u^\nu u^\ka.
\label{4acc}
\eeq
We then choose the post-newtonian frame 
to be the Sun-centered frame described in 
Sec.\ \ref{general considerations}.
In this frame,
the spacetime metric is given by Eq.\ \rf{scfmetric}.

To express the local acceleration 
in the Sun-centered frame,
the observer's trajectory $x^\Xi$ is required.
This can be adequately described to post-newtonian order by 
\beq
x^\Xi = e^{\Xi}_{\pt{\Xi}\tilde k} \xi^{\tilde k} + x^\Xi_{\oplus}.
\label{traj}
\eeq 
In this equation,
$x^\Xi_{\oplus}$ is the worldline of the Earth
and $\xi^{\tilde k}$ is the spatial coordinate location
of the observer in a frame centered on the Earth, 
for which indices are denoted with a tilde.
The components of the comoving spatial basis in that frame
are denoted $e^{\Xi}_{\pt{\Xi} \tilde k}$.
They can be obtained from Eq.\ \rf{ortho3} 
with $R^{jk}=0$ and 
the Earth velocity $v^{j}_\oplus$ 
given by Eq.\ \rf{velocity} with $V^J=V^J_{\oplus}$.
To sufficient post-newtonian accuracy,
the observer's coordinate location $\xi^{\tilde k}$ 
in the Earth frame 
can be taken as a vector pointing to the observer's location 
and rotating with the Earth.
Thus,
modeling the Earth as a sphere of radius $R_{\oplus}$,
we can write
\beq
\vec \xi = R_{\oplus} (\sin \ch \cos (\om_{\oplus}T+\ph), 
\sin \ch \sin (\om_{\oplus}T+\ph), \cos \ch) ,
\eeq
where $\ch$ is the colatitude of the observer
and $\ph$ is a standard phase fixing the time origin
(see Ref.\ \cite{km}, Appendix C):
$\ph = \om_{\oplus} (T_{\oplus}-T)$,
where $T_{\oplus}$ is measured from 
one of the times when the $\hat y$ and $Y$
axis coincide.

To obtain explicitly the local acceleration,
the potentials \rf{pots} must be evaluated for the Earth.
They can be written as functions 
of the spatial position $\vec X$ in the Sun-centered frame,
and the ones relevant here are given by
\bea
U &=& \fr {GM_{\oplus}}{|\vec X - \vec x_\oplus |},
\nonumber\\
U^{JK} &=& 
\fr {GM_{\oplus} (X-x_\oplus )^J (X-x_\oplus )^K }
{|\vec X - \vec x_\oplus |^3}
\nonumber\\
&&
-\fr {G I_{\oplus}}{3|\vec X - \vec x_\oplus|^5} 
[3 (X-x_\oplus)^J (X-x_\oplus)^K
\nonumber\\
&&
\pt{-\fr {G I_{\oplus}}{3|\vec X - \vec x_\oplus|^5}}
-\de^{JK} |\vec X - \vec x_\oplus|^2],
\nonumber\\
V^J &=& 
\fr {GM_{\oplus} V^J_{\oplus}}{|\vec X - \vec x_\oplus|}+ \ldots .
\label{epots}
\eea
In these equations,
$M_{\oplus}$ is the mass of the Earth.
The quantity $I_{\oplus}$ 
is the spherical moment of inertia of the Earth,
given by the integral 
\bea
I_{\oplus} = 
\int d^3 y \rh(\vec y) \vec y^{\hskip 2 pt 2}
\label{inertia}
\eea
over the volume of the Earth,
where $\vec y$ is the vector distance from the center of the Earth
and $\rh(\vec y)$ is the density of the Earth.
The ellipses in Eq.\ \rf{epots}
represent neglected gravitomagnetic effects.

Inserting Eq.\ \rf{traj} into Eq.\ \rf{4acc}
yields the following components of the observer's acceleration 
in the Sun-centered frame:
\bea
a^T &=& O(3),
\nonumber\\
a^J &=& g(1 + \frac 32 i_1 \sb^{TT}
+ \frac 32 i_2 \sb^{KL} 
\de^{K}_{\pt{K} \tilde k}
\de^{L}_{\pt{L} \tilde l} 
\hat \xi^{\tilde k} \hat \xi^{\tilde l} ) 
\de^{J}_{\pt{J} \tilde j} 
\hat \xi^{\tilde j} 
\nonumber\\
&&
\hskip -15pt
- g i_3 \sb^{JK} 
\de^{K}_{\pt{K} \tilde k} 
\hat \xi^{\tilde k} 
+ g i_3 
(\sb^{TJ} V_{\oplus}^K 
\de^{K}_{\pt{K} \tilde k} 
\hat \xi^{\tilde k}
+ \sb^{TK} V^J_{\oplus} 
\de^{K}_{\pt{K} \tilde k} 
\hat \xi^{\tilde k} )
\nonumber\\
&&
\hskip -15pt
-3 g (i_1 
\sb^{TK}V_{\oplus}^K 
\de^{J}_{\pt{J} \tilde j} 
\hat \xi^{\tilde j} 
+ i_2 \sb^{TK} V^L_{\oplus} 
\de^{K}_{\pt{K} \tilde k} 
\de^{L}_{\pt{L} \tilde l} 
\de^{J}_{\pt{J} \tilde j} 
\hat \xi^{\tilde k} \hat \xi^{\tilde l} \hat \xi^{\tilde j})
\nonumber\\
&&
\hskip -15pt
+ \de^{J}_{\pt{J} \tilde j} 
d^2 \xi^{\tilde j} / dT^2 + O(4).
\label{scfacc}
\eea
In these equations,
the reference gravitational acceleration $g$ is 
\beq
g=GM_{\oplus}/R^2_{\oplus},
\eeq
and the quantities $i_1$, $i_2$, $i_3$
are defined by 
\bea
i_{\oplus} &=& \fr {I_{\oplus}}{M_{\oplus} R^2_{\oplus}},
\nonumber\\
i_1 &=& 1 + \frac 13 i_{\oplus},
\nonumber\\
i_2 &=& 1 - \frac 53 i_{\oplus},
\nonumber\\
i_3 &=& 1 - i_{\oplus}.
\label{icombos}
\eea
Note that the analysis here disregards possible tidal changes in the 
Earth's shape and mass distribution
\cite{mel,cmw},
which under some circumstances
may enhance the observability 
of signals for Lorentz violation.

To obtain the corresponding expressions in the observer's frame,
it suffices to project Eq.\ \rf{scfacc} 
along the comoving basis vectors 
${\bf e}_{\hat \mu} = e^{\Xi}_{\pt{\Xi}\hat \mu} {\bf e}_{\Xi}$
of the accelerated and rotated observer on the Earth's surface.
Explicit expressions for these basis vectors
are given by Eq.\ \rf{ortho3},
with the observer velocity $v^j_{\rm o}$ 
and the rotation $R^{jk}$ taken as 
\bea
v^j_{\rm o} &=& \de_J^{\pt{J}j} V^J_{\oplus} 
+ \de^{j}_{\pt{j}\tilde j} d \xi^{\tilde j} / d T+ O(3),
\nonumber\\
R^{jk} &=& R^{jk}_Z ( \om_{\oplus} T).
\eea
To sufficient approximation,
we may use 
\beq
\vec V_{\oplus} = V_{\oplus} (\sin \Om_{\oplus} T,
-\cos \et \cos \Om_{\oplus} T, -\sin \et \cos \Om_{\oplus} T ),
\eeq
where $V_{\oplus}$ is the mean Earth orbital speed
and $\et$ is the inclination of the Earth's orbit,
shown in Fig.\ \ref{fig1}.

We are free to use the observer rotational invariance 
of the metric \rf{amet} 
to choose the observer's spatial coordinates 
$x^{\hat j}$ to coincide with the standard SME conventions 
for an Earth laboratory
(see Ref.\ \cite{km}, Appendix C).
Thus,
at a given point on the Earth's surface,
$\hat z$ points towards the zenith, 
$\hat y$ points east,
and $\hat x$ points south.
With these conventions,
we find the desired expression for 
the local acceleration to be 
\bea
a^{\hat j} &=& g(1 + \frac 32 i_1 \sb^{TT} 
+ \frac 32 i_2 
\sb^{\hat z \hat z} ) \de^{\hat j \hat z}
- g i_3 \sb^{\hat j \hat z} 
\nonumber\\
&& 
- \om^2 R_{\oplus} ( \sin^2 \ch \de^{\hat j \hat z} 
+ \sin \ch \cos \ch \de^{\hat j \hat x}) 
\nonumber\\
&&
+ g i_3 ( \sb^{T \hat j} V_{\oplus}^{\hat z}
+ \sb^{T \hat z} V^{\hat j}_{\oplus} )
\nonumber\\
&&
- 3g (i_1 \sb^{TJ}V_{\oplus}^J 
+ i_2 \sb^{T\hat z} V^{\hat z}_{\oplus}) 
\de^{\hat j \hat z}
+ \ldots .
\label{locacc}
\eea
The ellipses represent newtonian tidal corrections
from the Sun and Moon,
along with $O(4)$ terms. 
The various projections of quantities along the local axes 
appearing in Eq.\ \rf{locacc} 
are obtained
using the comoving basis vectors 
${\bf e}_{\hat \mu} = e^{\Xi}_{\pt{\Xi}\hat \mu} {\bf e}_{\Xi}$.
For example,
$\sb^{T \hat j} = e^{\pt{\Xi}\hat j}_\Xi \sb^{T\Xi}$.
Typically,
these projections introduce time dependence on $T$.
Note, however, that the coefficient $\sb^{TT}$ 
carries no time dependence up to $O(3)$
and hence represents an unobservable scaling.

\subsubsection{Gravimeter tests}
\label{gravimeter tests}

By restricting attention to motion along the $\hat z$ axis,
the expression \rf{locacc} for the local acceleration
can be used to obtain the dominant Lorentz-violating contributions 
to the apparent local variation $\de G$ 
of Newton's gravitational constant,
as would be measured with a gravimeter.
This gives
\bea
a^{\hat z} &=& g(1 + \frac 32 i_1 \sb^{TT} 
+ \frac 12 i_4 \sb^{\hat z \hat z} )  
- \om^2 R_{\oplus} \sin^2 \ch 
\nonumber\\
&&
- g i_4 \sb^{T \hat z} V_{\oplus}^{\hat z}
-  3 g i_1 \sb^{TJ} V_{\oplus}^J,  
\label{zacc}
\eea
where
\bea
i_4 &=& 1-3 i_{\oplus}.
\eea
The time dependence of the apparent variations in $G$
arises from the terms involving the indices $\hat z$ 
and the components $\sb^{TJ}$.
It can be decomposed in frequency according to
\bea
\fr {\de G}{G} &=& \sum_n [C_n \cos (\om_n T + \phi_n)
+ D_n \sin (\om_n T + \phi_n)].
\nonumber\\
\label{delg}
\eea
Explicit calculation of the time dependence 
using the comoving basis vectors yields the expressions 
for the amplitudes $C_n$, $D_n$ and phases $\ph_n$
listed in Table 3.
In this table,
the angle $\ch$ is the colatitude of the laboratory,
and the other quantities are defined 
in the previous subsection.

\begin{widetext}
\begin{center}
\begin{tabular}{lc}
\hline
\multicolumn{2}{c} 
{Gravimeter Amplitudes} \\
\hline
\hline
Amplitude & Phase \\ 
\hline
$C_{2\om} = \fr 14 i_4 (\sb^{XX}-\sb^{YY}) \sin^2 \ch$ & 
$2\phi$ \\
$D_{2\om} = \fr 12 i_4 \sb^{XY} \sin^2 \ch $ & $ 2\phi$  \\
$C_{\om} = \fr 12 i_4 \sb^{XZ} \sin 2 \ch $ & $\phi$ \\
$D_{\om} = \fr 12 i_4 \sb^{YZ} \sin 2 \ch $  & $\phi$ \\
$C_{2\om + \Om} 
= -\fr 14 i_4 V_{\oplus} \sb^{TY} (\cos \et-1) \sin^2 \ch $ 
& $2\phi$ \\
$D_{2\om + \Om} 
= \fr 14 i_4 V_{\oplus} \sb^{TX} (\cos \et-1 ) \sin^2 \ch $ 
& $2\phi$ \\
$C_{2\om - \Om} 
= -\fr 14 i_4 V_{\oplus} \sb^{TY} (1 + \cos \et ) \sin^2 \ch 
$ & $2\phi$ \\
$D_{2\om - \Om} 
= \fr 14 i_4 V_{\oplus} \sb^{TX} (1 + \cos \et ) \sin^2 \ch $ 
& $2\phi$ \\
$C_{\Om} = V_{\oplus} 
[\sb^{TY} \cos \et (\fr 12 i_4 \sin^2 \ch + 3 i_1)+ 
\sb^{TZ} \sin \et (3 i_1 + i_4 \cos^2 \chi)] $ 
\qquad \qquad 
& $0$ \\ 
$D_{\Om} = - V_{\oplus} \sb^{TX} (\fr 12 i_4 \sin^2 \chi + 3 i_1) $ & 
$0$ \\
$C_{\om + \Om} =  \fr 14 i_4 V_{\oplus} \sb^{TX} \sin 2\ch \sin \et $ & 
$\phi$ \\
$D_{\om + \Om} =  -\fr 14 i_4 V_{\oplus} [ \sb^{TZ}(1-\cos \et)- 
\sb^{TY} \sin \et ] \sin 2 \ch $ & $\phi$ \\
$C_{\om - \Om} =  \fr 14 i_4 V_{\oplus} \sb^{TX} \sin \et \sin 2 \ch$ & 
$\phi$ \\
$D_{\om - \Om} =  \fr 14 i_4 V_{\oplus} [ \sb^{TZ} (1 + \cos \et) + 
\sb^{TY} \sin \et ] \sin 2\ch $ & $\phi$ \\
\hline
\end{tabular}
\end{center}
\begin{center}
\label{terrosc}
Table 4.\ Gravimeter amplitudes.  
\end{center}
\end{widetext}

Assuming each amplitude in Table 4 can be separately
extracted from real data, 
it follows that
{\it gravimeter experiments can measure up 
to 4 independent coefficients in $\sb^{JK}$ and
all 3 coefficients in $\sb^{TJ}$.}
Moreover,
gravimeter experiments have attained impressive accuracies 
\cite{grvmet}
that would translate into stringent sensitivities
on the pure-gravity coefficients in the minimal SME. 
A crude estimate suggests that 
gravimeter experiments fitting data to Eq.\ \rf{delg}
could achieve sensitivity to 
$\sb^{JK}$ at parts in $10^{11}$ 
and to $\sb^{TJ}$ at parts in $10^7$.
Possible systematics that would require attention
include tidal influences from the Sun and Moon.
These can,
for example,
generate oscillations contributing to Eq.\ \rf{delg}
at frequencies $2\om_{\oplus}$ and $\om_{\oplus}$.

\subsubsection{Torsion-pendulum tests}
\label{torsion-pendulum tests}

A complementary class of laboratory tests
could exploit the anisotropy of the locally measured 
acceleration in the horizontal directions
instead of the vertical one.
Explicitly,
the accelerations in the $\hat x$ and $\hat y$ 
directions are given by
\bea
a^{\hat x} &=& - g i_3 \sb^{\hat x \hat z} 
- \om^2 R_{\oplus}\sin \ch \cos \ch 
\nonumber\\
&&
+ g i_3 \sb^{T \hat z} V_{\oplus}^{\hat x}
+ g i_3 \sb^{T \hat x} V^{\hat z}_{\oplus} ,
\nonumber\\
a^{\hat y} &=& - g i_3 \sb^{\hat y \hat z} 
+ g i_3 \sb^{T \hat z} V^{\hat y}_{\oplus} 
+ g i_3 \sb^{T \hat y} V^{\hat z}_{\oplus} 
\label{accy}
\eea
to order $O(3)$.
In effect,
the Lorentz-violating contributions to these expressions 
represent slowly varying accelerations 
in the horizontal directions
as the Earth rotates and orbits the Sun. 
In contrast,
the conventional centrifugal term in Eq.\ \rf{accy} is constant
in time.

\begin{figure}
\centerline{\psfig{figure=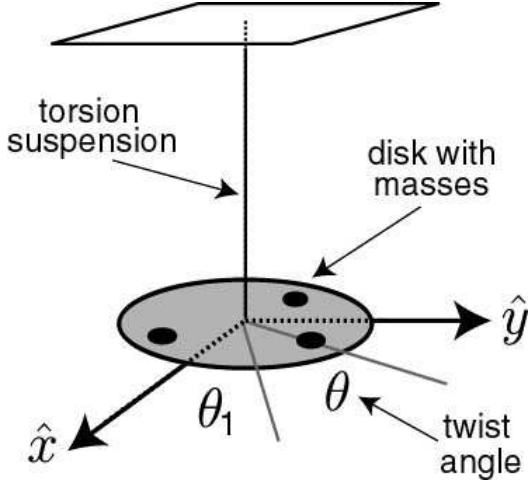,width=0.8\hsize}}
\caption{\label{fig4}
Idealized apparatus with $N=3$ masses.}  
\end{figure}

To gain insight into the experimental possibilities
for testing the effects contained in Eqs.\ \rf{accy}, 
consider an idealized scenario
with a disk of radius $R$ and mass $M$ 
suspended at its center
and positioned to rotate freely about its center of symmetry 
in the $\hat x$-$\hat y$ plane.
We suppose that $N$ spherical masses,
each of mass $m_j$,
$j = 1,2,\ldots, N$, 
are attached rigidly to the disk 
at radius $r_0<R$
and initially located at angular position $\th_j$
relative to the $\hat x$ axis.
For convenience,
we write the masses as $m_j = \mu_j m$,
where $m$ is a basic mass and $\mu_j$ are dimensionless numbers.
In what follows,
we assume a torsion suspension
and focus attention on small angular deviations $\th$ 
of the disk relative to its initial angular orientation.
The apparatus with $N=3$ is depicted in Fig.\ \ref{fig4}.
Note that the locations of the center of mass 
and the center of suspension are typically different.
In practice,
the technical challenge of keeping the disk level 
might be overcome in several ways,
perhaps involving rigid suspension or magnetic support,
but this is a secondary issue in the present context
and is disregarded here.

Using Eq.\ \rf{leff},
the effective potential energy for the apparatus 
can be constructed.
Up to an overall constant,
we find
\bea
V &=& -\frac 12 mr_0 [a^{\hat x} C_N + a^{\hat y} S_N 
+ r_0 \om_{\oplus}^2 \sin^2 \ch (2 C_N -1)] \th^2  
\nonumber\\
&&
- mr_0 (a^{\hat x} S_N - a^{\hat y} C_N 
+ \frac 12 a \om_{\oplus}^2 \sin^2 \ch S_{N2} ) \th.  
\label{effpot}
\eea
In this expression,
we have introduced the following quantities:
\bea
C_N &=& \sum_{j=1}^{N} 
\mu_j \cos \th_j ,
\nonumber\\
S_N &=& \sum_{j=1}^{N} 
\mu_j \sin \th_j ,
\nonumber\\
S_{N2} &=& \sum_{j=1}^{N} 
\mu_j \sin 2\th_j.
\label{cN}
\eea
These quantities depend on the number $N$ of masses used.
Note that the effective potential \rf{effpot}
depends only on the distribution of the $N$ masses 
and is independent of the disk mass $M$.

The first part of Eq.\ \rf{effpot} 
is proportional to $\th^2$.
This term contains contributions both from Lorentz violation 
and from the Earth's rotation.
It acts as an effective restoring force for the disk,
producing a time-dependent shift 
to the effective spring constant of the system
that primarily affects free oscillations.
The second part of Eq.\ \rf{effpot}
is proportional to $\th$,
and it contains 
a time-independent contribution from the Earth's rotation
along with a Lorentz-violating piece.
The latter produces a slowly varying 
Lorentz-violating torque $\ta = - dV/d\th$ on the disk.

The oscillations of the system are determined by
the second-order differential equation
\beq
I \fr{d^2\th}{dT^2} + 2 \ga I \fr{d\th}{dT} + \ka \th = \tau,
\label{ho} 
\eeq
where $I$ is the total moment of inertia 
of the disk and masses,
and $\ga$ is the damping parameter of the torsion fiber.
In writing this equation,
we have made use of the relation
$d/dx^{\hat 0} = (dT/dx^{\hat 0})d/dT = (1 + O(2))d/dT$. 

The presence of the damping term in Eq.\ \rf{ho}
ensures that free oscillations vanish in the steady-state solution.
In the absence of additional torques,
the steady-state solution is given by
\bea
\th(T) &=& \sum_n \fr {i_3 m r_0 g} 
{I \sqrt{(\om^2_0-\om^2_n)^2 + 4 \ga^2 \om^2_n} }
\nonumber\\
&&
\times 
\Big( [E_n \sin \al_n - F_n \cos \al_n] \sin (\om_n T + \be_n) 
\nonumber\\
&&
-[E_n \cos \al_n + F_n \sin \al_n] \cos (\om_n T + \be_n) \Big), 
\nonumber\\
\label{th}
\eea
where 
\beq
\be_n = 2 \ga \om_n / (\om^2_n - \om^2_0)
\eeq
and,
as before, 
$T$ is the time coordinate 
in the Sun-centered frame defined in Sec.\ \ref{framesetc}.
Table 5 provides
the amplitudes $E_n$, $F_n$ and the phases $\al_n$
in terms of quantities defined above.

\begin{widetext}
\begin{center}
\begin{tabular}{lc}
\hline
\multicolumn{2}{c} 
{Torsion-Pendulum Amplitudes} \\
\hline
\hline
Amplitude & Phase \\ 
\hline
$E_{2 \om} =  
\sb^{XY} C_N \sin \ch - \frac 14 (\sb^{XX}-\sb^{YY})
S_N \sin 2 \ch $ & $2\phi$ \\
$F_{2 \om} =  
-\fr 12 (\sb^{XX}-\sb^{YY}) C_N \sin \ch 
- \frac 12 \sb^{XY} S_N \sin 2\ch $ & $2\phi$ \\
$E_{\om} =   
\sb^{YZ} C_N \cos \ch  
- \sb^{XZ} S_N \cos 2\ch $ & $\phi$ \\
$F_{\om} =  
-\sb^{XZ} C_N \cos \ch  
- \sb^{ZY} S_N \cos 2\ch $ & $\phi$ \\
$E_{2\om + \Om} = 
- \frac 12 V_{\oplus} \sb^{TX} C_N (1-\cos \et) 
\sin \ch 
- \frac 14 V_{\oplus} \sb^{TY} S_N (1- \cos \et)  \sin 2\ch$ 
& $2\phi$ \\
$F_{2\om + \Om} = 
-\frac 12 V_{\oplus} \sb^{TY} C_N (1- \cos \et) 
\sin \ch
+ \frac 14 V_{\oplus} \sb^{TX} S_N (1-\cos \et) 
\sin 2\ch$ 
& $2\phi$ \\
$E_{2\om - \Om} = 
\frac 12 V_{\oplus} \sb^{TX} C_N (1 + \cos \et) 
\sin \ch
+ \frac 14 V_{\oplus} \sb^{TY} S_N (1+ \cos \et) \sin 2\ch $ 
& $2\phi$ \\
$F_{2\om - \Om} = 
\frac 12 V_{\oplus} \sb^{TY} C_N ( 1 + \cos \et) 
\sin \ch
-\frac 14 V_{\oplus} \sb^{TX} S_N (1+ \cos \et) \sin 2\ch $ 
& $2\phi$ \\
$E_{\om + \Om} =  
\frac 12 V_{\oplus} \sb^{TY} C_N \sin \et \cos \ch
+ V_{\oplus} \sb^{TX} S_N \sin \et (\frac 12 - \cos^2 \ch)$ 
& $\phi$ \\
$\pt{E_{\om + \Om} = \quad} 
+\frac 12 V_{\oplus} \sb^{TZ} C_N (\cos \et -1) \cos \ch $ 
& \\
$F_{\om + \Om} = 
- \frac 12 V_{\oplus} \sb^{TX} C_N \sin \et \cos \ch 
- V_{\oplus} \sb^{TZ} S_N (1 - \cos \et)(\frac 12-\cos^2 \ch)$
& $\phi$ \\
$\pt{F_{\om + \Om} = \quad }
+ V_{\oplus} \sb^{TY} S_N \sin \et (\frac 12-\cos^2 \ch) $ 
& \\
$E_{\om - \Om} =  
\frac 12 V_{\oplus} \sb^{TZ} C_N (1+\cos \et ) \cos \ch
+\frac 12 V_{\oplus} \sb^{TY} C_N \sin \et \cos \ch$
& $\phi$ \\
$\pt{E_{\om - \Om} = \quad}
+ V_{\oplus} \sb^{TX} S_N \sin \et (\frac 12-\cos^2 \ch) $ 
& \\
$F_{\om - \Om} =  
-\frac 12 V_{\oplus} \sb^{TX} C_N \sin \et \cos \ch 
+ V_{\oplus} \sb^{TY} S_N \sin \et (\frac 12-\cos^2 \ch)$
& $\phi$ \\
$\pt{F_{\om - \Om} = \quad}
+ V_{\oplus} \sb^{TZ} S_N (1+ \cos \et)(\frac 12-\cos^2 \ch)$ 
& \\
$E_\Om = 
- \frac 12 V_{\oplus} \sb^{TY} S_N \cos \et \sin 2\ch 
+ V_{\oplus} \sb^{TZ} S_N \sin \et \sin 2\ch $ & $0$ \\
$F_\Om = 
\frac 12 V_{\oplus} \sb^{TX} S_N \sin 2 \ch $ & $0$ \\
\hline
\end{tabular}
\end{center}
\begin{center}
\label{terrosc2}
Table 5.\ Torsion-pendulum amplitudes.
\end{center}
\end{widetext}

To estimate the attainable sensitivities to Lorentz violation,
we suppose the angular deflection of the apparatus
can be measured at the nrad level,
which in a different context has already been achieved 
with a torsion pendulum 
\cite{isql}.
For definiteness,
we suppose that the disk radius is $R=10$ cm
and that the $N=3$ masses have equal magnitude $m_j = M$
and are located at radius $r_0 \approx R$
at angular locations
$\th_1 = \pi/2$, $\th_2 = 4\pi/3$, $\th_3 = 5\pi/3$.
Assuming a natural oscillation frequency 
$\om_0 \approx 10^{-2}$ Hz
and a small damping parameter $\ga \approx 10^{-3}$,
and taking $i_3 = 2/5$ and $m\approx M$,
we find the apparatus could attain sensitivity
to the coefficients $\sb^{JK}$ 
at the level of parts in $10^{15}$ 
and to the $\sb^{TJ}$ coefficients 
at parts in $10^{11}$.
These are idealized sensitivities,
which are unlikely to be achieved in practice.
Nonetheless,
the result is of interest because it is about four orders 
of magnitude better than the (realistic) sensitivities 
that could be attained using current data from lunar laser ranging.

Provided the amplitudes in Table 5 can be separately extracted, 
and assuming measurements 
can be made with two different initial orientations $\th_1$,
it follows that 
{\it these types of torsion-pendulum experiments
can measure 4 coefficients in $\sb^{JK}$ and 
all 3 independent coefficients in $\sb^{TJ}$.}  
As before,
major systematics may include solar and lunar tidal effects.

Modifications of the apparatus proposed above
could lead to improved sensitivities.
One theoretically simple possibility is to decrease 
the effective spring constant of the system.
Different kinds of pendulum could also be used. 
For example,
the ring of point masses around the disk 
could be replaced with a ring of holes instead.
A more radical possibility is to change the geometry
of the apparatus.
For example,
one could replace the disk and masses 
with a thin rod lying in the horizontal plane
and hanging from a torsion fiber. 
If the rod is suspended off center by some means
while able to rotate freely about the vertical axis of the fiber,
the steady-state oscillations again depend on the
coefficients for Lorentz violation.
In fact,
the explicit results for this scenario can be obtained 
from Eq.\ \rf{th} and Table 5 by making the following replacements:
\bea
r_0 &\rightarrow& \half (L-2d),
\nonumber\\
C_N &\rightarrow& \cos \th_0,
\nonumber\\
S_N &\rightarrow& \sin \th_0.
\label{rod}
\eea 
Here,
$L$ is the total length of the rod,
and $d$ is the distance from the end of the rod 
to the point of suspension.
Other quantities still appear but are reinterpreted:
$m$ is the mass of the rod,
$I$ is its moment of inertia,
and $\theta_0$ is the initial angle the rod
subtends to the $\hat x$ axis. 
Again,
the practical issue of suspending a rod off center 
could conceivably be addressed with a rigid support system
or magnetic levitation.  
The former method might allow an increase in sensitivity
by reducing the effective torsion fiber frequency $\om_0$
of the system.
In any case, 
it would be of definite interest to investigate 
various experiments of this type.

Another potentially important issue is whether 
any of the existing torsion-pendulum experiments 
have sensitivity to Lorentz-violating effects. 
Consider, 
for example,
an experiment designed to search for deviations
from Newton's law of gravity at sub-millimeter distances 
\cite{isql}.
The basic component of this experiment is a pair of disks 
positioned one above the other
and with symmetric rings of holes in each.
The upper disk is suspended as a torsion pendulum,
in a manner analogous to that depicted in Fig.\ \ref{fig4},
while the lower one rotates uniformly.
A key effect of the holes is to produce 
a time-varying torque on the pendulum
that varies with the vertical separation between the two disks.
Measurements of the torques 
at the rotation frequency and its harmonics
offers high sensitivity to short-distance deviations 
from Newton's law of gravity.
The question of interest here is whether
the presence of Lorentz violation 
would change the predicted newtonian torque in an observable way.  
A detailed analysis of this experiment is involved
and lies beyond our present scope.
However,
some insight can be gained by using a simplified model
that treats the rings of holes 
as two rings of point masses and comparable radii
positioned one above the other,
with the lower ring rotating.

Consider first the case of two point masses $m_1$ and $m_2$
at coordinate locations $\vec x_1$ and $\vec x_2$.
The modified newtonian potential $V$ 
predicted by the pure-gravity sector of the minimal SME is 
\bea
V = - \fr {G m_1 m_2}{|\vec x_1 - \vec x_2|} 
(1+ \frac 12 \hat x^{\hat j} \hat x^{\hat k} \sb^{\hat j\hat k}) ,
\label{modpotV}
\eea
where $\hat x = (\vec x_1 - \vec x_2) /|\vec x_1 - \vec x_2|$.
The non-newtonian effects are controlled
by the SME coefficients for Lorentz violation $\sb^{\hat j\hat k}$.
Although Eq.\ \rf{modpotV} maintains the inverse-distance behavior 
of the usual newtonian potential,
the associated force is misaligned relative to the 
vector $\hat x$ between the two point masses.
It is therefore conceivable that
an unconventional vertical dependence of the torque 
might arise in the experiment discussed above.
To investigate this rigorously,
the result \rf{modpotV} would be inserted 
into the computer code that models the experiment 
and determines the predicted torque 
\cite{isql}.

In the context of the simplified model
involving two rings of point masses,
the difference between the newtonian and Lorentz-violating torques 
can be explored
assuming a single nonzero coefficient $\sb^{\hat j\hat k}$
in Eq.\ \rf{modpotV}.
We examined the Fourier components of the time-varying torque 
on the upper disk
at harmonics of the rotational frequency of the bottom disk.
The amplitudes of these Fourier components depend sensitively
on the number of masses used, 
on the geometry of the experiment,
and on the particular harmonic considered.
This indicates that a careful analysis involving 
a detailed model of the actual experiment
is necessary for definitive predictions. 
However,
the results suggest that Lorentz-violating effects 
from certain coefficients $\sb^{\hat j\hat k}$ of order 
$10^{-3}$ to $10^{-5}$
could be discernable at a separation of about 100 microns
in experiments of this type,
at the currently attainable torque sensitivities
of about $10^{-17}$ Nm.
Other experiments studying deviations
from short-range newtonian gravity
\cite{jl,sjs}
may also be sensitive to these types of effects.

\subsection{Gyroscope Experiment}
\label{gyroscope experiment}

In curved spacetime,
the spin of a freely falling test body typically
precesses relative to asymptotically flat spacetime
\cite{lis,lehspace}. 
The primary relativistic effects for a gyroscope
orbiting the Earth are the geodetic or de Sitter precession
about an axis perpendicular to the orbit 
and the gravitomagnetic frame-dragging 
or Lens-Thirring precession about the spin axis of the Earth.
In the present context of the post-newtonian metric 
for the pure-gravity sector of the minimal SME given in 
Eqs.\ \rf{g00}-\rf{gjk},
we find that precession also occurs due to Lorentz violation.
In particular,
there can be precession of the spin axis
about a direction perpendicular to both 
the spin axis of the Earth
and the angular-momentum axis of the orbit.

\subsubsection{Theory}
\label{gpbtheory}

To describe the motion of an orbiting gyroscope,
we adopt standard assumptions \cite{mtw} 
and use the Fermi-Walker transport equation given
in coordinate-independent form
by
\beq
{\bf \nabla}_{\bf u} {\bf S} = {\bf u} ({\bf a} \cdot {\bf S}), 
\label{fwt}
\eeq
where ${\bf u}$ is the four-velocity of the gyroscope,
${\bf S}$ is the spin four-vector and 
${\bf a}$ is the four-acceleration.
The dot denotes the inner product with the metric
two-tensor ${\bf g}$.

To obtain the equation of motion for the spin vector ${\bf S}$,
we first construct a local frame ${\bf e}_{\hat \mu}$ 
that is comoving with ${\bf S}$ but is nonrotating
with respect to the fixed stars 
in the asymptotically flat spacetime. 
The construction of this frame 
proceeds as in Sec.\ \ref{general considerations},
using the post-newtonian metric in Eqs.\ \rf{scfmetric}.
The basis can be taken from Eq.\ \rf{ortho3}
by setting the rotation terms to zero.
Using Eq.\ \rf{fwt},
we can then find the proper-time derivative 
of ${\bf S} \cdot {\bf e}_{\hat j}$.
It is given by
\beq
\fr {d {\bf S} \cdot {\bf e}_{\hat j}}{d \ta}
= {\bf S} \cdot ( {\bf \nabla}_{\bf u} {\bf e}_{\hat j} ).
\label{fwt2}
\eeq
This represents the proper time rate of change 
of the spatial spin vector as measured
in a comoving but nonrotating frame.

To display the dependence of the spin precession
on the underlying geometry,
which includes Lorentz-violating terms in the metric,
the right-hand side of Eq.\ \rf{fwt2}
can be expressed in the Sun-centered frame.
This requires expressing the components 
of ${\bf S}$ and ${\bf \nabla}_{\bf u} {\bf e}_{\hat j}$ 
to the appropriate post-newtonian order.
Of particular use for this purpose
are the Sun-centered frame components of 
${\bf e}_{\hat j}$,
which are given by
\bea
e^T_{\pt{T}\hat j} &=& \de^J_{\pt{J}\hat j}
[v^J (1 + h_{TT} + \frac 12 v^2) 
\nonumber\\
&&
\pt{\de^{\hat j J}}
+ \frac 12 h_{JK} v^K + h_{TJ}] + O(4),
\nonumber\\
e^J_{\pt{J}\hat j} &=& \de^K_{\pt{K}\hat j}
[\de_{JK} - \frac 12 h_{JK} + \frac 12 v^J v^K] + O(4).
\label{comp1}
\eea
We can then establish that
\bea
{\bf S}^T &=& v^J S^J + O(2),  
\nonumber\\
{\bf S}^J &=& 
\de^{J}_{\pt{J} \hat j} 
S^{\hat j} + O(2),  
\nonumber\\
({\bf \nabla}_{\bf u} {\bf e}_{\hat j})^T &=& 
\de^{J}_{\pt{J}\hat j} 
a^J + O(3),
\nonumber\\
({\bf \nabla}_{\bf u} {\bf e}_{\hat j})^K &=&  
\de^{J}_{\pt{J}\hat j} 
[ \frac 14 v^{K} \prt_{J} h_{TT} 
-\frac 14 v^{J} \prt_{K} h_{TT} 
+ a^{(J} v^{K)} 
\nonumber\\
&&
+ \prt_{[J} h_{K]T} 
+ v^L \prt_{[J} h_{K]L}]+ O(4).
\label{comp2}
\eea

Combining Eqs.\ \rf{comp1} and \rf{comp2}
via the expression
\beq
{\bf S} \cdot ({\bf \nabla}_{\bf u} {\bf e}_{\hat j})
= S^\Xi ( {\bf \nabla}_{\bf u} {\bf e}_{\hat j} )^{\Pi} g_{\Xi\Pi}
\eeq
yields the general expression 
for the rate of change of the spatial spin vector of a gyroscope,
valid to post-newtonian $O(3)$.
It is given by 
\bea
\fr {dS_{\hat j}}{d\tau} &=& S^{\hat k} 
\de^{K}_{\pt{K} \hat k } \de^{J}_{\pt{J} \hat j} 
\big( v^{[J} a^{K]} 
+ \frac 14 v^{K} \prt_{J} h_{TT}
- \frac 14 v^{J} \prt_{K} h_{TT}
\nonumber\\
&&
\pt{ S^{\hat k} \de^{\hat k K} \de^{\hat j J}}
+ \prt_{[J} h_{K]T} + v^L \prt_{[J} h_{K]L} \big).
\label{gensp}
\eea
In the limit of the usual general-relativistic metric,
this agrees with existing results 
\cite{cmw,mtw}.

\subsubsection{Gyroscope in Earth orbit}
\label{gyroscope in Earth orbit}

In this subsection,
we specialize to a gyroscope 
in a near-circular orbit around the Earth.
For this case,
the acceleration term vanishes.

The first step is to determine the contributions
to the metric potentials \rf{pots} 
from the various relevant bodies in the solar system.  
For simplicity,
we focus on the potentials due to the Earth,
neglecting any contributions from the Sun, Moon,
or other sources,
and we take the Earth as a perfect sphere 
rotating uniformly at frequency $\om$
and with spherical moment of inertia $I_{\oplus}$ 
give by Eq.\ \rf{inertia}.
For convenience in what follows,
we introduce a quantity
${\tilde i}_{\oplus}$ related to $I_{\oplus}$
along with corresponding quantities ${\tilde i}_{(\al)}$
defined for any real number $\al$
by the equations
\bea
{\tilde i}_{\oplus} &=& \fr {I_{\oplus}} {M_{\oplus} r^2_0} ,
\nonumber\\
{\tilde i}_{(\al)} &=& 1 + \al {\tilde i}_{\oplus},
\label{inertcombos}
\eea
where $r_0$ is the mean orbital distance to the gyroscope.

The explicit form is required for five
of the six types of metric potentials in Eq.\ \rf{pots}.
Some general-relativistic contributions from
the potential $W^J$ involving 
the Earth's angular momentum 
$J^K_{\oplus}= 2I_{\oplus}\om^K/3$ 
can be converted into ones involving $V^J$
by using the identity \rf{chiident}.
Also, contributions from the potential $Y^{JKL}$
can be converted into ones involving only the other potentials
by virtue of the antisymmetrization in Eq.\ \rf{gensp}
and the identity \rf{chijident}.
The five necessary Earth potentials 
can be written as functions 
of the spatial position $\vec X$ in the Sun-centered frame,
given by 
\bea
U &=& \fr {GM_{\oplus}}{|\vec X - \vec R|},
\nonumber\\
U^{JK} &=& \fr {GM_{\oplus} (X-R)^J (X-R)^K }{|\vec X - \vec R|^3}
\nonumber\\
&&
-\fr {G I_{\oplus}}{3|\vec X - \vec R|^5} [3 (X-R)^J (X-R)^K
\nonumber\\
&&
\pt{-\fr {G I_{\oplus}}{3|\vec X - \vec R|^5}}
-\de^{JK} |\vec X - \vec R|^2],
\nonumber\\
V^J &=& \fr {GM_{\oplus} V^J_{\oplus}}{|\vec X - \vec R|}
+ \fr {G \ep^{JKL} J^K_{\oplus} (X-R)^L}{2|\vec X - \vec R|^3},
\nonumber\\
W^J &=& V_{\oplus}^K U^{KJ} + \ldots,
\nonumber\\
X^{JKL} &=& V_{\oplus}^J U^{KL} + \ldots ,
\label{wjkle}
\eea
where $\vec R= R\hat R$ 
is the Earth's location in the Sun-centered frame
and $V^J_{\oplus}$ is its velocity.
Explicit expressions for these quantities
are given in Eq.\ \rf{R} of Appendix \ref{dlso}
and in the associated discussion.

The ellipses in Eqs.\ \rf{wjkle} 
represent omitted pieces of the potentials arising 
from the purely rotational component of the Earth's motion,
which generate gravitomagnetic effects involving Lorentz violation.
A crude estimate reveals that these gravitomagnetic terms 
are suppressed relative to
the associated geodetic terms displayed in Eqs.\ \rf{wjkle} 
by about two orders of magnitude.
As a result,
although gravitomagnetic frame-dragging terms are crucial 
for tests of conventional general relativity
\cite{mtw,cmw},
they can be neglected for studies of the
dominant Lorentz-violating contributions to the spin precession.
For simplicity,
the frame-dragging precession effects involving Lorentz violation
are omitted in what follows.

For use in Eq.\ \rf{gensp},
these potentials must be evaluated along the 
worldline of the gyroscope.
At post-newtonian order, 
this worldline is adequately described 
by the expressions for an arbitrary satellite orbit 
given in Appendix \ref{dlso}.
Thus,
the value of $(\vec X - \vec R)$ along the worldline
can be written as 
\beq
(\vec X -\vec R)_{\rm gyro} = r_0 \hat \rh,
\label{gyropos}
\eeq
where $\hat \rh$ is given by Eq.\ \rf{rhohat}.
To post-newtonian order,
the gyroscope velocity $v^J_{\rm gyro}$ 
can be taken as
$v^J_{\rm gyro} = V^J_{\oplus}+v^J_0$,
where $v^J_0$ is the mean orbital velocity 
defined by $\vec v_0 = d(r_0 \hat\rh)/dT = v_0 \hat \ta$.
The normal to the plane of the orbit is 
$\hat \si = \hat \rh \times \hat \ta$.

Inserting the potentials \rf{wjkle} into Eq.\ \rf{gensp}
and evaluating along the worldline
yields a result that can be separated into two pieces,
\bea
\fr {dS_{\hat j}}{d\ta} &=& 
\fr {dS^E_{\hat j}}{d\ta} 
+\fr {dS^{\sb}_{\hat j}}{d\ta} .
\label{dsdtsplit}
\eea
The first piece is the standard result
from general relativity, 
given by
\bea
\fr {dS^E_{\hat j}}{d\ta} &=& 
g \de^{J}_{\pt{J}\hat j} S^K 
[( -3 v^{[K}_0 \hat \rh^{J]} 
+ V^{[K}_{\oplus} \hat \rh^{J]}) 
\nonumber\\
&&
\qquad
- \fr{2 J_{\oplus}^L} {M_\oplus r_0} 
( \ep^{JKL} + 3 \ep^{LM[J} \hat \rh^{K]} \hat \rh^M )] ,
\nonumber\\
\eea
where the reference gravitational acceleration $g$ 
is now 
\beq
g=G M_{\oplus}/r^2_0.
\eeq
The second piece
is proportional to 
the coefficients for Lorentz violation $\sb^{JK}$,
and it takes the form
\bea
\fr {dS^{\sb}_{\hat j}}{d\ta} &=& 
g \de^{J}_{\pt{J}\hat j} S^K 
\Big[
2\sb^{T[K} \hat \rh^{J]}
+ 4 {\tilde i}_{(-1)} \sb^{L[J}v^{K]}_0 \hat \rh^L 
\nonumber\\
&&
\hskip -5 pt
+ v^{[K}_0 \hat \rh^{J]} 
(- \frac 52 {\tilde i}_{(3/5)} \sb^{TT} 
-\frac 92 {\tilde i}_{(-5/3)} \sb^{LM} 
\hat \rh^L \hat \rh^M) 
\nonumber\\
&&
\hskip -5 pt
+ {\tilde i}_{(1)} \sb^{L[K} \hat \rh^{J]} v^L_0
-2 \sb^{L[K}\hat \rh^{J]}V^L_{\oplus}
+ {\tilde i}_{(-1)} \sb^{L[K} V^{J]}_{\oplus} \hat \rh^L
\nonumber\\
&&
\hskip -5 pt
+ V^{[K}_{\oplus} \hat \rh^{J]} 
(-\frac 12 \sb^{TT} {\tilde i}_{(-1)}
+ \frac 32 {\tilde i}_{(-5/3)} \sb^{LM} \hat \rh^L \hat \rh^M )
\Big].
\nonumber\\
\eea

Among the effects described by the result \rf{dsdtsplit}
are oscillations of the orientation of the gyroscope spin vector 
that occur at frequencies $\om$, 
$\om \pm \Om_{\oplus}$, and higher harmonics.
At the frequency $\om$,
these oscillations change the angular orientation of a gyroscope 
by roughly $\de \th \approx 2\pi v_0^2$.
For a gyroscope in low Earth orbit,
this is $\de \th \approx 8 \times 10^{-4}$ arcseconds,
well below observable levels for existing sensitivity
to nonsecular changes.

We therefore focus instead on the dominant secular contributions 
to the motion of the gyroscope spin.
Averaging various key quantities over an orbital period
yields the following results:
\bea
<r^J_0 r^K_0> &=& \half r_0^2 
(\de^{JK} - \hat \si^J \hat \si^K), 
\nonumber\\
<r^J_0 v^K_0> &=& \half r_0 v_0 \ep^{JKL} \hat \si^L, 
\nonumber\\
<r^K_0 r^{[L}_0 v^{M]}_0 r^N_0> &=& 
\frac 14 r^3_0 v_0 \ep^{LMP} \hat \si^P 
(\de^{KN} - \hat \si^K \hat \si^N), 
\nonumber\\
<r^J_0 r^K_0 V_{\oplus}^L r^M_0> &\approx& 0,
\nonumber\\
<r^J_0 V_{\oplus}^K> &\approx & 0,
\nonumber\\
<r^J_0> &=& 0.
\label{avgs} 
\eea
In these equations,
the vector $\hat \si^J$
is given explicitly as
\bea
\hat \si^J &=& (\sin \al \sin \be,
-\cos \al \sin \be,
\cos \be),
\eea
where the angles $\al$ and $\be$ are those defined 
in Sec.\ \ref{llr} and Fig.\ \ref{fig3}.
Note that the averaging process
eliminates the terms involving $V^J_{\oplus}$.
In the conventional case,
such terms can be shown to be unobservable in principle 
by a consideration of the physical precession referenced 
to the fixed stars 
\cite{cmw,cmwgyro}.
It is an open question whether this stronger result 
remains true in the presence of Lorentz violation.

Combining the above results yields the vector equation 
for the secular evolution of the gyroscope spin.
It can be written in the form
\beq
\fr {d\vec S}{dt} = g v_0 \vec \Om \times \vec S .
\label{spprec}
\eeq 
The secular precession frequency $\vec \Om$ 
is comprised of two pieces,
\bea 
\Om^J &=& 
\Om^J_E + \Om^J_{\sb}.
\label{om}
\eea
The first term $\Om^J_E$ contains precession 
due to conventional effects in general relativity,
and it is given by
\bea
\Om^J_E &=& \frac 32 \hat \si^J
+\fr{J_\oplus^K}{2M_\oplus r_0 v_0}
(\de^{JK} - 3 \hat \si^J \hat \si^K).
\label{omE}
\eea
The second term $\Om^J_{\sb}$
contains contributions from 
the coefficients for Lorentz violation $\sb^{JK}$,
and it has the form
\bea
\Om^J_{\sb} &=& 
\frac 98 ({\tilde i}_{(-1/3)} \sb^{TT} 
-{\tilde i}_{(-5/3)} \sb^{KL}
\hat \si^K \hat \si^L) \hat \si^J
\nonumber\\
&&
+ \frac 54 {\tilde i}_{(-3/5)} \sb^{JK} \hat \si^K .
\label{oms}
\eea
In the limit of no Lorentz violation,
Eq.\ \rf{om} reduces as expected 
to the conventional result \rf{omE} of general relativity 
involving the geodetic precession 
arising from the spacetime curvature near the Earth
and the frame-dragging precession 
arising from the rotation of the Earth. 

\subsubsection{Gravity Probe B}
\label{gpb}

Gravity Probe B (GPB) 
\cite{gpb}
is an orbiting-gyroscope experiment
that has potential sensitivity to certain combinations of coefficients
contained in Eq.\ \rf{oms}.
Our general results \rf{spprec} for the spin precession
can be specialized to this experiment.
In terms of the angles in Fig.\ \ref{fig3},
the GPB orbit is polar with 
$\al=163^{\circ}$ and $\be=90^{\circ}$.
One consequence of the polar orbit 
is that the normal vector $\hat \si^J$ to the orbital plane 
appearing in Eq.\ \rf{oms} reduces to the simpler form
\bea
\hat \si^J &=& (\sin \al, -\cos \al, 0).
\eea

The terms in Eq.\ \rf{spprec} proportional
to $\hat \si \times \vec S$
act to precess the gyroscope spin in the plane of the orbit.
They include the geodetic precession in Eq.\ \rf{omE}
predicted by general relativity,
along with modifications due to Lorentz violation.
The magnitude of the precession is determined by the 
projection $\hat \si \cdot \vec \Om$.
The explicit form of the relevant projection of $\vec \Om$
arising due to Lorentz violation is found to be 
\bea
\Om^{\si}_{\sb} &=&
\frac 98 {\tilde i}_{(-1/3)} \sb^{TT} 
+ \frac 18 {\tilde i}_{(9)} \sb^{JK} \hat \si^J \hat \si^K.
\eea

The terms in Eq.\ \rf{spprec} proportional to $\hat J \times \vec S$,
where $\vec J$ is the Earth's angular momentum,
cause precession of the gyroscope spin
in the plane perpendicular to $\vec J$.
Since $\vec J$ lies along $\hat Z$,
this is the $XY$ plane of Fig.\ \ref{fig3}.
These terms include the frame-dragging effect in Eq.\ \rf{omE} 
predicted by general relativity,
together with effects due to Lorentz violation.
These have magnitude determined by the projection 
$\hat Z \cdot \vec \Om$.
For the contributions due to Lorentz violation,
we find the explicit form
\bea
\Om^{Z}_{\sb} &=& 
\frac 54 {\tilde i}_{(-3/5)} \sb^{ZK} \hat \si^K.
\eea

In addition to the above,
however,
Lorentz violation induces a third type of precession
that is qualitatively different from those predicted by 
general relativity.
This is a precession of the gyroscope spin 
in the plane with normal $\hat n$
along $\hat \si \times \hat J$.
With the definitions shown in Fig.\ \ref{fig3},
this normal is given by $\hat n = (\cos \al , \sin \al, 0)$.
The magnitude of the associated precession
is given by $\hat n \cdot \vec \Om$.
We obtain 
\bea 
\Om^{n}_{\sb} &=& 
\frac 54 {\tilde i}_{(-3/5)} \sb^{JK} \hat n^J \hat \si^K .
\eea

The GPB experiment is expected to attain sensitivity 
to secular angular precessions 
of the order of $5 \times 10^{-4}$ arcseconds per year.
If the spin precession in the three orthogonal directions
can be disentangled,
GPB could produce the first measurements
of three combinations of the $\sb^{JK}$ coefficients,
including one involving precession about the $\hat n$ direction.
The attainable sensitivity to $\sb^{JK}$ coefficients
is expected to be at the $10^{-4}$ level.

\subsection{Binary pulsars}
\label{binary pulsars}

Binary pulsars form a useful testing ground 
for general relativity 
\cite{bp1,cmw01}.
This subsection examines the possibility
of probing SME coefficients for Lorentz violation 
using pulsar timing data
\cite{gl2}.
We find that Lorentz violation induces secular variations 
in the orbital elements of binary-pulsar systems,
along with modifications to the standard pulsar timing formula.
The discussion is separated into five parts:
one establishing the basic framework and assumptions,
a second presenting the secular changes,
a third describing the timing formula,
a fourth determining some experimental effects,
and a final part explaining the transformation
between the frame used to study the properties of the binary pulsar
and the standard Sun-centered frame.

The reader is cautioned that
the notation for the binary-pulsar systems 
in this subsection
is chosen to match the literature
and differs in certain respects from the notation
adopted elsewhere in the paper.
This notation is described 
in Sec.\ \ref{basicsbp}.

\subsubsection{Basics}
\label{basicsbp}

The general problem of describing the dynamics of binary pulsars
is complicated by the strong-field gravitational 
properties of the spacetime regions
close to the pulsar and its companion.
Nonetheless, 
provided the typical radius of the pulsar and companion 
are much smaller than the average orbital distance between them,
the orbital dynamics can be modeled 
with an effective post-newtonian description.

One standard approach is to construct a modified 
Einstein-Infeld-Hoffman (EIH) lagrangian 
describing the post-newtonian dynamics of the system 
\cite{cmw}.
In principle,
an analysis of the effects of Lorentz violation 
could be performed via a modified EIH approach 
based on the action of the pure-gravity sector of the minimal SME.
This would involve generalizing Eq.\ \rf{lpp} 
to allow for the usual EIH effects,
along with modifications to the Lorentz-violating behavior 
arising from the structure of the pulsar and its companion.
A detailed analysis along these lines
would be of interest but lies beyond the scope of the present work.

In fact, 
a simpler approach suffices to extract 
key features arising from Lorentz violation.
Instead of a full EIH-type treatment,
we adopt in what follows a simplified picture and 
consider only the effects arising 
in a point-mass approximation 
for each of the two bodies in the system.
To retain some generality and an EIH flavor in the treatment,
we interpret the gravitational constant,
the masses,
and the coefficients for Lorentz violation
as effective quantities 
\cite{bp3}.
However,
the results disregard details of possible 
strong-field gravitational effects 
or consequences of nonzero multipole moments 
and tidal forces across the pulsar and companion. 
The reader is cautioned that 
the corresponding conventional effects from newtonian theory
can be crucial for studies of binary pulsars,
so they should be included in a complete analysis.
Nonetheless,
the results obtained here offer a good baseline
for the sensitivities to Lorentz violation 
that might be achieved 
through observations of binary pulsars. 

\begin{figure}
\centerline{\psfig{figure=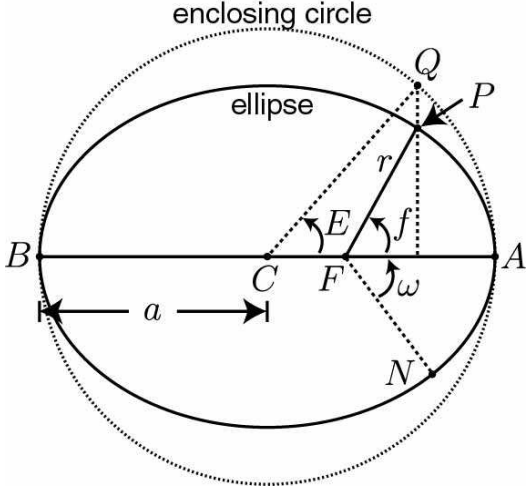,width=0.8\hsize}}
\caption{\label{eccentricanomaly} 
Definitions for the elliptical orbit.}
\end{figure}

The analysis that follows makes use of 
the standard keplerian characterization of an elliptical orbit
\cite{bp2}.
Some of the key quantities are displayed in 
Fig.\ \ref{eccentricanomaly}.
In this figure,
the point P is the location of the pulsar at time $t$,
and it lies a coordinate distance $r$ from the focus F
of the elliptical orbit.
The line segments CA and CB are the semimajor axes,
each of length $a$.
The eccentric anomaly $E$ 
is defined as the angle ACQ, 
where Q is the point on the enclosing circle
with the same abcissa as P.
The true anomaly $f$ is
defined as the angle AFP,
while $\om$ 
is the angle subtended between 
the line of ascending nodes FN and the major axis
of the ellipse.

The general solution of the elliptic two-body problem
is given in terms of the time $t$
and six integration constants
called the orbital elements.
Three of the latter are associated with 
inherent properties of the ellipse:
the semimajor axis length $a$,
the eccentricity $e$,
and a phase variable $l_0$
called the mean anomaly at the epoch.
The other three describe the position
of the ellipse in the reference coordinate system:
the inclination $i$,
the longitude of the ascending node $\Om$,
and the angle $\om$ defined above.  
The orbital frequency is denoted $n$,
and it is related as usual to the period $P_b = 2\pi/n$.
Standard expressions relate these various quantities.
For example,
the eccentric anomaly $E$ is related 
to the mean anomaly $l_0$ at the epoch $t = t_0$ by 
\beq
E-e \sin E = l_0 + n (t-t_0).
\label{mean anomaly}
\eeq
The reader is referred to 
Refs.\ \cite{bp2} 
for further details.

\subsubsection{Secular changes}
\label{secular changes}
 
We begin the analysis with the derivation 
of the secular changes in the orbital elements 
of a binary-pulsar system 
that arise from Lorentz violation.
Adopting the point-mass limit as described above,
Eq.\ \rf{nbody} can be used to determine 
the relative coordinate acceleration 
of the pulsar and companion,
which are taken to be located at
$\vec r_1$ and $\vec r_2$,
respectively.
The result can be written in the form of Eq.\ \rf{esacc},
with the quadrupole moment set to zero 
and the effects from external bodies omitted.

It is convenient for the analysis 
to choose a post-newtonian frame 
with origin at the center of mass of the binary-pulsar system.
The newtonian definition of the center of mass suffices
here because the analysis is perturbative.
The frame is taken to be asymptotically inertial as usual,
so that the coefficients for Lorentz violation 
obey Eqs.\ \rf{conds}.
The coordinates in this frame are denoted $x^\mu = (t,x^j)$,
and the corresponding basis is denoted ${\bf e}_\mu$. 
The issue of matching the results to the Sun-centered frame
is addressed in Sec.\ \ref{transformation} below.

In this post-newtonian binary-pulsar frame,
the relative acceleration of the pulsar and companion 
can be written
\bea
\fr {d^2 r^j}{dt^2} &=& -\fr {GM}{r^3}[(1+ \frac 32 \sb^{00})r^j
-\sb^{jk} r^k + \frac 32 \sb^{kl} \hat r^k \hat r^l r^j]
\nonumber\\
&&
+ \fr{2G \de m}{r^3} [\sb^{0k} v^k r^j-\sb^{0j} v^k r^k]
+ \ldots ,
\label{bpeqns}
\eea
where the definitions of $\vec r$, $r$, $\vec v$, 
$M$ and $\de m$ parallel those adopted 
in Table 1 of Sec.\ \ref{llr}.
The ellipses here represent corrections from multipole moments, 
from general relativistic effects at $O(4)$,
and from Lorentz-violating effects at $O(4)$. 
Note that enhancements of secular effects 
stemming from general relativistic corrections at $O(4)$ 
are known to occur under some circumstances 
\cite{bp3},
but the issue of whether 
$O(4)$ enhancements can occur in the SME context 
is an open question lying beyond our present scope.

To establish the observational effects of Lorentz violation
that are implied by Eq.\ \rf{bpeqns},
we adopt the method of osculating elements 
\cite{bp2}.
This method assumes the relative motion of 
the binary pulsar and companion 
is instantaneously part of an ellipse.
The subsequent motion is obtained by letting 
the orbital elements vary with time 
while keeping the equations of the ellipse fixed.
In the present instance,
the unperturbed ellipse solves the equation
\beq
\fr {d^2 r^j}{dt^2} = -\fr {GM}{r^3}(1+ \frac 32 \sb^{00})r^j.
\label{unpertacc}
\eeq
Note that this implies the frequency and the semimajor axis 
of the orbit are related by
\beq
n^2 a^3 = GM(1 + \frac 32 \sb^{00}).
\label{period}
\eeq
The remaining terms 
are then viewed as a perturbing acceleration $\al'^j$,
given in this case by 
\bea
\al'^j &=& 
\fr {GM}{r^3}\sb^{jk} r^k 
- \frac 32 \fr {GM}{r^3}\sb^{kl} \hat r^k \hat r^l r^j
\nonumber\\
&&
+ \fr{2G \de m}{r^3} [\sb^{0k} v^k r^j-\sb^{0j} v^k r^k].
\label{pertacc}
\eea
The idea is to extract from this equation the variations 
in the six orbital elements 
$a$, $e$, $l_0$, $i$, $\Om$, $\om$
describing the instantaneous ellipse for the relative motion.

\begin{figure}
\centerline{\psfig{figure=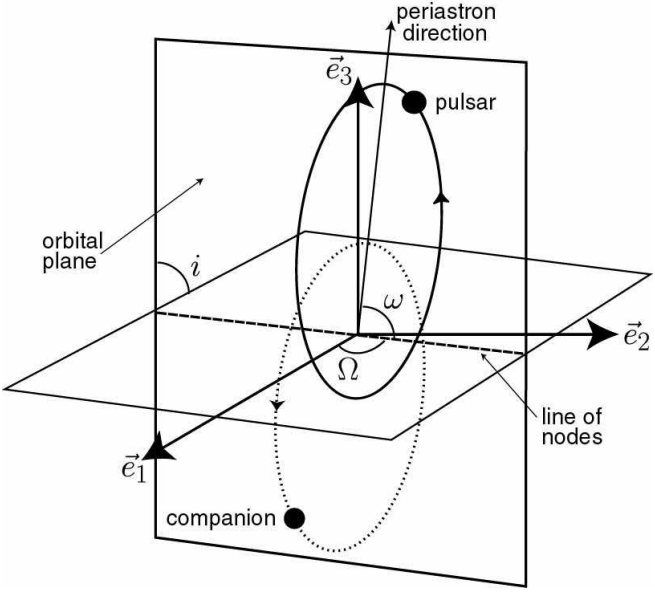,width=\hsize}}
\caption{\label{bp} 
Elliptical orbits in the post-newtonian frame of the binary pulsar.}
\end{figure}

The elliptical orbits of the pulsar and companion,
referred to the chosen post-newtonian frame,
are illustrated in Fig.\ \ref{bp}.
The relative motion can also be represented by
an ellipse in the same plane.
To specify the orientation of the orbit,
three linearly independent unit vectors are needed.
A natural set includes
one vector $\vec k$ normal to the orbital plane, 
a second vector $\vec P$ 
pointing from the focus to the periastron,
and a third vector $\vec Q = \vec k \times \vec P$
perpendicular to the first two.  
This triad of vectors can be expressed in terms 
of the spatial basis 
${\bf e}_j$ 
for the post-newtonian frame.
Explicitly,
these vectors are given by
\bea
\vec P &=& (\cos \Om \cos \om - \cos i \sin \Om \sin \om) \vec e_1
\nonumber\\
&&
+ (\sin \Om \cos \om + \cos i \cos \Om \sin \om) \vec e_2
\nonumber\\
&&
+ \sin i \sin \om \vec e_3,
\nonumber\\
\vec Q &=& -(\cos \Om \sin \om + \cos i \sin \Om \cos \om) \vec e_1
\nonumber\\
&&
+ (\cos i \cos \Om \cos \om -\sin \Om \sin \om) \vec e_2
\nonumber\\
&&
+ \sin i \cos \om \vec e_3,
\nonumber\\
\vec k &=& \sin i \sin \Om \vec e_1
- \sin i \cos \Om \vec e_2
\nonumber\\
&&
+ \cos i \vec e_3.
\label{k}
\eea
In terms of these vectors,
the unperturbed elliptical orbit can be expressed as 
\bea
\vec r &=& \fr {a(1-e^2)}{1+e \cos f} (\vec P \cos f + \vec Q \sin f).
\label{r}
\eea

Following standard methods,
we insert the solution $\vec r$ for the elliptic orbit
given by Eq.\ \rf{r} 
into the expression \rf{pertacc} 
for the perturbative acceleration,
project the result as appropriate to extract the orbital elements,
and average over one orbit.
The averaging is performed via integration over 
the true anomaly $f$,
which serves as the phase variable.
Details of this general procedure are discussed 
in Refs.\ \cite{bp2}. 

After some calculation,
we find the secular evolutions 
of the semimajor axis and the eccentricity
are given by 
\bea
\langle \fr {da}{dt} \rangle &=& 0,
\nonumber\\
\langle \fr {de}{dt} \rangle &=& 2 n (1-e^2)^{1/2}
\Big[\fr {(e^2 - 2\ff )}{2e^3}\sb_{PQ} 
- \fr {\de m}{M} \fr {na \ff } {e^2}\sb_P  \Big],
\nonumber\\
\label{edot}
\eea
where the eccentricity function $\ff(e)$ 
is given by
\bea
\ff(e) &=& 1-(1-e^2)^{1/2}.
\label{eccfns}
\eea
The secular evolutions of
the three orbital elements associated with the orientation
of the ellipse relative to the post-newtonian frame are 
\bea
\langle \fr {d \om}{dt} \rangle &=& 
- \fr {n}{\tan i (1-e^2)^{1/2}}
\Big[ \fr {\ff} {e^2} \sb_{kP} \sin \om 
\nonumber\\
&&
\qquad
+ \fr {(e^2- \ff)} {e^2} \sb_{kQ} \cos \om 
- \fr {\de m}{M} \fr {2n a \ff} {e} \sb_k \cos \om 
\Big]
\nonumber\\
&&
- n \Big[ 
\fr {(e^2-2\ff)} {2e^4} (\sb_{PP}-\sb_{QQ}) 
\nonumber\\
&&
\qquad\qquad
+ \fr {\de m}{M} \fr { 2na (e^2-\ff)}{e^3 (1-e^2)^{1/2}} \sb_Q  \Big],
\nonumber\\
\langle \fr {di}{dt} \rangle &=& 
\fr {n \ff} {e^2 (1-e^2)^{1/2}}
\sb_{kP} \cos \om 
- \fr {n \ff} {e^2} \sb_{kQ} \sin \om 
\nonumber\\
&&
+ \fr {\de m}{M} \fr {2 n^2 ae \ff} {e^2 (1-e^2)^{1/2}} \sb_k \sin \om,
\nonumber\\
\langle \fr {d \Om }{dt} \rangle &=& 
\fr {n}{\sin i (1-e^2)^{1/2}}
\Big[ \fr {\ff} {e^2} \sb_{kP} \sin \om 
\nonumber\\
&&
+ \fr {(e^2-\ff)} {e^2} \sb_{kQ} \cos \om 
-  \fr {\de m}{M} \fr {2 na \ff } {e} \sb_k \cos \om \Big].
\nonumber\\
\label{Omdot}
\eea
Finally,
the secular evolution of the mean anomaly at the epoch is
\bea
\langle \fr {d l_0 }{dt} \rangle &=& 
\frac 12 n (\sb_{PP} + \sb_{QQ})
\nonumber\\
&&
+
\fr {n (e^2 - 2\ff ) (1 - e^2 -\ff )} {2e^4} (\sb_{PP}-\sb_{QQ}) 
\nonumber\\
&&
+  \fr {\de m}{M} \fr { 2n^2a (e^2-\ff )}{e^3 } \sb_Q .
\label{l0dot}
\eea
In all these expressions,
the coefficients for Lorentz violation 
with subscripts $P$, $Q$, and $k$
are understood to be appropriate projections of 
$\sb^{\mu\nu}$ along the unit vectors 
$\vec P$, $\vec Q$, and $\vec k$,
respectively.
For example,
$\sb_P \equiv \sb^{0j}P^j$,
while $\sb_{QQ} \equiv \sb^{jk} Q^j Q^k$.

The analysis of the timing data from binary-pulsar systems
permits the extraction of information
about the secular changes in the orbital elements.
In the present context,
performing this analysis for a given binary pulsar
would permit measurements 
of the projected combinations of coefficients for Lorentz violation
appearing in the equations above.
This is discussed further in Sec.\ \ref{bp experiment}.
Matching the projections of the coefficients
to standard components in the Sun-centered frame 
can be performed with a Lorentz transformation,
described in Sec.\ \ref{transformation}.

\subsubsection{Timing formula}
\label{timing formula}

Accompanying the above secular changes in the orbital elements 
are modifications to the standard pulsar timing formula
arising from Lorentz violation in 
the curved spacetime surrounding the binary pulsar.
The basic structure of the pulsar timing formula 
can be derived by considering the trajectory 
of a photon traveling from the emitting pulsar to the Earth.
Together with the relation between proper time 
at emission and arrival time on Earth,
this trajectory is used to determine 
the number of pulses as a function of arrival time.  
For simplicity,
the derivation that follows neglects 
any effects on the photon trajectory from gravitational influences 
of bodies in the solar system 
and effects from the motion of the detector during measurement.

The photon trajectory is determined 
by the null condition $ds^2 = 0$.
The relevant metric is the post-newtonian result 
of Eqs.\ \rf{g00}, \rf{g0j}, \rf{gjk} taken at $O(2)$
and incorporating effects of the perturbing companion 
of mass $m_2$ at $\vec r_2$.
In practice,
the time-delay effects from the pulsar itself
produce only a constant contribution to the result
and can be disregarded.
The null condition can then be solved for $dt$
and integrated from the emission time $t_{\rm em}$
at the pulsar located at $\vec r_1$
to the arrival time $t_{\rm arr}$ at the Earth at $\vec r_E$.
Assuming
$|\vec r_E| \gg |\vec r_1|$ and $|\vec r_E| \gg |\vec r_2|$,
we find the time delay $t_{\rm arr} - t_{\rm em}$ 
for the photon is given by 
\bea
t_{\rm arr} - t_{\rm em} &=& |\vec r_E - \vec r_1|
+ G m_2 \De^{jk} \hat n^j \hat n^k \ln \left[ \fr {2 r_E}
{r+ \hat n \cdot \vec r} \right].
\nonumber\\ 
\label{bptimedelay}
\eea
This result is a simplification of the 
general time-delay formula,
which is derived in Sec.\ \ref{time-delay effect} below.
In Eq.\ \rf{bptimedelay},
$r_E = |\vec r_E|$ is the coordinate distance to the Earth,
$\vec r = \vec r_1 - \vec r_2$ is the pulsar-companion separation
as before,
and $\hat n$ is a unit vector along $\vec r_E- \vec r_1$.
The quantity $\De^{jk}$ contains combinations 
of coefficients for Lorentz violation and $\hat n$,
given explicitly in Eq.\ \rf{Dejk} of
Sec.\ \ref{time-delay effect}.

We seek an expression relating the photon arrival time
to the proper emission time.
The differential proper time $d\ta$ as measured by an ideal clock 
on the surface of the pulsar is given in terms of 
the differential coordinate time $dt$ by the expression
\beq
d\ta = dt [ 1 - (1 + \frac 32 \sb^{00})U  
- \frac 12 \sb^{jk} U^{jk}-\frac 12 \vec v^2].
\label{proptime1}
\eeq
The potentials appearing in this expression 
include those from both the pulsar and its companion,
and they are to be evaluated along the worldline 
of the ideal clock. 
It suffices here to display the explicit potentials 
for the companion,
so that
\bea
U &=& \fr {Gm_2}{r} + U_1,
\nonumber\\
U^{jk} &=& \fr {Gm_2 r^j r^k }{r^3} + U^{jk}_1,
\label{bppots}
\eea
where $U_1$ and $U^{jk}_1$
are the contributions from the pulsar.
In Eq.\ \rf{proptime1},
$\vec v$ is the velocity of the clock,
which can be identified with the velocity $\vec v_1$
of the pulsar.
Using the time derivative of Eq.\ \rf{r} to obtain $\vec v_1$
and explicitly evaluating Eq.\ \rf{proptime1} 
along the pulsar trajectory,
we obtain
\bea
\fr {d\ta}{dt} &=& 1
- (1+ \frac 32 \sb^{00}) \fr {G m^2_2}{Mr}
\nonumber\\
&&
- G m_2 \Big[ (1+ \frac 32 \sb^{00}) \fr {1}{r}
+ \frac 12 \fr {\sb^{jk} r^j r^k}{r^3} \Big].
\nonumber \\
\label{proptime2}
\eea
All effects from the pulsar itself contribute to constant
unobservable shifts and have been neglected in this result.
Note that the factors of $(1+ \frac 32 \sb^{00})$
appearing in Eq.\ \rf{proptime2} 
merely act to scale $G$ and hence are unobservable
in the present context.
For simplicity in what follows,
we absorb them into the definition of $G$.

To integrate Eq.\ \rf{proptime2},
it is convenient to change variables 
from the coordinate time
to the eccentric anomaly $E$.
This makes it possible to insert the expression 
\beq
\vec r= a(\cos E - e) \vec P + a (1-e^2)^{1/2} \sin E ~\vec Q
\eeq
into Eq.\ \rf{proptime2} and perform the integration over $E$.
Up to constants,
we obtain
\bea
\ta &=& t - \fr {G m_2 P_b}{2 \pi a} E
- \fr {G m_2 P_b}{4 \pi a} \fr {\sb_{PP}-\sb_{QQ}(1-e^2)}{e^2} E
\nonumber\\
&&
- \fr {G m_2 P_b}{4 \pi a}(\sb_{PP}-\sb_{QQ}) \cF_1 (E)
\nonumber\\
&&
- \fr {G m_2 P_b}{2\pi a} (1-e^2)^{1/2} \sb_{PQ} \cF_2 (E). 
\label{proptime3}
\eea
The functions $\cF_1$ and $\cF_2$ appearing 
in the last two terms are given by
\bea
\cF_1 (E) &=& \fr {(1-e^2)\sin E}{e(1-e \cos E)} 
-\fr {2(1-e^2)^{1/2}}{e^2} 
\nonumber\\
&&
\times
\arctan \Big[ \Big( \fr {1+e}{1-e}\Big)^{1/2} \tan \frac 12 E \Big],
\nonumber\\
\cF_2 (E) &=& \fr {e- \cos E}{e(1-e \cos E)}
-\fr {\ln (1 - e \cos E)}{e^2}.
\label{f2}
\eea
Note that in the circular limit $e\to 0$ 
the result \rf{proptime3} remains finite,
despite the seeming appearance of poles in $e$.
The constant in the definition \rf{mean anomaly}
is fixed by requiring 
\bea
E-e \sin E = n (t_{\rm arr}-r_E).
\label{E}
\eea
The integration assumes that the positions of the pulsar
and its companion are constant over the photon travel time
near the binary-pulsar system. 

The desired expression for the proper time $\ta$ is obtained
by combining the result \rf{bptimedelay}
with Eq.\ \rf{proptime3}
to yield an expression as a function of the arrival time.
It turns out that the Lorentz-violating corrections to the 
Shapiro time delay, 
which arise from the last term of Eq.\ \rf{bptimedelay},
are negligibly small for the cases of interest
and so can be disregarded.
We find
\bea
\ta &=& t_{\rm arr} - \cA (\cos E-e) 
- (\cB+\cC) \sin E 
\nonumber\\
&&
- \fr{2\pi}{P_b}
\fr{(\cA \sin E- \cB \cos E)}
{(1-e \cos E)}
[\cA(\cos E-e) + \cB \sin E] 
\nonumber\\
&&
\hskip -17 pt
- \fr {\pi a a_1}{P_b}[(\sb_{PP}-\sb_{QQ})\cF_1(E)
+2\sqrt{(1-e^2)} ~\sb_{PQ} \cF_2(E)].
\nonumber\\ 
\label{tarr}
\eea
In this expression,
$a_1=a m_2/M$
and the quantities $\cA$, $\cB$, and $\cC$ are given by
\bea
\cA &=& a_1 \sin i \sin \om,
\nonumber\\
\cB &=& \sqrt{(1-e^2)} ~a_1 \sin i \cos \om,
\nonumber\\
\cC &=& \fr {2 \pi a a_1}{P_b} \Big[e 
+ \fr {\sb_{PP}-\sb_{QQ}(1-e^2)}{2e}\Big].
\label{C}
\eea
The reader is reminded that all these equations
are to be used in conjunction with the method 
of osculating elements,
for which the orbital elements vary in time.

In practice,
the application of the above results 
involves inserting the result \rf{tarr} into 
an appropriate model for the number $N(\ta)$ of emitted pulses 
as a function of pulsar proper time $\ta$.
The model for $N(\ta)$ is usually of the form 
\beq
N(\ta) = N_0 + \nu \tau + \dot{\nu} \tau^2/2 + \ldots,
\label{timing}
\eeq
where $\nu$ is the frequency of the emitted pulses.
The result of the substitution is an expression 
representing the desired timing formula 
for the number of pulses $N(t_{\rm arr})$
as a function of arrival time.

\subsubsection{Experiment}
\label{bp experiment}

We are now in a position to consider some 
experimental implications of our results.
Consider first the timing formula \rf{timing}
with the substitution of the expression \rf{tarr}
for the proper time.
Modifications to this formula
arising from Lorentz violation can be
traced to two sources.
First,
the terms in Eqs.\ \rf{tarr} and \rf{C} 
proportional to $\sb_{PP}$, $\sb_{QQ}$, and $\sb_{PQ}$
represent direct corrections to the proper time 
arising from Lorentz violation.  
Second,
the structural orbital elements 
entering the expression for the proper time $\ta$ 
acquire secular Lorentz-violating corrections,
given in Eqs.\ \rf{edot} and \rf{Omdot}.
Both these sources of Lorentz violation
affect the predictions of the timing formula. 

In applications,
the timing formula $N$ can be regarded as a function of 
a set of parameters 
\beq
N= N(P_b, \dot{P_b}, e,\dot{e},\om,\dot{\om},\ldots ).
\eeq
One approach to measuring 
the coefficients for Lorentz violation
would be to fit the observed number of pulses $N$
as a function of arrival time $t_{\rm arr}$
using a least-squares method.
In principle,
an estimate of the sensitivity of this approach 
could be performed
along the lines of the analysis 
for the conventional timing formula 
in Ref.\ \cite{bp4}.
However,
we adopt here a simpler method 
that nonetheless provides
estimates of the sensitivities 
that might be attainable in a detailed fitting procedure. 

To obtain these simple estimates,
we limit attention to the secular changes 
in the orbital elements $e$ and $\om$
given in Eqs.\ \rf{edot} and \rf{Omdot}.
Values for $\dot{e}$ and $\dot{\om}$ 
obtained from the timing analysis
can be used to probe combinations of coefficients in $\sb^\mn$.
Assuming only measurements from the variations 
of the orbital elements are considered,
the number of independent coefficients 
for Lorentz violation to which binary-pulsar systems are sensitive
can be quantified.
Through the measurement of $\dot{\om}$ and $\dot{e}$, 
a given binary-pulsar system
is sensitive to at least 2 combinations of coefficients.
These are
\bea
\sb_e &=& \sb_{PQ} 
- \fr {\de m}{M} \fr {2nae \ff} {(e^2 - 2\ff )} \sb_P, 
\nonumber\\
\sb_{\om} &=& 
\sb_{kP} \sin \om 
+ (1-e^2)^{1/2} \sb_{kQ} \cos \om 
\nonumber\\
&&
- \fr {\de m}{M} 2n a e ~\sb_k \cos \om
\nonumber\\
&&
+ \tan i 
\fr { (1-e^2)^{1/2} (e^2 - 2\ff ) } {2 e^2 \ff }
(\sb_{PP}-\sb_{QQ}) 
\nonumber\\
&&
+ \fr {\de m}{M} {2na \tan i} 
\fr {(e^2 - \ff )} {e \ff } \sb_Q .
\label{som}
\eea
In principle,
sensitivities to the secular changes in $i$ and $\Om$
might increase the number of independent coefficients
that could be measured in a given system.

Since there are 9 independent coefficients in $\sb^\mn$,
it is worth studying more than one binary-pulsar system.
Typically,
distinct binary-pulsar systems are oriented differently in space.
This means the basis vectors $\vec P$, $\vec Q$, $\vec k$ 
differ for each system,
and hence the dependence on the coefficients for
Lorentz violation in the Sun-centered frame
also differs.
Examination of Eq.\ \rf{som}
reveals that the combination $\sb_{PP}+\sb_{QQ}$ is inaccessible
to any orientation
and that
\it 5 independently oriented binary-pulsar systems can in principle 
measure 8 independent coefficients in $\sb^\mn$. \rm 
If information from $i$ and $\Om$ can also be extracted,
then fewer than 5 binary-pulsar systems might suffice.

The error bars from existing timing analyses offer 
a rough guide to the sensitivities 
of binary pulsars to Lorentz violation.
As an explicit example,
consider the binary pulsar PSR $1913+16$.
The errors on 
$\dot{e}$ and $\dot{\om}$
are roughly
$10^{-14}$ and $10^{-7}$, 
respectively 
\cite{bp5}.
These imply the following sensitivities
to Lorentz violation: 
$\sb_e \lsim 10^{-9}$
and
$\sb_{\om} \lsim 10^{-11}$. 
Interesting sensitivities could also be achieved
by analysis of data from other pulsar systems
\cite{drl,doublepulsar}. 

\subsubsection{Transformation to Sun-centered frame}
\label{transformation}

In the preceding discussions of tests with binary pulsars,
all the coefficients for Lorentz violation 
enter as projections along the triad of vectors
$\vec P$, $\vec Q$, $\vec k$
defining the elliptical orbit.
These projected coefficients
involve the post-newtonian frame of the binary-pulsar system.
The explicit construction of the transformation 
that relates these projections 
to the standard coefficients 
$\sb^{\Xi\Pi}$
in the Sun-centered frame is given next.
 
The first step is to express
the various projections of the coefficients
directly in terms of the orbital angles $i$, $\Om$ and $\om$
and the coordinate system $x^\mu$ of the binary pulsar. 
This involves substitution for the projection vectors
using Eqs.\ \rf{k}.
For example,
for the projection $\sb_{PP}$ we have 
\bea
\sb_{PP} &\equiv & P^j P^k \sb^{jk}
\nonumber\\
&=& \sb^{11} (\cos \Om \cos \om - \cos i \sin \Om \sin \om)^2
\nonumber\\
&&
+ \sb^{22} (\sin \Om \cos \om + \cos i \cos \Om \sin \om)^2
\nonumber\\
&&
+\sb^{33} \sin^2 i \sin^2 \om 
\nonumber\\
&&
+ 2 \sb^{12} (\cos \Om \cos \om - \cos i \sin \Om \sin \om)
\nonumber\\
&&
\quad \times (\sin \Om \cos \om + \cos i \cos \Om \sin \om)
\nonumber\\
&&
+ 2 \sb^{13} \sin i \sin \om 
(\cos \Om \cos \om - \cos i \sin \Om \sin \om)
\nonumber\\
&&
+ 2 \sb^{23} \sin i \sin \om 
(\sin \Om \cos \om + \cos i \cos \Om \sin \om).
\nonumber\\
\label{spp}
\eea
The resulting expressions give the projected coefficients 
in terms of the coefficients $\sb^{jk}$,
with components in the basis ${\bf e}_\mu$
of the binary pulsar.

The conversion to coefficients 
$\sb^{\Xi\Pi}$
in the Sun-centered frame
can be accomplished with a Lorentz transformation.
We have
\beq
\sb^\mn
= \La^\mu_{\pt{\mu}\Xi} \La^\nu_{\pt{\nu}\Pi} \sb^{\Xi\Pi}.
\label{bpscf} 
\eeq
The Lorentz transformation consists
of a rotation to align the axes 
of the binary-pulsar and Sun-centered frames,
along with a boost to correct for the relative velocity.
The latter is taken as approximately constant
over the time scales of relevance for the experimental observations.

The components of the Lorentz transformation
are given by
\bea
\La^0_{\pt{0} T} &=& 1 , 
\nonumber\\  
\La^0_{\pt{0}J} &=& -\be^J,
\nonumber\\  
\La^j_{\pt{j}T} &=& -(R \cdot \vec \be)^j, 
\nonumber\\  
\La^j_{\pt{j}J} &=& R^{jJ}.
\label{lambda}
\eea 
In this expression,
$\be^J$ is the boost of the binary-pulsar system
as measured in the Sun-centered frame.
The matrix with components $R^{jJ}$ 
is the inverse of the rotation 
taking the basis 
${\bf e}_\Xi$
for the Sun-centered frame 
to the basis 
${\bf e}_\mu$
for the binary-pulsar frame.
Explicitly, 
these components are 
\bea
R^{xX} &=& \cos \ga \cos \al - \sin \ga \sin \al \cos \be,
\nonumber\\
R^{xY} &=& \cos \ga \sin \al + \sin \ga \cos \al \cos \be,
\nonumber\\
R^{xZ} &=& \sin \ga \sin \be,
\nonumber\\
R^{yX} &=& -\sin \ga \cos \al - \cos \ga \sin \al \cos \be,
\nonumber\\
R^{yY} &=& -\sin \ga \sin \al + \cos \ga \cos \al \cos \be,
\nonumber\\
R^{yZ} &=& \cos \ga \sin \be,
\nonumber\\
R^{zX} &=& \sin \be \sin \al,
\nonumber\\
R^{zY} &=& -\sin \be \cos \al,
\nonumber\\
R^{zZ} &=& \cos \be.
\label{bprot}
\eea
In these expressions,
the angles $\al$, $\be$, and $\ga$ 
are Euler angles defined for the rotation
of ${\bf e}_J$ to ${\bf e}_j$ 
as follows:
first, rotate about ${\bf e}_Z$ by $\ga$; 
next, rotate about the new $Y$ axis by $\al$;
finally, rotate about the new $Z$ axis by $\ga$.

\subsection{Classic Tests}
\label{classic}

Among the classic tests of general relativity
are the precession of the perihelia
of the inner planets,
the time-delay effect,
and the bending of light around the Sun.
The sensitivities to Lorentz violation attainable 
in existing versions of these tests 
are typically somewhat less than those discussed 
in the previous sections.
Nonetheless,
an analysis of the possibilities reveals
certain sensitivities attaining interesting levels,
as described in this subsection.

\subsubsection{Perihelion shift}
\label{Perihelion shift}

To obtain an estimate of the sensitivity to Lorentz violation
offered by studies of perihelion shifts,
the results from Sec.\ \ref{binary pulsars} 
can be adapted directly.
The reference plane can be chosen as the ecliptic,
so the inclination $i$ is small for the inner planets.

We seek the perihelion shift with respect to the equinox.
The angular location $\tilde\om$ of the perihelion
with respect to the equinox
is given by $\tilde \om = \om + \Om \cos i $.
The change in $\tilde \om$ per orbit 
can be obtained by appropriate use of 
Eqs.\ \rf{Omdot} with $\de m/M \approx 1$.
Neglecting terms proportional to $\sin i$,
we find that the shift in perihelion angle per orbit is 
\bea
\De \tilde \om &=& -{2 \pi}  
\Big[ 
\fr {(e^2 - 2\ff )}{2e^4}
(\sb_{PP}-\sb_{QQ}) 
+ \fr {2na (e^2 -\ff )}{e^3(1-e^2)^{1/2}} \sb_Q 
\Big].
\nonumber\\
\label{per1}
\eea

Substituting the values of $e$ and $na$ 
for Mercury and the Earth in turn,
we find the corresponding perihelion shifts
in arcseconds per century (C) are given by
\bea
\dot {\tilde \om }_{\mercury} & \approx & 
7 \times 10^{7 \prime \prime} \, \sb_{\mercury} \, {\rm C}^{-1},
\nonumber\\
\dot {\tilde \om }_{\oplus} & \approx & 
2 \times 10^{7 \prime \prime} \, \sb_{\oplus} \, {\rm C}^{-1}.
\label{per2}
\eea
In these expressions,
the perihelion combinations of SME coefficients 
for Mercury and the Earth are defined by
\bea
\sb_{\mercury} & \approx & (\sb_{PP}-\sb_{QQ})- 6 \times 10^{-3} \sb_Q,
\nonumber\\
\sb_{\oplus} & \approx & (\sb_{PP}-\sb_{QQ}) - 5 \times 10^{-2} \sb_Q.
\label{sbE}
\eea
The reader is cautioned that the vector projections
of $\sb^\mn$ along $\vec P$ and $\vec Q$ differ
for Mercury and the Earth.
Indeed,
Eqs.\ \rf{k} reveal that they differ by the values 
of $i$, $\om$ and $\Om$,
once again indicating the value of considering more than one system
for measurements separating the independent values of $\sb^\mn$.

The error bars in experimental data for 
the perihelion shifts of Mercury and the Earth
provide an estimated attainable sensitivity
to Lorentz violation.
For example,
adopting the values from Refs.\ \cite{cmw,cmw01},
the error bars on the perihelion shift 
of Mercury and the Earth 
are roughly given by 
$0.043^{\prime \prime}$ C$^{-1}$ 
and $0.4^{\prime \prime}$ C$^{-1}$,
respectively.
Taking these values as an upper bound on the size 
of the observable shifts in Eq.\ \rf{per2},
we obtain attainable sensitivities of 
$\sb_{\mercury} \lsim 10^{-9}$
and 
$\sb_{\oplus} \lsim 10^{-8}$.

For simplicity in the above, 
attention has been focused on Mercury and the Earth.  
However,
the varying orientations $i$, $\om$, $\Om$
of other bodies in the solar system 
suggests examining the corresponding perihelion shifts
might yield interesting sensitivities
on independent coefficients for Lorentz violation.
For example,
observations of the asteroid Icarus have been used to verify
to within $20$ percent
the general-relativistic prediction
of $10^{\prime \prime}$ C$^{-1}$ 
for the perihelion shift 
\cite{icarus}.
Translated into measurements of Lorentz-violating effects,
this would correspond to a sensitivity 
at the $10^{-7}$ level 
for the combination 
$\sb_{\rm Ic} = (\sb_{PP} - \sb_{QQ}) - 7 \times 10^{-4} \sb_Q$. 

\subsubsection{Time-delay effect}
\label{time-delay effect}

For the binary pulsar,
the Lorentz-violating corrections to the Shapiro time-delay effect 
are negligible compared to other effects,
as described in Sec.\ \ref{binary pulsars}.
However,
solar-system tests involving the time-delay effect 
have the potential to yield interesting sensitivities.
To demonstrate this,
we derive next the generalization 
of the usual time-delay result in general relativity
to allow for Lorentz violation.

We seek an expression for the time delay of a light signal
as it passes from an emitter to a receiver
in the presence of a massive body such as the Sun.
Assuming that light travels along a null geodesic in spacetime,
the photon trajectory satisfies the condition $ds^2=0$.
In terms of the underlying post-newtonian coordinates,
this can be written in the form 
\bea
\left(\fr {dx^j}{dt} \cdot \fr {dx^j}{dt}\right)^{1/2} 
&=& 1 - \frac 12 h_{00} - h_{0j}\fr {dx^j}{dt} 
- \half h_{jk}\fr {dx^j}{dt}\fr {dx^k}{dt}.  
\nonumber\\
\label{nullgeod}
\eea
The metric components are taken to be the $O(2)$ forms 
of Eqs.\ \rf{g00}, \rf{g0j}, and \rf{gjk}
in the presence of Lorentz violation
and with a perturbing point mass $m_2$ located at $\vec r_2$.

Solving Eq.\ \rf{nullgeod} for $dt$,
we integrate from an initial emission spacetime point 
$(t_{\rm em}, \vec r_1)$ 
to a final arrival spacetime point 
$(t_{\rm arr}, \vec r_E)$.
At order $O(2)$,
the zeroth-order straight-line solution 
\beq
\vec r (t) = \hat n (t-t_{\rm em}) + \vec r_1 (t_{\rm em})
\eeq
can be inserted into the metric components in Eq.\ \rf{nullgeod}.
Here,
$\hat n$ is a unit vector in the direction $\vec r_E - \vec r_1$.
In performing the integration,
the positions of the emitter at $\vec r_1$,
receiver at $\vec r_E$,
and perturbing body at $\vec r_2$ 
are all taken to be constant during 
the travel time of the light pulse.

The resulting expression for the time difference
$t_{\rm arr} - t_{\rm em}$
is found to be 
\beq
t_{\rm arr} - t_{\rm em} = |\vec r_E - \vec r_1| + \De_1 
+ \De_2(t_{\rm arr}) - \De_2(t_{\rm em}).
\label{timedelay}
\eeq
The terms $\De_1$ and $\De_2$ on the right-hand side are
given by 
\bea
\De_1 &=& Gm_2 \De^{jk} \hat n^j \hat n^k 
\nonumber\\
&&
\times
\ln \left[\fr {|\hat n t_{\rm arr} + \vec \ze(t_{\rm arr})|+ t_{\rm arr} 
+ \hat n \cdot \vec \ze(t_{\rm arr})} 
{|\hat n t_{\rm em} + \vec \ze(t_{\rm em})|
+ t_{\rm em} + \hat n \cdot \vec \ze(t_{\rm em})}\right]
\label{De1} 
\nonumber\\
\eea
and 
\bea
\De_2 (t) &=& 
\fr{Gm_2 \De^{jk} } {\vec d^2 |\hat n t +\vec \ze|}
\Big[ {\hat n^j \hat n^k [((\hat n \cdot \vec \ze)^2 -\vec d^2)t 
+ \hat n \cdot \vec \ze \vec \ze^2]}
\nonumber\\
&&
\pt{Gm \De^{jk}}
\hskip-10pt
-{2 \hat n^j \ze^k (\vec \ze^2 + \hat n \cdot \vec \ze t)}
+{\ze^j \ze^k (t + \hat n \cdot \vec \ze)}
\Big],
\nonumber\\
\label{De2}
\eea
where $\vec\ze$ and $\vec d$ are functions of a time argument
and are defined as 
\bea
\vec \ze (t) &=& \vec r_1 (t_{\rm em}) - \vec r_2 (t) 
- \hat n t_{\rm em},
\nonumber\\
\vec d (t) &=& \hat n \times ( \vec \ze \times \hat n).
\label{d}
\eea
The combination of coefficients $\De^{jk}$,
which includes the effects due to Lorentz violation,
is given by
\bea
\De^{jk} &=& 2 \de^{jk} + \De_{\sb}^{jk},
\label{Dejk}
\eea
where
\bea
\De_{\sb}^{jk} &=& 
(\sb^{00} - \sb^{0l}\hat n^l ) \de^{jk} 
\nonumber\\
&&
+ \sb^{00} \hat n^j \hat n^k + \sb^{jk}
- \frac 12 \sb^{0j} \hat n^k - \frac 12 \sb^{0k} \hat n^j  
\nonumber\\
&&
-\frac 12 \sb^{jl} \hat n^l \hat n^k
-\frac 12 \sb^{kl} \hat n^l \hat n^j . 
\eea
Note that the trace part of $\De^{jk}$
cancels in Eq.\ \rf{De2}.

The result \rf{timedelay} is the generalization 
to the pure-gravity minimal SME
of the standard time-delay formula of general relativity,
which can be recovered as the limit $\sb^\mn=0$.
In principle,
Eq.\ \rf{timedelay}
could be applied to solar-system experiments including,
for example,
reflection of a signal from a planet or spacecraft.
A detailed analysis along these lines would be of definite
interest but lies beyond our present scope.
However,
the form of Eq.\ \rf{timedelay} indicates that 
differently oriented experiments 
would be sensitive to different coefficients for Lorentz violation.

At present,
the most sensitive existing data on the time-delay effect
within the solar system 
are obtained from tracking the Cassini spacecraft 
\cite{td}.
Assuming a detailed analysis of these data,
we estimate that sensitivities to various combinations 
of the coefficients $\sb^{JK}$
could attain at least the $10^{-4}$ level.
This is comparable to the sensitivities of GPB
discussed in Sec.\ \ref{gpb},
but the differences between the two spacecraft trajectories
makes it likely that the attainable sensitivities
would involve different combinations of the coefficients $\sb^{JK}$.

We remark in passing that a generalization 
of the standard formula for the bending of light 
should also exist,
following a derivation along lines similar to those 
leading to Eq.\ \rf{timedelay}.
Current measurements of the deflection of light by the Sun 
have attained the $10^{-4}$ level
\cite{cmw01},
which suggests sensitivities to Lorentz violation
at this level might already be attainable 
with a complete data analysis.
Future missions such as GAIA 
\cite{gaia}
and LATOR 
\cite{lator}
may achieve sensitivities to coefficients for Lorentz violation 
at the level of $10^{-6}$ and $10^{-8}$,
respectively.

\section{Summary}
\label{summary}

In this work,
we investigated the gravitational sector of the minimal SME
in the Riemann-spacetime limit, 
with specific focus on the associated post-newtonian physics.
The relevant basics for this sector of the SME are reviewed in 
Sec.\ \ref{basics}.
There are 20 independent background coefficient fields
$u$, $s^\mn$, and $t^{\ka\la\mu\nu}$,
with corresponding vacuum values 
$\ub$, $\sb^\mn$, and $\tb^{\ka\la\mu\nu}$.

The primary theoretical challenge 
to the study of post-newtonian physics 
is the extraction of the linearized field equations for the metric
while accounting for the dynamics of the coefficients 
for Lorentz violation.
At leading order,
this challenge is met in Sec.\ \ref{linearization}.
Under a few mild conditions,
the linearized effective field equations 
applicable for any theory with leading-order Lorentz violation 
are obtained in the practical form \rf{fleq}.

Section \ref{post-newtonian expansion}
is devoted to the post-newtonian metric.
Its derivation from the effective linearized field equations
is described in Sec.\ \ref{metric}.
Under the simplifying assumption $\ub = 0$,
the result is given in Eqs.\ \rf{g00}-\rf{gjk}.
Section \ref{dynamics} obtains
the equations of motion \rf{euler} and \rf{accdens} 
describing the corresponding post-newtonian behavior 
of a perfect fluid.
The results are applied to the gravitating many-body system,
with the point-mass equations of motion and lagrangian given 
by Eqs.\ \rf{nbody} and \rf{lpp}.
The post-newtonian metric obtained here
is compared and contrasted with
other existing post-newtonian metrics
in Sec.\ \ref{othermetrics}.
Finally,
to complete the theoretical discussion,
illustrative examples of the practical application 
of our methodologies and results to specific theories
are given in Sec.\ \ref{bbmodel}
in the context of various bumblebee models.

The largest part of the paper,
Sec.\ \ref{experimental applications},
is devoted to exploring 
the implications of these theoretical results 
for a variety of existing and proposed gravitational experiments.
It begins with a generic discussion of coordinate frames
and existing bounds in Sec.\ \ref{general considerations}.
Section \ref{llr} considers
laser ranging both to the Moon and to artificial satellites,
which offers promising sensitivity 
to certain types of Lorentz violation 
controlled by the coefficients $\sb^{\Xi\Pi}$
in the Sun-centered frame.
The key result for the relative acceleration 
of the Earth-satellite system is given in Eq.\ \rf{esacc} and
the associated definitions.
The technical details of some of the derivations in this
section are relegated to Appendix \ref{dlso}.

Section \ref{Earth laboratory experiments}
considers Earth-based laboratory experiments.
The expression for the local terrestrial acceleration 
as modified by Lorentz violation is obtained as Eq.\ \rf{locacc}
in Sec.\ \ref{Earth laboratory theory}.
Applications of this result to gravimeter measurements 
are treated in Sec.\ \ref{gravimeter tests},
while implications for experiments with torsion pendula
are described in Sec.\ \ref{torsion-pendulum tests}.
The latter include some proposed high-sensitivity tests
to the coefficients $\sb^{\Xi\Pi}$
based on the local anisotropy of gravity.

Possible sensitivities to Lorentz violation in 
measurements of the spin precession of orbiting gyroscopes
are considered in Sec.\ \ref{gyroscope experiment}.
The generalized formula describing the spin precession is
found in Eq.\ \rf{gensp}, 
and the secular effects are isolated in Eq.\ \rf{spprec}.
The results are applied to the Gravity Probe B experiment
\cite{gpb}.
This experiment is shown to have potential sensitivity 
to effects involving spin precession 
in a direction orthogonal to the usual geodetic
and frame-dragging precessions predicted in general relativity.

Timing studies of signals from binary pulsars
offer further opportunities to measure certain
combinations of coefficients for Lorentz violation.
Section \ref{binary pulsars}
derives the secular evolution of the six orbital elements
of the binary-pulsar motion
and calculates the modifications to the pulsar timing formula
arising from Lorentz violation.
Fits to observational data for several binary-pulsar systems
have the potential to achieve high sensitivity
to coefficients of the type 
$\sb^{\Xi\Pi}$.

Section \ref{classic} discusses the attainable sensitivities
in some of the classic tests of general relativity,
involving the perihelion precession,
time-delay, and bending of light. 
Perihelion shifts for the Earth, Mercury, and other bodies 
are considered in Sec.\ \ref{Perihelion shift},
while versions of the time-delay and light-bending tests
are studied in Sec.\ \ref{time-delay effect}.

Table 6 collects the estimated attainable sensitivities 
for all the experiments considered 
in Sec.\ \ref{experimental applications}.
Each horizontal line contains estimated sensitivities
to a particular combination 
of coefficients for Lorentz violation 
relevant for a particular class of experiment.
Most experiments are likely to have some sensitivity
to Lorentz violation,
but the table includes only the better estimated sensitivities 
for each type of experiment.

Relatively few sensitivities at these levels 
have been achieved to date.
In Table 6,
estimated sensitivities that might be attainable in principle
but that as yet remain to be established 
are shown in brackets. 
Values without brackets represent estimates of bounds
that the theoretical treatment in the text
suggests are already implied by existing data.
All estimates are comparatively crude
and should probably be taken to be valid
only within about an order of magnitude.
They are based on existing techniques or experiments,
and some planned or near-future experiments
can be expected to supersede them.

In conclusion,
both astrophysical observations 
and experiments on the Earth and in space
have the potential to offer high sensitivity to signals 
for Lorentz violation of purely gravitational origin.
A variety of tests is required to span
the predicted coefficient space 
for the dominant Lorentz-violating effects.
These tests are currently feasible,
and they represent a promising direction
to follow in the ongoing search for
experimental signals of Lorentz violation from the Planck scale. 

\begin{widetext}
\begin{center}
\begin{tabular} { |c|c|c|c|c|c|c|c|c|c| }
\hline
Coefficient & Lunar & Satellite & Gravimeter & Torsion 
& Gravity & Binary & Time & Perihelion & Solar-spin \\
combinations & ranging & ranging & & pendulum 
& Probe B & pulsar & delay & precession & precession \\
\hline
\hline
$\sb^{12}$ & $[10^{-11}]$ & $[10^{-10}]$ & - & - & - & - &-&- &-\\
$\sb^{11}-\sb^{22}$ & $[10^{-10}]$ 
& $[10^{-9}]$ & - & - & - & -&- &- &-\\
$\sb^{01}$ & $[10^{-7}]$ & $[10^{-8}]$ & - & - & - & - &-&- &-\\
$\sb^{02}$ & $[10^{-7}]$ & $[10^{-8}]$ & - & - & - & - &-&- &-\\
$\sb_{\Om_{\oplus} c}$ & $[10^{-7}]$ & - 
& - & - & - & - &-&- &-\\
$\sb_{\Om_{\oplus} s}$ & $[10^{-7}]$ & - & - 
& - & - & - & - & - &-\\
$\sb^{XX}-\sb^{YY}$ & - & - & $[10^{-11}]$ & $[10^{-15}]$ & -&-&-&- &-\\
$\sb^{XY}$ & - & - & $[10^{-11}]$ & $[10^{-15}]$ & -&-&-&- &-\\ 
$\sb^{XZ}$ & - & - & $[10^{-11}]$ & $[10^{-15}]$ & -&-&-&- &-\\
$\sb^{YZ}$ & - & - & $[10^{-11}]$ & $[10^{-15}]$ & -&-&-&- &-\\
$\sb^{TY}$ & - & - & $[10^{-7}]$ & $[10^{-11}]$ & -&-&-&- &-\\
$\sb^{TX}$ & - & - & $[10^{-7}]$ & $[10^{-11}]$ & -&-&-&- &-\\
$\sb^{TZ}$ & - & - & $[10^{-7}]$ & $[10^{-11}]$ & -&-&-&- &-\\
$\Om^\si_{\sb}$ & -&-&-&-& $[10^{-4}]$ & -& -&- &- \\
$\Om^{n}_{\sb}$ & -&-&-&-& $[10^{-4}]$ & -& -&- &- \\
$\Om^Z_{\sb}$ & -&-&-&-& $[10^{-4}]$ & - & -&- &- \\
$\sb_e$ & - & - & - & - & - & $[10^{-9}]$ & - & - &- \\
$\sb_\om$ & - & - & - & - & - & $[10^{-11}]$ & - & - &- \\
$\De_{\sb}^{jk}$ & - & - & - & - & - & - & $[10^{-4}]$ &- &-\\
$\sb_{\mercury}$ & - & - & -& - & - & - &-  & $10^{-9}$ &-\\
$\sb_{\oplus}$ & - & - & - & - & - & - &- & $10^{-8}$ &-\\
$\sb_{\rm Ic}$ & - & - & - & - & - & - &- & $[10^{-7}]$ &-\\
$\sb_{\rm SSP}$ & - & - & - & - & - & - & - & - & $10^{-13}$ \\
\hline
\end{tabular}
\end{center}
\begin{center}
Table 6.\ 
Crude estimates of attainable experimental sensitivities.
\label{sens}
\end{center}
\end{widetext}

\section*{Acknowledgments}
\label{Acknowledgments}

This work is supported in part 
by DOE grant DE-FG02-91ER40661
and by NASA grant NAG3-2914.

\appendix

\section{Satellite oscillations}
\label{dlso}

This appendix is devoted to calculating 
the Lorentz-violating corrections to the 
observable relative Earth-satellite distance 
$r = |\vec r_1 - \vec r_2|$
in the Sun-centered frame.
These corrections are needed for the analysis of lunar
and satellite laser ranging in Sec.\ \ref{llr}.
For simplicity,
the only perturbing body is taken to be the Sun.
Also,
we limit attention to the dominant frequencies
of direct relevance for the text.
A more detailed analysis could reveal other frequencies
that would permit sensitivity to different combinations 
of coefficients for Lorentz violation.

The derivation is separated into four parts.
The first discusses generalities
of the perturbative expansion
and establishes notational conventions.
Some of the results in this part 
are also applied in other contexts  
elsewhere in the text.
The second and third parts
are devoted to forced and unforced oscillations,
respectively.
The final part extracts the radial oscillations
needed for the analysis in Sec.\ \ref{llr}.

\subsection{Generalities}
\label{general setup}

The procedure adopted here involves
perturbation techniques similar to those presented 
in Refs.\ \cite{nicarus,nanisllr}. 
The equation \rf{esacc} for the coordinate acceleration
of the Earth-satellite distance
is expanded in a vector Taylor series 
about an unperturbed circular-orbit solution $r^J_0$.
This unperturbed circular orbit satisfies 
\beq
\ddot r^J_0 = \al^J_0 (r_0) ,
\label{unpert} 
\eeq
where 
\beq
\al^J_0 (r) =
- \fr {GM r^J}{r^3} + \fr {\Om_{\oplus}^2 r^J}{2} - \fr {3Q r^J}{2r^5} ,
\label{al0}
\eeq
with
\beq
\Om^2_{\oplus} = \fr {G M_{\odot}}{R^3}.
\label{OmSun}
\eeq
The frequency $\om$ of the circular orbit is defined as 
\bea
\om^2 &=& \fr {GM}{r_0^3}- \half \Om_{\oplus}^2 + \fr {3Q}{2r_0^5}.
\label{circfreq}
\eea
Contributions from perturbing bodies 
and the Earth's quadrupole potential
are included to mute resonant effects 
\cite{nllr,al1llr,nanisllr},
arising in this instance from Lorentz-violating accelerations.

Denoting the perturbation by $\ep^J$,
we find  
\bea
\ddot \ep^J - (\ep^K \prt_K) \al^J_0 (r_0) &=&
\frac 12  (\ep^K \ep^L \prt_K \prt_L) \al^J_0 (r_0) +
\ldots 
\nonumber\\
& & 
\hskip -10pt
+ \al^J_{\rm T'} (r_0)+ (\ep^K \prt_K) \al^J_{\rm T'} (r_0) 
+ \ldots 
\nonumber\\
& & 
\hskip -10pt
+ \al^J_{\rm Q'} (r_0)+ (\ep^K \prt_K) \al^J_{\rm Q'} (r_0)+
\ldots 
\nonumber\\
& & 
\hskip -10pt
+ \al^J_{\rm LV} (r_0)+ (\ep^K \prt_K) \al^J_{\rm LV} (r_0)+
\ldots .
\nonumber\\
\label{taylor}
\eea
The ellipses here represent 
higher-order terms in the vector Taylor series.
The quantities $\al^J_{\rm T'}$ and $\al^J_{\rm Q'}$ 
are given by modifications of Eqs.\ \rf{T} and \rf{Q},
respectively,
in which the tidal and quadrupole terms for Eq.\ \rf{al0}
are removed.
The term $\al^J_{\rm LV} (r)$ is defined in Eq.\ \rf{lv}. 

It is convenient to decompose Eq.\ \rf{taylor} 
using a basis of 3 orthonormal unit vectors 
$\hat \rh$, $\hat \ta$, $\hat \si$.
These vectors are,
respectively,
normal, tangential, and perpendicular 
to the unperturbed orbit.
In the basis for the Sun-centered frame, 
the explicit form for $\hat\rh$ is 
\bea
\hat \rh &=& \big(\cos \al \cos (\om T + \th) 
- \sin \al \cos \be \sin (\om T + \th),
\nonumber\\
&&
\pt{\big(} \sin \al \cos (\om T + \th) 
+ \cos \al \cos \be \sin (\om T + \th),
\nonumber\\
&&
\pt{\big(} \sin \be \sin (\om T + \th) \big),
\label{rhohat}
\eea
where $\th$ is the phase of the orbit
with respect to the line of ascending nodes.
Explicit forms for the other two basis vectors
can be obtained from the equations
\beq 
\om \hat \tau = d\hat \rh/dT 
\eeq
and 
\beq
\hat \si = \hat \rh \times \hat \ta .
\eeq
Note also that $r^J_0 = r_0 \hat \rh^J$,
where $r_0$ is the radius of the circular orbit.

For our purposes, 
it suffices to assume that $m_2 \gg m_1$ 
and to take $R^J$ as the vector pointing 
to the Earth's location.
The Earth's motion can also be approximated 
as a circular orbit around the Sun.
This means that 
\bea
\hat R &=& \big(-\cos \Om_{\oplus} T, 
- \cos \et \sin \Om_{\oplus} T,
- \sin \et \sin \Om_{\oplus} T \big)
\nonumber\\
\label{R}
\eea
in the Sun-centered frame,
so we can write 
$R^J = R \hat R^J$ and $V^J_{\oplus} = dR^J/dT$.

We write the perturbative oscillations $\ep^J$ 
in the general form
\beq
\vec \ep_{\nu,\th_\nu} = X_{\nu} \vec \rh_{\nu} 
+ Y_{\nu} \vec \ta_{\nu}
+ Z_{\nu} \vec \si_{\nu},
\label{oscbasis}
\eeq
where the oscillating basis vectors 
$\vec \rh_{\nu}$,
$\vec \ta_{\nu}$,
$\vec \si_{\nu}$
are given by 
\bea
\vec \rh_{\nu} &=& \hat \rh \cos (\nu T + \th_{\nu}),
\nonumber\\
\vec \ta_{\nu} &=& \hat \ta \sin (\nu T + \th_{\nu}),
\nonumber\\
\vec \si_{\nu} &=& \hat \si \cos (\nu T + \th_{\nu}).
\label{vecsi}
\eea
With respect to this oscillating basis,
Eq.\ \rf{taylor} can be expressed as a matrix equation.  
For example,
$\vec \ep_{\nu}$ takes the column-vector form:
\beq
\vec \ep_{\nu,\th_\nu} = 
\left(
\begin{array}{c} 
X_{\nu} \\ Y_{\nu} \\ Z_{\nu}
\end{array} 
\right)
\label{epcolumn}.
\eeq

The types of oscillations contained in Eq.\ \rf{taylor}
can be split into two categories: 
forced or inhomogeneous oscillations,
and unforced or homogeneous oscillations
\cite{nicarus,nanisllr}.
The forced oscillations arise primarily from the terms 
$\vec \al_{\rm T'}$, 
$\vec \al_{\rm Q'}$,
$\vec \al_{\rm LV}$.
The unforced oscillations 
are the natural oscillations of the Earth-satellite system 
in the presence of the perturbative forces.
They are controlled by the matrix structure 
of the derivative terms in Eq.\ \rf{taylor},
such as $\ep^K \ep^L \prt_K \prt_L \al^J_0 (r_0) /2$.
Next, 
we study each of these categories of oscillations in turn.

\subsection{Forced oscillations}
\label{forced oscillations}

The forced oscillations due to Lorentz violation 
emanate primarily from the terms in $\vec \al_{\rm LV}$.
For oscillations near the resonant frequency $\om$, 
however, 
the tidal and quadrupole derivative terms in Eq.\ \rf{taylor} 
can play a significant role.
Keeping all the relevant terms,
the lowest-order contributions arise from the equation 
\bea
\ddot \ep^J - (\ep^K \prt_K) \al^J_0 (r_0) &=& \al^J_{\rm LV} (r_0) 
+ (\ep^K \prt_K) \al^J_{\rm T'} (r_0)
\nonumber\\ 
& & 
+ (\ep^K \prt_K) \al^J_{\rm Q'} (r_0).
\label{forced}
\eea
In the oscillating basis of Eq.\ \rf{epcolumn},
the relevant part of this equation takes the matrix form 
\bea
\lrvec K (\nu) \cdot \vec \ep_{\nu,\th_\nu} &=& 
(\lrvec \al_{\rm T} + \lrvec \al_{\rm Q}) 
\cdot \vec \ep_{\nu,\th_\nu} 
+ \vec \al_{\rm LV} (\nu,\th_\nu)
\nonumber\\
\label{forcedmatrix}   
\eea
for each frequency $\nu$.
Here,
we define $\lrvec K(\nu)$ to be the matrix operator  
with components
$\de_K^{\pt{K}J} (d^2/dT^2) - \prt_K \al^J_0 (r_0)$
in the Sun-centered frame,
while 
$\lrvec \al_{\rm T}$ and $\lrvec \al_{\rm Q}$
denote the diagonal parts of $\prt_K \al^J_{\rm T'}$ 
and $\prt_K \al^J_{\rm Q'}$,
respectively.
 
By evaluating all terms in Eq.\ \rf{forced}
in the oscillating basis,
we find that the components of these matrix operators
for a given frequency $\nu$ are given by 
\bea
\lrvec K (\nu)&=& 
-\left(
\begin{array}{ccc} 
\nu^2 + 3\om^2 + \frac 32 \tilde \Om_{\oplus}^2 & 2\om\nu & 0 \\
2\om\nu & \nu^2 & 0 \\
0 & 0 & \nu^2-\om^2 \\
\end{array} 
\right) ,
\nonumber\\
\lrvec \al_{\rm T} &=& 
3 \Om_{\oplus}^2 \left(
\begin{array}{ccc} 
a_1 - \frac 12 & 0 & 0 \\
0 & a_1 - \frac 12 & 0 \\
0 & 0 & a_2-\frac 12 \\
\end{array} 
\right),
\nonumber\\
\lrvec \al_{\rm Q} &=& 
\fr{Q}{r^5_0}\left(
\begin{array}{ccc} 
21 \sin^2 \be & 0 & 0 \\
0 & -\frac {21}{4} \sin^2 \be & 0 \\
0 & 0 & -3-\frac 34 \sin^2 \be \\
\end{array} 
\right).
\nonumber\\
\label{GQ}
\eea
Here, 
the quantity $\tilde \Om_{\oplus}^2$
is defined by
\bea
\tilde \Om_{\oplus}^2 &=& \Om_{\oplus}^2 + \fr {2Q}{r^5_0},
\eea
while $a_1$ and $a_2$ are given by
\bea
a_1 &=& \frac 1 4 [\cos^2 \al (1+\cos^2 \be \cos^2 \et) 
+ \sin^2 \be \sin^2 \et
\nonumber\\
&&
\quad 
+ \sin^2 \al (\cos^2 \be + \cos^2 \et)
\nonumber\\
&&
\quad
+ 2 \cos \al \cos \be \sin \be \cos \et \sin \et ],
\nonumber\\
a_2 &=& \frac 12 ( \sin^2 \al \sin^2 \be 
+ \cos^2 \al \sin^2 \be \cos^2 \et 
\nonumber\\
&&
\quad
+ \cos^2 \be \sin^2 \et)
- \cos \al \cos \be \sin \be \cos \et \sin \et.
\nonumber\\
\label{a2}
\eea

Inspection of the form of $\lrvec \al_{\rm LV}$
given in Eq.\ \rf{lv} reveals that 
the dominant oscillations 
forced by the Lorentz violation
occur at the frequencies 
$2\om$, 
$\om$,
and 
$\Om_{\oplus}$.
The amplitudes at these frequencies
are determined by the following matrix equations:
\bea
\lrvec K (2\om ) \cdot \vec \ep_{2\om,2 \th - \si_{2 \om}} &=& 
(\lrvec \al_{\rm T} 
+ \lrvec \al_{\rm Q}) \cdot \vec \ep_{2\om,2 \th - \si_{2\om}} 
\nonumber\\
&&
+ \vec \al_{\rm LV} (2\om, 2\th-\si_{2\om}), 
\nonumber\\
\lrvec K (\om ) \cdot \vec \ep_{\om,\th+\si_{\om, 1}} &=& 
(\lrvec \al_{\rm T} 
+ \lrvec \al_{\rm Q}) \cdot \vec \ep_{\om,\th+\si_{\om, 1}} 
\nonumber\\
&&
+ \vec \al_{\rm LV} (\om, \th+\si_{\om, 1}),
\nonumber\\
\lrvec K (\om ) \cdot \vec \ep_{\om,\th-\si_{\om, 2}} &=& 
(\lrvec \al_{\rm T} 
+ \lrvec \al_{\rm Q}) \cdot \vec \ep_{\om,\th-\si_{\om, 2}} 
\nonumber\\
&&
+ \vec \al_{\rm LV} (\om, \th-\si_{\om, 2}),
\nonumber\\
\lrvec K (\Om_{\oplus}) \cdot 
\vec \ep_{\Om_{\oplus}, \si_{\Om_{\oplus}, 1}} &=& 
(\lrvec \al_{\rm T} + \lrvec \al_{\rm Q}) \cdot 
\vec \ep_{\Om_{\oplus}, \si_{\Om_{\oplus},1}} 
\nonumber\\
&&
+ \vec \al_{\rm LV} (\Om_{\oplus}, \si_{\Om_{\oplus}, 1}),
\nonumber\\
\lrvec K (\Om_{\oplus}) \cdot 
\vec \ep_{\Om_{\oplus}, \si_{\Om_{\oplus}, 2}} &=& 
(\lrvec \al_{\rm T} + \lrvec \al_{\rm Q}) \cdot 
\vec \ep_{\Om_{\oplus}, \si_{\Om_{\oplus}, 2}} 
\nonumber\\
&&
+ \vec \al_{\rm LV} (\Om_{\oplus}, \si_{\Om_{\oplus}, 2}).
\label{Om_matrix}
\eea

In these equations,
the relevant terms in $\vec \al_{\rm LV}$ are given 
by the expressions 
\bea
\vec \al_{\rm LV} (2\om, 2\th - \si_{2\om}) &=& 
-\fr {GM \sb_{2\om}}{r^2_0} 
\left(
\begin{array}{c} 
\frac 12 \\ 1 \\ 0 \\
\end{array} 
\right),
\nonumber\\
\vec \al_{\rm LV} (\om, \th + \si_{\om, 1} ) &=& 
\fr {2G \de m \sb_{\om, 1} v_0 }{r^2_0} 
\left(
\begin{array}{c} 
1 \\ 0 \\ 0 \\
\end{array} 
\right),
\nonumber\\
\vec \al_{\rm LV} (\om, \th-\si_{\om, 2}) &=& 
\fr {G M \sb_{\om, 2}} {r^2_0} 
\left(
\begin{array}{c} 
0 \\ 0 \\ 1 \\
\end{array} 
\right),
\nonumber\\
\vec \al_{\rm LV} (\Om_{\oplus}, \si_{\Om_{\oplus}, 1}) &=& 
\fr {-3GM V_{\oplus} {\sb_{\Om_{\oplus}, 1}}}{r^2_0} 
\left(
\begin{array}{c} 
1 \\ 0 \\ 0 \\
\end{array} 
\right),
\nonumber\\
\vec \al_{\rm LV} (\Om_{\oplus}, \si_{\Om_{\oplus}, 2}) &=& 
\fr {-GM V_{\oplus} {\sb_{\Om_{\oplus}, 2}}}{2 r^2_0} 
\left(
\begin{array}{c} 
1 \\ 0 \\ 0 \\
\end{array} 
\right).
\label{al2Om}
\eea
The various combinations of the coefficients 
for Lorentz violation have magnitudes and phases defined by
\bea
\sb_{2 \om} \cos \si_{2\om} &=& \frac 12 (\sb^{11}-\sb^{22}),
\nonumber\\
\sb_{2 \om} \sin \si_{2\om} &=& \sb^{12},
\nonumber\\
\sb_{\om, 1} \cos \si_{\om, 1} &=& \sb^{02},
\nonumber\\
\sb_{\om, 1} \sin \si_{\om, 1} &=& \sb^{01},
\nonumber\\
\sb_{\om, 2} \cos \si_{\om, 2} &=& \sb^{13},
\nonumber\\
\sb_{\om, 2} \sin \si_{\om, 2} &=& \sb^{23},
\nonumber\\
\sb_{\Om_{\oplus},1} \cos \si_{\Om_{\oplus},1} 
&=& \cos \et \sb^{TY} + \sin \et \sb^{TZ},
\nonumber\\
\sb_{\Om_{\oplus},1} \sin \si_{\Om_{\oplus},1} 
&=& \sb^{TX},
\nonumber\\
\sb_{\Om_{\oplus},2} \cos \si_{\Om_{\oplus},2} 
&=& \sin \al \cos \et \sb^{01} 
\nonumber\\
&&
+ (\sin \be \sin \et + \cos \al \cos \be \cos \al) \sb^{02},
\nonumber\\
\sb_{\Om_{\oplus},2} \sin \si_{\Om_{\oplus},2} 
&=& \cos \al \sb^{01} - \sin \al \cos \be \sb^{02}.
\label{phases}
\eea
In these expressions,
the quantities $\sb^{\mu\nu}$ are combinations of
the coefficients for Lorentz violation 
$\sb^{\Xi\Pi}$
in the Sun-centered frame.
For $\sb^{11}- \sb^{22}$, $\sb^{12}$, $\sb^{01}$, and $\sb^{02}$,
these combinations can be found 
in Eqs.\ \rf{sb02} of Sec.\ \ref{llr},
while for
$\sb^{13}$ and $\sb^{23}$
they are given by 
\bea
\sb^{13} &=& \frac 12 \sin \be \sin 2 \al (\sb^{XX}-\sb^{YY})
-\sin \be \cos 2\al \sb^{XY} 
\nonumber\\
&&
+ \cos \be (\cos \al \sb^{XZ} 
+ \sin \al \sb^{YZ}),
\nonumber\\
\sb^{23} &=& -\frac 12 \sin 2\be [\frac 12 (\sb^{XX} +\sb^{YY}) - 
\sb^{ZZ}]
\nonumber\\
&&
+ \frac 12 \sin 2\be \cos 2\al (\sb^{XX}-\sb^{YY}) 
\nonumber\\
&&
+ \frac 12 \sin 2\be \sin 2\al \sb^{XY} 
\nonumber\\
&&
-\sin \al \cos 2\be \sb^{XZ}
+\cos \al \cos 2\be \sb^{YZ}.
\label{s23}
\eea

Using the appropriate inverse matrix,
the solutions for $2\om$ and $\Om_{\oplus}$ can be written as 
\bea
\vec \ep_{2\om,\th-\si_{2\om}} &=& 
\fr {\sb_{2\om}r_0}{12} 
\left(
\begin{array}{c} 
-2 \\ 5 \\ 0 \\
\end{array} 
\right),
\nonumber\\
\vec \ep_{\Om_{\oplus}, \si_{\Om_{\oplus}, 1}} &=& 
-3 V_{\oplus} r_0 \sb_{\Om_{\oplus}, 1} 
\left(
\begin{array}{c} 
b_1/b_2 \\ 2\om/\Om_{\oplus} b_2 \\ 0 \\
\end{array} 
\right),
\nonumber\\
\vec \ep_{\Om_{\oplus}, \si_{\Om_{\oplus}, 2}} &=& 
- \fr {V_{\oplus} r_0}{2} \sb_{\Om_{\oplus}, 2} 
\left(
\begin{array}{c} 
b_1/b_2 \\ 2\om/\Om_{\oplus} b_2 \\ 0 \\
\end{array} 
\right),
\label{epOm}
\eea
where
\bea
b_1 &=& \half -3 a_1 
+ \fr {21 Q \sin^2 \be}{4 \Om_{\oplus}^2 r_0^5},
\nonumber\\
b_2 &=& - \frac{11}{2} + 9 a_1 
-\fr {63Q \sin^2 \be}{4 r^5_0 \Om_{\oplus}^2} ,
\label{b2}
\eea
with $a_1$ defined in Eq.\ \rf{a2}.

Special consideration is required for the near-resonant
frequency $\om$.
The matrix to be inverted for this case is
$K'(\om) = K(\om) -\al_{\rm T} -\al_{\rm Q}$.
To sufficient approximation,
the determinant of this matrix 
can be expanded in terms 
of the anomalistic frequencies $\om_0$ and $\om'_0$.
These are associated with eccentric motions 
in the plane of the orbit and perpendicular to the orbit,
respectively.
The motion involving the phase $\th+\si_{\om, 1}$ 
is confined to the plane of the orbit
and hence is at frequency $\om_0$,
while that involving the phase $\th-\si_{\om, 2}$ 
is perpendicular to the orbit 
and is at frequency $\om'_0$.
We make use of the approximation
\bea
\det K'(\om) &=& \det K'(\om_0 + \om-\om_0) 
\nonumber\\
& \approx & \det K'(\om_0) 
+ (\om-\om_0) \fr {d \det K'(\om_0)}{d\om_0},
\nonumber\\
\label{det}
\eea
along with the similar relation for the frequency $\om'_0$.
This permits the inverse of $K'(\om)$ to be determined as
\bea
[K'(\om)]^{-1} & \approx & 
\fr {1}{2 \om_0 (\om-\om_0)} \left(
\begin{array}{ccc} 
-1 & 2  & 0 \\
2 & -4 & 0 \\
0 & 0 & -\ga \\
\end{array} 
\right) , \quad
\label{ominv}
\eea
where
\beq
\ga = \fr{\om_0 (\om-\om_0)}{ \om'_0 (\om-\om'_0)}.
\label{ratio}
\eeq
In consequence,
the $\om$ amplitudes now read
\bea
\vec \ep_{\om, \th + \si_{\om, 1}} &=& 
\fr {\om}{(\om-\om_0)} \fr {\de m \sb_{\om, 1} v_0 r_0}{M} 
\left(
\begin{array}{c} 
-1 \\ 2 \\ 0 \\
\end{array} 
\right),
\nonumber\\
\vec \ep_{\om, \th-\si_{\om, 2}} &=& 
\fr {\om}{(\om - \om'_0)} \fr {\sb_{\om, 2} r_0}{2} 
\left(
\begin{array}{c} 
0 \\ 0 \\ -1 \\
\end{array} 
\right).
\label{epom2}
\eea

\subsection{Unforced oscillations}
\label{unforced oscillations}

Among the unforced oscillations satisfying Eq.\ \rf{taylor}
are the natural eccentric oscillations of the system
occurring in the absence of Lorentz violation.
These satisfy a set of coupled matrix equations 
that can be derived from Eq.\ \rf{taylor}.
Here,
we are interested in unforced oscillations at the
anomalistic frequency $\om_0$.

In a simplified scenario, 
where contributions from the right-hand side of Eq.\ \rf{taylor}
can be ignored, 
the anomalistic oscillation satisfies
\beq
\lrvec K (\om_0) \cdot \vec \ep_{\om_0} =
(\lrvec \al_{\rm T} + \lrvec \al_{\rm Q}) \cdot \vec \ep_{\om_0}.
\label{unforced1}
\eeq
Requiring the vanishing of the determinant of the matrix
$K'=K-\al_{\rm T}-\al_{\rm Q}$ allows the extraction 
of approximate solutions for $\om_0$.
This gives three solutions,
two of which are physical.
One is denoted $\om_0$ and corresponds to eccentric motion 
in the plane of the orbit.
The other is denoted $\om'_0$ and 
corresponds to eccentric motion 
perpendicular to the orbit.
Explicitly, 
we find the approximate solutions
\bea
\om-\om_0 &\approx& \fr {\Om^2_1}{2\om^2} \om ,
\nonumber\\
\om-\om'_0 &\approx& \fr {\Om^2_2}{2\om^2} \om ,
\label{omp}
\eea
where
\bea
\Om^2_1 &=& \Om_{\oplus}^2 [15 a_1 - 6] + \fr {3Q}{r^5_0} ,
\nonumber\\
\Om^2_2 &=& \Om_{\oplus}^2 [3 a_2 -\frac 32] 
- \fr {3Q}{r^5_0}(1+\frac 14 \sin^2 \be).
\label{Om2}
\eea

More accurate solutions require a detailed study 
of a system of coupled equations involving other frequencies.
For example, 
the eccentric oscillations in the lunar case
are at frequencies $\om_0$ and $2(\om-\Om)-\om_0$
\cite{nanisllr}.
Also,
in the case of a generic satellite orbit,
the Earth's quadrupole terms
can be expected to play a significant role.

For our purposes,
the primary interest is in the eigenvector for $\om_0$,
which is given by
\bea
\vec \ep_{\om_0} &\approx& 
e r_0 
\left(
\begin{array}{c} 
1 \\ -2 \\ 0 \\
\end{array} 
\right).
\label{epom0}
\eea
The overall normalization of this eigenvector 
is defined in terms of the eccentricity of the orbital motion.

Using Eq.\ \rf{epom0},
the leading unforced oscillation due to Lorentz violation
can be established. 
Typically,
the dominant terms occur at frequencies near the resonant frequency,
which here is the circular orbit frequency $\om$.
In this case,
we find that the oscillation at frequency $2\om-\om_0$ 
is particularly enhanced.
In the Sun-centered frame,
the dominant terms contributing to the amplitude
are the frequency-mixing terms in 
$\ep^K \ep^L \prt_K \prt_L \al^J_0 (r_0)/2$
and $\ep^K \prt_K \al^J_{\rm LV}$ 
that lower the frequency by $2\om$.
The former is obtained by substitution of 
the forced-oscillation solution for 
$\ep_{2\om,\th-\si_{2\om}}^K$ 
in Eq.\ \rf{epOm},
while the latter is obtained from
\bea
(\al^{-2\om}_{\sb})_J^{K} &=& \prt_J \Big( \fr {GM \sb^{KL} r^L}{r^3}
- \fr {3G M \sb^{LM} r^L r^M r^K}{2 r^5}\Big).
\nonumber\\
\eea

The resulting matrix equation takes the form 
\bea
\lrvec K' (2\om-\om_0 ) \cdot \vec \ep_{2\om-\om_0,2 \th 
- \si_{2\om}} &=& 
\nonumber\\ 
&& 
\hskip -80 pt
\left(2\lrvec \al^{-2\om}_0 + \lrvec \al^{-2\om}_{\sb} \right) 
\cdot \vec \ep_{2\om-\om_0, 2 \th - \si_{2\om}} ,
\nonumber\\
\label{2om-om0_matrix}
\eea
where the operators on the right-hand side
are defined in the oscillating basis by
\bea
2\lrvec \al^{-2\om}_0 & \approx & 
\fr { \sb_{2\om} \om^2 }{2}
\left(
\begin{array}{ccc} 
1 & \frac 54 & 0 \\
\frac 54 & \frac 12 & 0 \\
0 & 0 & -\frac 12 \\
\end{array} 
\right)
\label{al-2om_0},
\\
\lrvec \al^{-2\om}_{\sb} & \approx & 
\sb_{2\om} \om^2 
\left(
\begin{array}{ccc} 
\frac 12 & 1 & 0 \\
1 & \frac 54 & 0 \\
0 & 0 & -\frac 34 \\
\end{array} 
\right)
\label{al-2om_s}.
\eea

Since the frequency $2\om-\om_0$ is near resonant, 
the procedure to find the inverse of $K'(2\om-\om_0)$ 
can be modeled on that used for the frequency $\om$ 
in Eqs.\ \rf{det}-\rf{ominv}.
We find 
\bea
[K'(2\om-\om_0)]^{-1} & \approx & 
\fr {1}{ 4 \om_0 (\om-\om_0)} \left(
\begin{array}{ccc} 
-1 & 2  & 0 \\
2 & -4 & 0 \\
0 & 0 & -\ga \\
\end{array} 
\right),
\nonumber\\
\label{2om-om0_inv}
\eea
where $\ga$ is defined in Eq.\ \rf{ratio}.
It then follows that the solution 
for the oscillations at frequency $2\om-\om_0$ 
is given by
\bea
\vec \ep_{2\om-\om_0,2\th-\si_{2\om} } &=& 
\fr { \sb_{2\om} \om e r_0}{4(\om-\om_0)} 
\left(
\begin{array}{c} 
-\frac 12 \\ 1 \\ 0 \\
\end{array} 
\right)
\label{ep_2om-om0}.
\eea

\subsection{Radial oscillations}
\label{radial oscillations}

At this stage,
we can extract the information appearing in Table 2 
for radial oscillations 
at the frequencies $\om$, $\Om_{\oplus}$, $2\om$, $2\om-\om_0$.
The procedure is to take 
the $\hat \rh$ component of each column vector 
$\vec \ep_{\nu, \th_\nu}$, 
to multiply by $\cos (\nu T + \th_\nu)$,
and to expand the result in terms of sines and cosines 
using the phase definitions in Eq.\ \rf{phases}.

After some calculation,
we find the following radial oscillations:
\bea
(\ep_{2\om})_\rh &=& -\frac {1}{12} (\sb^{11}-\sb^{22})r_0 
\cos (2\om T + 2\th) 
\nonumber\\
&&
-\frac 16 \sb^{12} r_0 \sin  (2\om T + 2\th),
\nonumber \\
(\ep_{\om})_\rh &=& -\fr {\om}{(\om-\om_0)}\fr {(\de m) v_0 r_0}{M}
[\sb^{02} \cos (\om T + \th)
\nonumber\\
&&
\pt{-{(\om-\om_0)}{(\de m) v_0 r_0}}
-\sb^{01} \sin  (\om T + \th) ],
\nonumber\\
\hskip -10pt
(\ep_{\Om_{\oplus}})_\rh = && 
\hskip -14pt
\fr {b_1}{b_2} V_E r_0 
[\sb_{\Om_{\oplus}c} \cos \Om_{\oplus} T
+\sb_{\Om_{\oplus}s} \sin \Om_{\oplus} T],
\nonumber\\
(\ep_{2\om-\om_0})_\rh &=& -\fr {\om}{(\om-\om_0)}\fr {e r_0}{8} 
\nonumber\\
&&
\times \big(\frac 12 (\sb^{11}-\sb^{22}) \cos [(2\om-\om_0)T + 2\th]
\nonumber\\
&&
\pt{\times}
+\sb^{12} \sin [(2\om-\om_0)T + 2\th] \big).
\label{2om-om0_rad}
\eea
The new combinations of coefficients for Lorentz violation
appearing in these equations are defined by
\bea
\sb_{\Om_{\oplus}c} &=& 
-3 \sb_{\Om_{\oplus},1} \cos \si_{\Om_{\oplus},1} 
- \frac 12  \sb_{\Om_{\oplus},2} \cos \si_{\Om_{\oplus},2} ,
\nonumber\\
\sb_{\Om_{\oplus}s} &=& 
-3 \sb_{\Om_{\oplus},1} \sin \si_{\Om_{\oplus},1} 
- \frac 12  \sb_{\Om_{\oplus},2} \sin \si_{\Om_{\oplus},2}.
\nonumber\\
\eea

\end{document}